\documentclass[11pt]{article}
\pdfoutput=1
\usepackage{jcapmod}

\usepackage{amsmath, amsfonts}
\usepackage{booktabs}
\usepackage[english]{babel}
\usepackage{graphicx}
\usepackage{multirow}
\usepackage{subfig}
\usepackage{hyperref}
\usepackage[range-units=brackets, range-phrase=-, per-mode=reciprocal, mode=math]{siunitx}
\usepackage[babel=true]{microtype}

\numberwithin{equation}{section}
\allowdisplaybreaks[1]
\def\beq{\begin{equation}}
\def\eeq{\end{equation}}

\def\d{\mathrm{d}}
\def\ee{\mathrm{e}}
\def\ii{\mathrm{i}}

\def\k{{\vec{k}}}
\def\p{{\vec{p}}}
\def\q{{\vec{q}}}
\def\r{{\vec{r}}}
\def\x{{\vec{x}}}
\def\va{{\vec{\alpha}}}

\def\As{A_\mathrm{s}}
\def\ns{n_\mathrm{s}}
\def\kp{k_\star}

\def\Pnw{P^\mathrm{nw}}
\def\Pw{P^\mathrm{w}}
\def\L{\mathcal{L}}

\def\fsky{{f_\mathrm{sky}}}
\def\lmax{\ell_\mathrm{max}}
\def\kmin{k_\mathrm{min}}
\def\kmax{k_\mathrm{max}}
\def\zmax{z_\mathrm{max}}

\def\BAO{\mathrm{BAO}}
\def\Asin{A^\mathrm{sin}}
\def\Acos{A^\mathrm{cos}}
\def\lin{\mathrm{lin}}
\def\omegalin{\omega_\lin}
\def\Alin{A_\lin}
\def\Alinsin{\Alin^\mathrm{sin}}
\def\Alincos{\Alin^\mathrm{cos}}
\def\phase{\varphi}
\def\phaselin{\phase_\lin}
\def\Log{\mathrm{log}}
\def\omegalog{\omega_\Log}
\def\Alog{A_\Log}
\def\Alogsin{\Alog^\mathrm{sin}}
\def\Alogcos{\Alog^\mathrm{cos}}
\def\phaselog{\phase_\Log}

\DeclareSIUnit{\parsec}{pc}
\DeclareSIUnit{\Mpc}{\mega\parsec}
\DeclareSIUnit{\Gpc}{\giga\parsec}
\DeclareSIUnit{\h}{\mathit{h}}
\DeclareSIUnit{\hPerMpc}{\h\per\Mpc}
\DeclareSIUnit{\MpcPerh}{\per\h\Mpc}

\definecolor{Blue}{rgb}{0.25, 0.41, 0.88}
\definecolor{Red}{rgb}{0.92,0.,0.}

\DeclareRobustCommand{\SkipTocEntry}[4]{}

\begin{document}

\pagenumbering{roman}
\begin{titlepage}
	\baselineskip=15.5pt \thispagestyle{empty}
	
	\bigskip\
	
	\vspace{1cm}
	\begin{center}
		{\fontsize{20.74}{24}\selectfont \sffamily \bfseries Primordial Features\\[10pt]from Linear to Nonlinear Scales}
	\end{center}
	
	\begin{center}
		{\fontsize{12}{30}\selectfont Florian Beutler,$^{1,2}$ Matteo Biagetti,$^{3}$ Daniel Green,$^{4}$\\[4pt] An\v{z}e Slosar$^{5}$ and Benjamin Wallisch$^{6,7,3,4}$} 
	\end{center}
	
	\begin{center}
		\vskip8pt
		\textsl{$^1$ Institute of Cosmology \& Gravitation, University of Portsmouth, Portsmouth, PO1 3FX, UK}

		\vskip8pt
		\textsl{$^2$ Lawrence Berkeley National Laboratory, 1 Cyclotron Road, Berkeley, CA 94720, USA}
		
		\vskip8pt
		\textsl{$^3$ Institute for Theoretical Physics, University of Amsterdam, 1098 XH Amsterdam, NL}
		
		\vskip8pt
		\textsl{$^4$ Department of Physics, University of California, San Diego, La Jolla, CA 92093, USA}
		
		\vskip8pt
		\textsl{$^5$ Physics Department, Brookhaven National Laboratory, Upton, NY 11973, USA}
		
		\vskip8pt
		\textsl{$^6$ School of Natural Sciences, Institute for Advanced Study, Princeton, NJ 08540, USA}
		
		\vskip8pt
		\textsl{$^7$ DAMTP, University of Cambridge, Cambridge CB3 0WA, UK}
	\end{center}

	\vspace{1.2cm}
	\hrule \vspace{0.3cm}
	\noindent {\sffamily \bfseries Abstract}\\[0.1cm]
	Sharp features in the primordial power spectrum are a powerful window into the inflationary epoch. To date, the cosmic microwave background~(CMB) has offered the most sensitive avenue to search for these signatures. In this paper, we demonstrate the power of large-scale structure observations to surpass the~CMB as a probe of primordial features. We show that the signatures in galaxy surveys can be separated from the broadband power spectrum and are as robust to the nonlinear evolution of matter as the standard baryon acoustic oscillations. As a result, analyses can exploit a significant range of scales beyond the linear regime available in the datasets. We develop a feature search for large-scale structure, apply it to the final data release of the Baryon Oscillation Spectroscopic Survey and find new bounds on oscillatory features that exceed the sensitivity of Planck for a significant range of frequencies. Moreover, we forecast that the next generation of galaxy surveys, such as~DESI and Euclid, will be able to improve current constraints by up to an order of magnitude over an expanded frequency range.
	\vskip10pt
	\hrule
	\vskip10pt
\end{titlepage}

\thispagestyle{empty}
\setcounter{page}{2}
\tableofcontents

\clearpage
\pagenumbering{arabic}
\setcounter{page}{1}

\clearpage
%%%%%%%%%%%%%%%%
\section{Introduction}
\label{sec:introduction}
%%%%%%%%%%%%%%%%

Characterizing the nature of inflation is one of the major challenges in cosmology. While current data are compatible with the simplest incarnation of inflation, a single weakly-coupled scalar field on a very flat potential, the space of possibilities for inflation are vast and should ultimately be settled by data. An appealing aspect of the simplest models is that the observed (near) scale invariance of the power spectrum of fluctuations is easily explained by the flatness of this potential~\cite{Akrami:2018odb}. However, attempts to realize inflation from a more fundamental starting point can lead to much more complicated models where many interconnected pieces are needed to achieve scale invariance~\cite{Chen:2010xka, Chluba:2015bqa}. This dichotomy between the simplest models and ultraviolet-complete examples originates from quantum gravity itself: sufficiently flat potentials can be engineered using symmetries, but quantum gravity famously abhors them~\cite{Baumann:2009ds, Baumann:2014nda}. As a result, a wide variety of models gives rise to non-trivial deviations from canonical slow-roll and scale invariance (i.e.~features)~\cite{Chluba:2015bqa, Slosar:2019gvt}. Other mechanisms avoid this picture altogether by invoking non-trivial interactions that can lead to non-Gaussian $n$-point correlation functions (primordial non-Gaussianity)~\cite{Meerburg:2019qqi}.

\vskip4pt
Our ability to test these ideas directly with data relies on separating the primordial signatures of interest from a broad range of processes at late times. This challenge is particularly acute for constraints on inflation from large-scale structure~(LSS) surveys since the observed objects, i.e.~galaxies, owe their existence to the nonlinear gravitational evolution of matter fluctuations in the late universe. While both current and future galaxy surveys have the raw statistical power to compete with other cosmological probes, such as the cosmic microwave background~(CMB), the useful information is greatly diminished if we restrict our analyses to modes that are sufficiently linear to use forward modeling in order to isolate the primordial information. There have been steady improvements in this direction, but eventually these techniques are expected to be limited by the complexity of (astro)physics at short distances~\cite{Alvarez:2014vva, Baldauf:2016sjb}.

\vskip4pt
An alternate and already successful approach is to look for special observables that are (at least partially) immune to the complications presented by LSS~data. The best-known example of this type are the baryon acoustic oscillations~(BAO). Although they manifest themselves as an oscillation in the power spectrum in a range of wavenumbers sensitive to nonlinearities, it is more usefully understood as a sharp peak in the two-point correlation function at the size of the sound horizon, which is a scale that is much larger than the scale where nonlinear evolution dominates~\cite{Eisenstein:2006nj}. More recently, a constant phase of the baryon acoustic oscillations was shown to be immune to nonlinear evolution~\cite{Baumann:2017lmt}. These are useful examples as they show that smooth and oscillatory power spectra are not sensitive to the same nonlinear effects.
 
\vskip4pt
Nevertheless, it has proven challenging to constrain the most common inflationary parameters (e.g.~the scalar spectral index~$\ns$ or its running~$\alpha_\mathrm{s}$) by~LSS. Changes to these parameters typically lead to smooth variations in the power spectrum (as a function of wavenumber~$k$) and are therefore degenerate with other contributions such as galaxy biasing and baryonic effects. Furthermore, most inflationary observables get their constraining power from the smallest physical scales accessible in a given survey where gravitational nonlinearities dominate. With current data, large-scale structure is most competitive with the~CMB as a probe of inflation for constraints on local primordial non-Gaussianity~\cite{Slosar:2008hx, Leistedt:2014zqa, Castorina:2019wmr}. In this case, the non-Gaussian signature in the initial conditions manifests itself in the biasing of galaxies on the largest scales where nonlinearities are negligible~\cite{Dalal:2007cu} (see~\cite{Biagetti:2019bnp} for a recent review). Future surveys will search these large scales with increasing sensitivity and have the potential to ultimately exceed the~CMB~\cite{Dore:2014cca, Schmittfull:2017ffw}.

\begin{figure}
	\centering
	\includegraphics{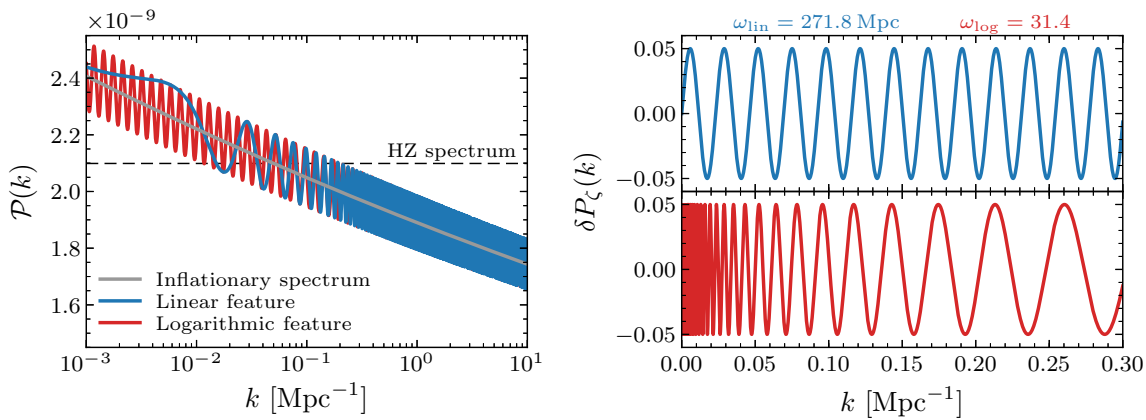}
	\caption{Illustration of linear and logarithmic oscillations in the primordial power spectrum~$\mathcal{P}(k)$. The employed frequencies~$\omega_X$ are comparable in the range of scales $k=\SIrange{0.1}{0.2}{\hPerMpc}$ where~BOSS has the largest signal-to-noise ratio. We see that the effective frequency of the logarithmic oscillations decreases as we go to larger wavenumbers~$k$.}
	\label{fig:featureVisualisation}
\end{figure}

\vskip4pt
In this paper, we demonstrate that features in the primordial spectra, much like the standard BAO~signal itself, are immune to short-distance nonlinear processes of the late universe and the effects of large-scale bulk flows can be captured analytically. We can therefore test these models with the full statistical power of LSS~surveys. Phenomenologically, primordial features are generally characterized by significant deviations from scale invariance over a narrow range of scales, usually in the power spectrum. The most canonical examples, shown in Fig.~\ref{fig:featureVisualisation}, %
are written as oscillations in either~$k$ or~$\log k$~\cite{Slosar:2019gvt}:
\beq
P_\zeta(k) = P^s(k)\, \big[ 1 + \Alin \sin(\omegalin k + \phaselin) + \Alog \sin(\omegalog \log(k/\kp) + \phaselog) \big]\, ,	\label{eq:Pzeta}
\eeq
where~$P_\zeta(k)$ is the primordial power spectrum of adiabatic density fluctuations~($\zeta$) and $P^s(k)$~is a smooth function of~$k$. For a linear oscillation~($\Alin$), the insensitivity to nonlinear effects is identical to the case of the baryon acoustic oscillations: if we Fourier transform the signal, the linear oscillation is a sharp peak at the scale~$\omegalin$ in the two-point correlation function. For sufficiently large~$\omegalin$, one is effectively looking for a second BAO~peak. For logarithmic oscillations~($\Alog$), there is not a simple description in terms of scales, but we will show that local nonlinear evolution is incapable of producing the same oscillation for sufficiently large frequencies~$\omegalog$.

While the analogy with the BAO~signal is useful, it is worth observing that the primordial features arise directly in the initial conditions of the dark matter and baryons, and are suppressed only by the amplitude of the oscillation. By contrast, the baryon acoustic oscillations themselves are a consequence of the physics of baryons. This means that their impact on LSS~data is suppressed by $\omega_b/\omega_m$ and the growth of dark matter density fluctuations prior to recombination. As a result, the amplitude of the BAO~spectrum is suppressed relative to a primordial feature and is roughly equivalent to a linear feature amplitude of $\Alin = 0.05$. The fact that we do not see an additional oscillation beyond the BAO~signal by eye already suggests that $\Alin \ll 0.1$ without any analysis.

In addition to being protected quantities in~LSS, primordial features of this form are a well-motivated probe of the early universe in their own right~\cite{Starobinsky:1992ts, Adams:2001vc, Ashoorioon:2006wc, Bean:2008na, Chen:2008wn, Ashoorioon:2008qr, Barnaby:2009dd, Flauger:2009ab, Achucarro:2010da, Chen:2011zf, Miranda:2012rm, Achucarro:2012fd, Bartolo:2013exa, Green:2014xqa, Chen:2015lza, Slosar:2019gvt}. They notably arise in axion monodromy inflation~\cite{McAllister:2008hb} as a direct consequence of the fundamental symmetry structure needed to produce large-field inflation. The periodic, nonperturbative potential generated for axions gives rise to an oscillation in the inflationary potential and manifests itself as a logarithmic oscillation in the power spectrum~\cite{Flauger:2009ab}. In addition, there are a number of scenarios where particles are excited from the vacuum at a specific time through non-adiabatic evolution and can give rise to linear oscillations. Moreover, most features in the power spectrum can be efficiently decomposed in this basis of functions which therefore captures large parts of model space.

\vskip10pt
The outline of the paper is as follows. In Section~\ref{sec:theory}, we show that features are robust to small-scale nonlinearities and compute the nonlinear damping effect due to long-wavelength modes. In Section~\ref{sec:featureSearch}, we introduce our new analysis to search for these features in LSS~data and verify that these oscillatory signals can be reliably constrained. In Section~\ref{sec:dataAnalysis}, we apply this pipeline to BOSS~DR12 data and present a new constraint that exceeds Planck over a significant range of frequencies. Moreover, we forecast the sensitivity of future observational surveys (cf.\ e.g.~\cite{Chen:2016zuu, Chen:2016vvw, Ballardini:2016hpi, Xu:2016kwz, Finelli:2016cyd, Fard:2017oex, Hazra:2017joc, Palma:2017wxu, LHuillier:2017lgm, Ballardini:2017qwq, Ballardini:2018noo, Sohn:2019rlq} for previous work). We conclude in Section~\ref{sec:conclusions}. Additional details on the theoretical calculation, the employed forecasts, and the performed LSS~and CMB~analyses are provided in a set of four appendices.

%%%%%%%%%%%%%%%%
\section{Primordial Features and Galaxy Surveys}
\label{sec:theory}
%%%%%%%%%%%%%%%%

In this section, we determine how a primordial feature will appear in the nonlinear (low-redshift) universe. We first characterize the signals in the linear matter power spectrum. Then, we will use a linear-response argument to show that nonlinear evolution on small scales does not change the amplitude of the feature in the nonlinear power spectrum. Finally, we will use infrared resummation to determine the nonlinear damping of the features from large-scale modes.

\subsection{Oscillatory Features in the Primordial Spectrum}
\label{sec:features}

The physics of inflation determines the primordial power spectrum~$P_\zeta(k)$. In the simplest versions of inflation, this power spectrum arises from the freeze-out of quantum mechanical fluctuations when the physical wavelength reaches the Hubble radius, $k = a H(t_\star)$, where~$k$ is the constant comoving wavenumber, $a(t)$~is the scale factor and $H(t)$~is the Hubble parameter. (This equation defines the freeze-out time~$t_\star$.) As a result, the amplitude of fluctuations for a comoving wavenumber~$k$ is determined by the physics of inflation around the time~$t_\star$. Furthermore, the near scale invariance of the resulting spectrum is a consequence of the weak time dependence in the evolution of the perturbations during inflation.

Scale-dependent features in the primordial spectrum therefore arise from strongly time-dependent physics during inflation.\footnote{We focus on the inflationary origin of features, but note that they might also appear in alternatives to inflation~\cite{Chen:2018cgg}.} This may be due to sharp features in the underlying potential for a scalar field, or special locations in the field space of the inflaton where other particles become light and can be excited from the vacuum. Despite the wide range of possibilities, the signatures do not significantly depend on the details of the model since the nature of the time dependence controls the deviations from scale invariance.

\vskip4pt
We will assume that the smooth spectrum~$P^s(k)$ of~\eqref{eq:Pzeta} is the almost scale-invariant power spectrum of curvature perturbations in vanilla models of inflation, 
\beq
P_{\zeta,0}(k) = \frac{2\pi^2}{k^3} \mathcal{P}_{\zeta,0}(k) = \frac{2\pi^2 \As}{k^3} \left(\frac{k}{\kp}\right)^{\!\ns-1}\, ,
\eeq
where~$\As$ and~$\ns$ are the scalar amplitude and spectral index at the pivot scale~$\kp$, which we generally take to be $\kp = \SI{0.05}{\per\Mpc}$. We then write the full power spectrum~$P_\zeta(k)$ including the contribution from features,~$\delta P_\zeta(k)$, as
\beq
P_\zeta(k) = P_{\zeta,0}(k) \left[ 1 + \delta P_\zeta(k) \right] .	\label{eq:primordialSpectrum}
\eeq
As suggested in~\eqref{eq:Pzeta}, we will consider oscillatory features with linearly-spaced oscillations,~$\delta P_\zeta^\lin$, and logarithmically-spaced oscillations,~$\delta P_\zeta^\Log$. We phenomenologically parameterize the former, which we refer to as linear features, as follows:\footnote{Note that the feature frequency is sometimes defined in the literature as $\tilde{\omega}_\lin = \omegalin/2$ or with respect to a pivot scale~$\kp$. More generally, we highlight that~$\omegalin$ is a frequency in Fourier space which corresponds to a physical scale in real space.}
\beq
\begin{aligned}	
	\delta P_\zeta^\lin(k)	&= \Alinsin \sin\!\left(\omegalin k\right) + \Alincos \cos\!\left(\omegalin k\right)	\\
							&= \Alin \sin\!\left(\omegalin k + \phaselin\right) ,
\end{aligned} \label{eq:linearFeatures}
\eeq
with the feature frequency~$\omegalin$, and the amplitudes of the sine and cosine contributions~$\Alinsin$ and~$\Alincos$, respectively, or the overall feature amplitude~$\Alin$ and corresponding phase~$\phaselin$. The so-called logarithmic features are similarly defined as
\beq
\begin{aligned}
	\delta P_\zeta^\Log(k) 	&= \Alogsin \sin\!\left[\omegalog \log(k/\kp)\right] + \Alogcos \cos\!\left[\omegalog \log(k/\kp)\right]	\\
							&= \Alog \sin\!\left[\omegalog \log(k/\kp) + \phaselog\right] ,
\end{aligned} \label{eq:logarithmicFeatures}
\eeq
with the feature frequency~$\omegalog$, and the amplitudes of the sine and cosine contributions~$\Alogsin$ and~$\Alogcos$, respectively, or the overall feature amplitude~$\Alog$ and corresponding phase~$\phaselog$. We refer to the parameterization in terms of two amplitudes as `amplitude parameterization' and the one in terms of the overall amplitude and a phase as `phase parametrization'. We note that it has been customary in the literature to define the linear feature frequency~$\omegalin$ as a dimensionful quantity in units of~\si{\Mpc}, whereas the logarithmic feature frequency~$\omegalog$ is dimensionless. In addition, we remark that the feature amplitudes give the contribution relative to the standard power spectrum~$P_{\zeta,0}$ and the feature contribution~$\delta P_\zeta(k)$ oscillates around zero by construction.

\begin{figure}
	\centering
	\includegraphics{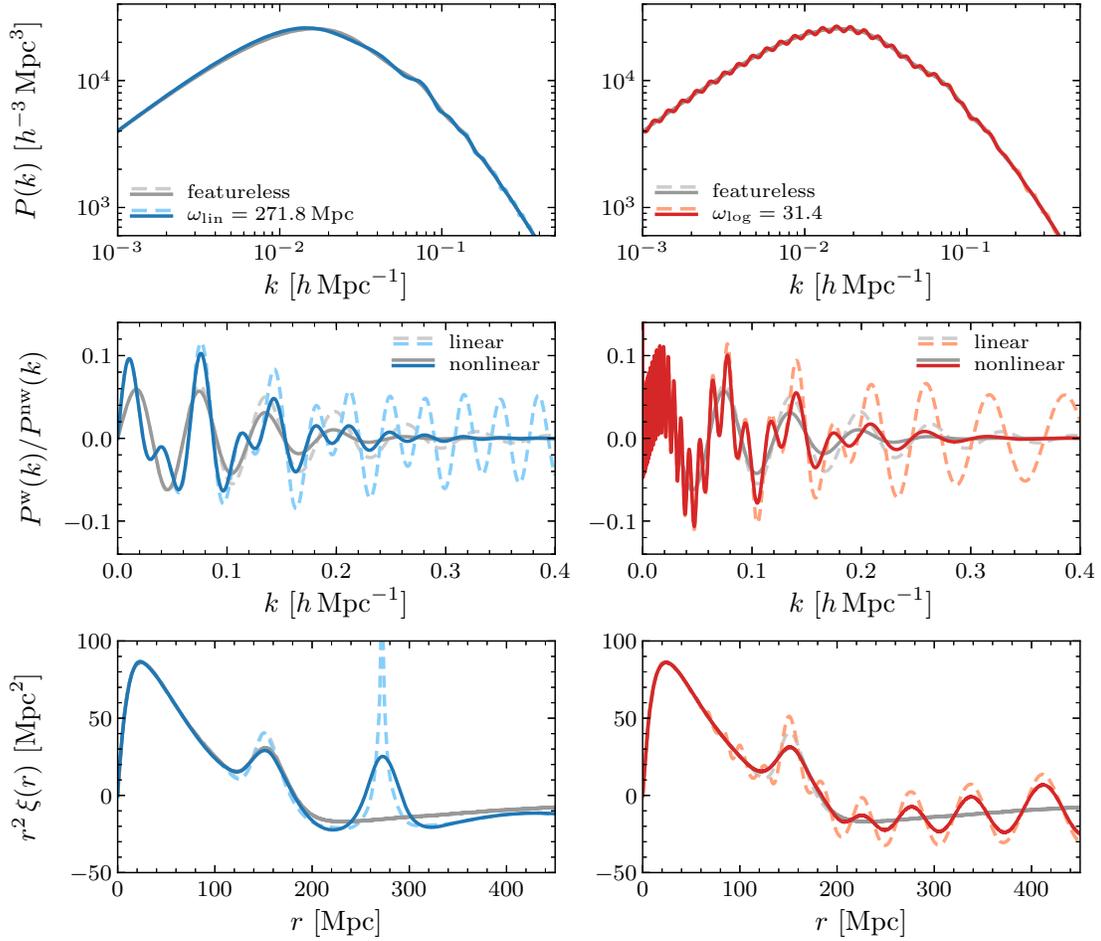}
	\caption{Imprint of primordial features in several large-scale structure observables at redshift \mbox{$z=0$}: the matter power spectrum~$P(k)$~(\textit{top}) and the relative wiggle spectrum~$\Pw(k)/\Pnw(k)$~(\textit{middle})\ in Fourier space, and the (rescaled) two-point correlation function~$\xi(r)$ in real space~(\textit{bottom}). We compare a featureless model~(gray) to scenarios involving a linear feature~(\textit{left}) and a logarithmic feature~(\textit{right}) with $\Asin_X = 0.05$ and $\Acos_X=0$, $X=\lin,\Log$. In addition to the predictions in linear theory~(dashed), we also show the observables including nonlinear corrections~(solid) from a theoretical calculation of the damping.}
	\label{fig:observablesComparison}
\end{figure}

\vskip4pt
The information in the primordial power spectrum is transferred to the matter power spectrum, as illustrated in Fig.~\ref{fig:observablesComparison}, %
through the usual linear evolution from initial conditions,
\beq
P(k,z) = k^4\, T(k)^2\, D(z)^2\, P_\zeta(k)\, ,
\eeq
where~$D(z)$ is the linear growth rate and $T(k)$~the transfer function. For simplicity, we omit the redshift dependence of the matter power spectrum, $P(k) = P(k,z)$, from now on. For large enough feature frequencies, these oscillations can be distinguished from the broadband shape of the power spectrum, similar to the baryon acoustic oscillations. It is therefore natural to constrain the feature models as contributions to the BAO~spectrum, which is why we split the power spectrum into a smooth (`no-wiggle') part and an oscillatory (`wiggle') part,
\beq
P(k) \equiv \Pnw(k) + \Pw(k)\, .	\label{eq:powerSpectrumDecomposition}
\eeq
Since primordial features with large enough frequencies are only contained in the second term, it is useful to further decompose the wiggle spectrum as follows:
\beq
\Pw(k) \equiv \Pw_\BAO(k)+ \Pw_X(k)+ \Pw_\BAO(k)\, \delta P^X_\zeta(k) \, ,	\label{eq:pw}
\eeq
where~$\Pw_\BAO(k)$ is the standard BAO~spectrum in a featureless $\Lambda$CDM~cosmology, the auto-spectrum of possible primordial features is
\beq
\Pw_X(k) = \Pnw(k) \delta P_\zeta^X(k) \, ,	\label{eq:pwx}
\eeq
with $X=\lin,\Log$, and the third term is the BAO-feature cross-correlation power spectrum. Since the BAO~signal itself is only a small (five-percent) contribution to~$P(k)$, we will be able to generally neglect this cross-spectrum term in our theoretical considerations. Given that~\eqref{eq:powerSpectrumDecomposition} is the linear matter power spectrum, we will show in the following two sections that small-scale modes do not affect~$\Pw(k)$ for large enough feature frequencies~(\textsection\ref{sec:smallScaleNonlinearities}), but that each of its oscillatory components in~\eqref{eq:pw} is affected by gravitational large-scale nonlinearities and exponentially damped~(\textsection\ref{sec:largeScaleNonlinearities}).

\subsection{Robustness of Features to Small-Scale Nonlinearities}
\label{sec:smallScaleNonlinearities}

The smallest scales in an LSS~survey carry most of the statistical power, but are also the most prone to nonlinear corrections, including galaxy bias and baryonic effects. For this reason, a typical analysis might cut at $k=\SIrange{0.1}{0.2}{\hPerMpc}$ to avoid the complications of modeling and marginalizing over these effects. In the case of the baryon acoustic oscillations, it has long been known that they are robust to those effects which change the power spectrum only by a smooth window function and that it is possible to aggressively marginalize over these smooth corrections without losing any information. In the following, we show that high-frequency oscillatory features in the power spectrum are protected from small-scale nonlinearities in precisely the same fashion as the BAO~signal. We first give an intuitive argument that this is indeed the case which we then confirm more rigorously.

\subsubsection{Intuitive argument}
\begin{figure}
	\centering
	\includegraphics{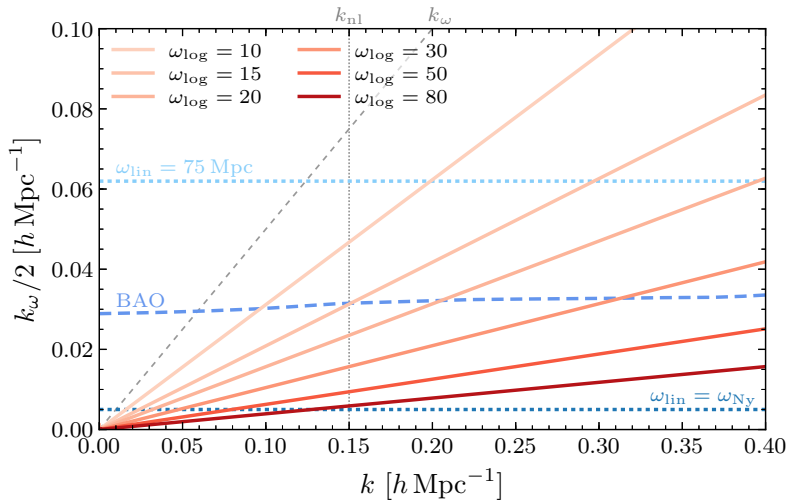}
	\caption{Separation of zeros (or, equivalently, peaks and troughs) in the spectrum of logarithmically-spaced oscillations~\eqref{eq:logarithmicFeatures} as a function of wavenumber~$k$ for a range of frequencies~$\omegalog$. For comparison, the separations in the standard BAO~spectrum and for linear oscillations with $\omegalin=\SI{75}{\Mpc}$ and $\omegalin=\omega_\mathrm{Ny} \approx \SI{929}{\Mpc}$, which is the Nyquist frequency in our BOSS~analysis below, are also shown. The dotted and dashed gray lines indicate the nonlinear scale~$k_\mathrm{nl} \approx \SI{0.15}{\hPerMpc}$ and $k_\omega = k$, respectively. The former marks the approximate wavenumber where nonlinearities are expected to become large, $k \gtrsim k_\mathrm{nl}$, while these nonlinearities do not alter features with $k_\omega \lesssim k$. The zero separations are denoted by~$k_\omega/2$ since they are equivalent to half of a feature oscillation period.}
	\label{fig:logsep}
\end{figure}

The power spectrum of linear oscillations~$\Pw_\lin(k)$ as defined in~\eqref{eq:linearFeatures} and~\eqref{eq:pwx} is the same as the approximate form of the BAO~signal, where the linear feature frequency~$\omegalin$ corresponds to the sound horizon~$r_s$ and the feature phase~$\phaselin$ is the phase shift due to free-streaming neutrinos~\cite{Baumann:2017lmt, Baumann:2017gkg, Baumann:2018qnt}.\footnote{The real form of the BAO~spectrum has an additional $k$-dependence in the phase and amplitude~\cite{Baumann:2017gkg} which are however unimportant for this analogy.} It is a long-established fact that the baryon acoustic oscillations are essentially immune from nonlinearities on small scales because the sound horizon is a large-distance scale. The same will be true for any linear oscillation with~$\omegalin$ larger than the BAO~scale~$r_s$. Smaller frequencies will remain immune to nonlinearities provided that they are sufficiently large compared to the nonlinear scale. Conservatively, we will use $\omegalin \gtrsim \SI{75}{\Mpc}$ as a reference value. As illustrated in Fig.~\ref{fig:logsep}, %
this frequency is consistent with our more rigorous constraint on the scales where small-scale nonlinearities are not important (cf.~\textsection\ref{sec:smallScaleNonlinearities_rigorous}). More recently, it has been shown that the phase~$\phaselin$ is also protected~\cite{Baumann:2017lmt} which implies that this argument will hold for both the sine and cosine contributions.

\vskip4pt
A priori, it is less clear that logarithmic oscillations will share the same nice properties. Unlike the linear frequency~$\omegalin$, the frequency~$\omegalog$ of the logarithmically-spaced oscillations is a dimensionless parameter and, therefore, does not refer to any fixed physical scale in either Fourier or configuration space. As a result, the signal must appear on all scales in both descriptions and, consequently, is not clearly distinct from nonlinear effects. Fortunately, in practice, our information only comes from a limited range of scales which is given by $k \sim \SIrange{0.1}{0.3}{\hPerMpc}$ for~BOSS. We can always (Fourier) decompose any feature into a sum over linear oscillations. For sufficiently large~$\omegalog$, the signal in this range of wavenumbers is reliably reproduced keeping only the linear oscillations with frequencies large enough to be protected by our previous argument. We illustrate this argument in Fig.~\ref{fig:logsep}, where we show the peak-trough separation for the logarithmically-oscillating power spectrum $\Pw_\Log(k)$ defined in~\eqref{eq:logarithmicFeatures} and~\eqref{eq:pwx} as a function of~$k$ in our range of interest. For $\omegalog \gtrsim 20$, the peak-trough separation is at the level of the BAO~spectrum, while nonlinearities are expected to affect the power spectrum at smaller scales. For the purposes of an analysis of BOSS~data,\footnote{In future surveys covering a larger range of~$k$, more work may be required to account for the fact that the effective frequency evolves with~$k$ and may not be protected over the full range of scales. Moreover, these surveys will generally require more theoretical control since they will be sensitive to smaller signals and, therefore, subleading effects.} the logarithmic features therefore do not present any significant complications.

\subsubsection{Rigorous argument}
\label{sec:smallScaleNonlinearities_rigorous}
We can rigorously derive these results using the techniques developed in~\cite{Baumann:2017lmt}. Since the amplitude is known to be small (percent level at best), we can find the nonlinear density fluctuations as the linear response in the feature amplitude ($\Alin$~or~$\Alog$). On general grounds, this response must take the form~\cite{Baumann:2017lmt}
\beq
\delta^\mathrm{w}(\x,\tau) =\int\! \d^3 x' \, G({\color{Red}\x}, {\color{Blue}\x-\x\hskip 1pt'}; \tau) \,\delta_\mathrm{in}^\mathrm{w}(\x\hskip 1pt')\, ,	\label{eq:linear_pos}
\eeq
where~$G({\color{Red}\x}, {\color{Blue}\x-\x\hskip 1pt'}; \tau)$ is the response function and~$\delta_\mathrm{in}^\mathrm{w}(\x')$ is the contribution to the initial density contrast at linear order in~$A_X$, $X=\lin,\Log$. The response function is independent of the oscillatory signal and has previously been studied for more general applications~\cite{Neyrinck:2013iza, Nishimichi:2014rra, Barreira:2017kxd}. The first index in red,~${\color{Red}\x}$, arises from the underlying inhomogeneity of the universe and would be absent in a translation-invariant system. The second index in blue,~${\color{Blue}\x-\x\hskip 1pt'}$, characterizes the propagation of information from one point to another.

Unlike for the BAO~spectrum, we are not concerned that nonlinear evolution will make a small change to the frequency or phase. Since nonlinear evolution is not expected to create such a frequency out of nothing, the primary concern is that nonlinear evolution will make a large change to the amplitude by some incalculable amount, making it impossible to relate bounds on the nonlinear spectrum to the amplitude in the initial conditions. To be dangerous, this change must specifically alter the amplitude of a high-frequency oscillation relative to the amplitude of the underlying smooth nonlinear matter power spectrum. To proceed, we Fourier transform~\eqref{eq:linear_pos} to arrive at
\beq
\delta^\mathrm{w}(\k, \tau) = \int\! \frac{\d^3 q}{(2 \pi)^3}\, G({\color{Red}\k-\q}, {\color{Blue}\q\hskip 2pt}; \tau)\, \delta_\mathrm{in}^\mathrm{w}(\q)\, .
\eeq
Note that it is the wavevector ${\color{Red}\k-\q} \equiv {\color{Red}\p}$ that characterizes the scale of the inhomogeneities. Following~\cite{Baumann:2017lmt}, we will define a scale $k_\omega \ll k$ as the approximate period of oscillations in the power spectrum (e.g.~$k_\omega = 2\pi/\omegalin$). We will separate the domains $p > k_\omega$ and $p < k_\omega$ to distinguish the ``small-scale'' and ``large-scale'' inhomogeneities, respectively.

Using an argument from~\cite{Baumann:2017lmt}, it is easy to see that small-scale inhomogeneities do not contribute an oscillatory signal in the nonlinear matter power spectrum.\footnote{We note that we would have $G({\color{Red}\p}, {\color{Blue}\q\hskip 2pt}; \tau) = \delta({\color{Red}\p}\hskip1pt) \tilde G({\color{Blue}q})$ if the matter distribution was in the linear regime. As a result, the absence of an oscillatory signal from $p > k_\omega$ in the nonlinear case is not a suppression of the oscillation in the initial conditions, but the absence of a correction.} Suppose now that $p > k_\omega$ did contribute a high-frequency oscillatory signal. This would imply that we should see a large change in $\delta^\mathrm{w}(k)$ if we shifted $\k \to \k + \va k_\omega$, where $|\vec{\alpha}| \sim \mathcal{O}(1)$. However, recall that $G({\color{Red}\p}, {\color{Blue}\q\hskip 2pt}; \tau)$ is determined from the nonlinear density field in a universe \textit{without} the oscillatory signals and should therefore be a smooth function of~$\p$ and~$\q$. As a result, we can Taylor expand~$G({\color{Red}\k + \va k_\omega-\q}, {\color{Blue}\q\hskip 2pt}; \tau)$ in~$\va$ to find
\begin{align}
\delta^\mathrm{w}_{p > k_\omega}(\k + \alpha \k, \tau)	&= \int\! \frac{\d^3 q}{(2 \pi)^3}\, G({\color{Red}\k +\va k_\omega -\q}, {\color{Blue}\q\hskip 2pt}; \tau)\, \delta_\mathrm{in}^\mathrm{w}(\q) \nonumber	\\ 
											&\approx \int\! \frac{\d^3 q}{(2 \pi)^3} \left[ G({\color{Red}\k -\q}, {\color{Blue}\q\hskip 2pt}; \tau) + k_\omega \va \cdot \vec{\nabla}_{\k} G({\color{Red}\k -\q}, {\color{Blue}\q\hskip 2pt}; \tau) \right] \delta_\mathrm{in}^\mathrm{w}(\q)	\\ 
											&\approx \delta^\mathrm{w}_{p >k_\omega}(\k, \tau) + \mathcal{O}\left(\frac{\alpha k_\omega}{k} \right) ,	\nonumber
\end{align}
where we used $\nabla_{\k} \sim k^{-1}$ because~$G$ is a smooth function of~${\color{Red}\k -\q}$. From this argument, we see that~$\delta^\mathrm{w}_{p > k_\omega}(\k, \tau)$ is a smooth function of~$k$. This means that it does not change rapidly over one period of the initial oscillations and, therefore, does not contribute to the oscillatory signal. In other words, small-scale nonlinearities do not change the amplitude of features in the power spectrum provided that they have a large enough frequency.

\vskip4pt
To conclude the discussion of small-scale nonlinearities, we note that this result is independent of the precise shape of the initial feature. It only requires that the feature, $\delta^\mathrm{w}(k)$, changes by order one over a very small range of~$\k$. We do not require that it is sinusoidal or that it is associated with a large physical scale in configuration space. Furthermore, the result that $k_\omega / k \ll 1$ is not altered by small-scale nonlinearities is the same condition which requires that the oscillation is distinct from a smooth polynomial. The power spectrum on a scale~$k$ is ``smooth'' if
$\frac{\partial \log P}{\partial \log k} \lesssim 1$. This implies that our features will be ``sharp'' if
\beq
\frac{\partial \log \Pw_X(k)}{\partial \log k}\gg 1 \quad \to \quad k \omegalin \gg 1\, , \quad \omegalog \gg 1 \, .
\eeq
These conditions are illustrated in the left panel of Fig.~\ref{fig:featureVisualisation}, where the linear oscillations at small wavenumbers are smooth, while the logarithmic feature is sharp on all scales. Of course, implicit in this discussion is that~$k$ is a scale where nonlinearities are important. In our universe, nonlinear effects are strongly $k$-dependent and, therefore, primarily affect modes near the nonlinear scale, $k \approx k_\mathrm{nl}$.

\subsection{Damping from Large-Scale Nonlinearities}
\label{sec:largeScaleNonlinearities}

In the previous section, we established that short-wavelength inhomogeneities cannot alter the appearance of an oscillatory signal in the initial conditions. We now turn to the long-wavelength modes. In general, it is hard to compute the consequences of nonlinearities on the matter power spectrum from first principles. Nevertheless, it has been shown that the nonlinear effects of large-scale modes on the BAO~spectrum can be computed and resummed in perturbation theory, resulting in a damping of the amplitude and shape of the standard BAO~signal (cf.\ e.g.~\cite{Eisenstein:2006nj, Crocce:2007dt, Creminelli:2013poa, Baldauf:2015xfa, Blas:2016sfa, Senatore:2017pbn, Ivanov:2018gjr, Lewandowski:2018ywf}). In the following, we generalize this calculation to a generic linear feature and further extend it to the case of logarithmically-spaced oscillations.

\subsubsection{Perturbative Treatment}
\label{sec:damping_perturbative}
Our aim is to compute the damping of a generic oscillatory feature due to long-wavelength modes. It is well known that a simple perturbative treatment is not enough to capture the full damping effect in the case of the standard oscillatory features in the matter power spectrum, the BAO~signal~\cite{Crocce:2007dt, Creminelli:2013poa, Baldauf:2015xfa, Blas:2016sfa, Senatore:2017pbn, Ivanov:2018gjr, Lewandowski:2018ywf}. As in that case, it is however also useful to start with a perturbative treatment for generic features and then include nonperturbative effects in the calculation.

\vskip4pt
The full one-loop power spectrum is given by
\beq
P_{1\text{-loop}}(k) = \int\!\frac{\d^3q}{(2\pi)^3} \left[6 F_3(\q,-\q,\k\hskip1pt) P(k)+2F_2^2(\q,\k-\q\hskip1pt) P(|\k-\q\hskip1pt|)P(q)\right] ,	\label{eq:1loopPk}
\eeq
where~$F_n$ are the usual perturbation theory kernels (see \cite{Bernardeau:2001qr} for a review). For a generic oscillatory component~$\Pw(k)$ of the matter power spectrum, the effects of long modes~$q$, with $q < \Lambda \equiv \epsilon k$,\footnote{In this case, long modes~$q$ are defined as those modes with wavelengths much longer than the typical width~$\sigma$ of the feature, $q < \Lambda \ll 2\pi/\sigma$. This implies $\Lambda_\BAO \lesssim \SI{0.6}{\hPerMpc}$ for the standard BAO~signal. However, in practice, we want to predict the matter power spectrum at any wavenumber~$k$, which implies that the prescription for long modes should also satisfy $q\ll k$. Since the latter requirement is stronger than the former for the range of scales under consideration in this work, we choose the separation scale to be $\Lambda=\epsilon k$, with $\epsilon\ll1$~\cite{Baldauf:2015xfa}.} can be captured by 
\beq
\Pw_{1\text{-loop}}(k) = \frac{1}{2}\int^\Lambda\! \frac{\d^3q}{(2\pi)^3}\, \frac{(\q\cdot\k\hskip1pt)^2}{q^4}\,\Pnw(q) \left[ \Pw(|\k+\q\hskip1pt|)+\Pw(|\k-\q\hskip1pt|)-2\Pw(k)\right] .	\label{eq:loopl}
\eeq
Since features break the scale invariance of the matter power spectrum, we cannot keep only the first few orders in the Taylor expansion $\Pw(|\k+\q\hskip1pt|) = \Pw(k) + \q\cdot\nabla_{\k} \Pw(k) + \mathcal{O}(q^2/k^2)$, but have to resum the entire series into the exponential
\beq
\Pw(|\k+\q\hskip1pt|) = \ee^{\q\cdot\nabla_{\k}} \Pw(k)\, .
\eeq
We can therefore rewrite~\eqref{eq:loopl} as 
\beq
\Pw_{1\text{-loop}}(k) = \int^\Lambda\! \frac{\d^3q}{(2\pi)^3}\, \frac{(\q\cdot\k\hskip1pt)^2}{q^4}\,\Pnw(q) \left[\cosh\hskip-1pt \left(\q\cdot\nabla_{\k}\right)-1\right] \Pw(k)\, .	\label{eq:1loop}
\eeq
We stress that we did not assume any particular form of~$\Pw(k)$ in this expression. It is only based on \mbox{(\hskip-0.5pt\textit{i})}~calculating the contribution of long modes to the one-loop power spectrum~\eqref{eq:1loopPk} and \mbox{(\hskip-0.5pt\textit{ii})}~the matter power spectrum having an oscillatory component of any kind. Since the calculation proceeds by applying the operator~$\cosh\hskip-1pt\left(\q\cdot\nabla_{\k}\right)$ to the wiggle power spectrum~$\Pw(k)$, we now consider the two oscillatory feature models separately.

\paragraph{Linear features.} 
As a consequence of the baryon acoustic oscillations in the early universe, there is an enhanced probability to find pairs of galaxies at a separation given by the size of the sound horizon at the drag epoch, $r_s \approx \SI{150}{\Mpc}$. We therefore find a peak in the galaxy two-point correlation function and linearly-spaced oscillations in Fourier space, with the location and frequency given by~$r_s$, respectively. An enhanced probability of finding galaxy pairs at another distance scale~$\omegalin$ would produce the same signatures. This is why we can compute the damping of such feature oscillations in exactly the same way as for the BAO~spectrum. We will review them here for a generic scale $\omegalin \gtrsim \SI{75}{\Mpc}$ in order to not be affected by small-scale nonlinearities.

Since the BAO~signal is itself a small contribution to the overall matter power spectrum, we neglect its contribution and simply use $\Pw(k)\equiv~\Pw_\lin(k)$ as defined in~\eqref{eq:pwx}. Applying $2n$~gradients to~$\Pw_\lin(k)$ results in
\beq
q^{i_1}\dots q^{i_{2n}}\nabla_{k_{i_1}}\dots\nabla_{k_{i_{2n}}}\Pw_\lin(k) \approx (-1)^n q^{i_1}\dots q^{i_{2n}} \hat k_{i_1} \dots \hat k_{i_{2n}}\,\omegalin^{2n}\, \Pw_\lin(k)\, ,	\label{eq:coshlin}
\eeq
where we used the series expansion of the hyperbolic cosine and neglected small corrections that arise from acting with the derivative operators on the smooth envelope~$\Pnw(k)$. Plugging this result into~\eqref{eq:1loop}, we get
\beq
\Pw_{1\text{-loop}}(k) = -k^2\, \Sigma^2_\lin\,\Pw_\lin(k)\, ,	\label{eq:1looplin}
\eeq
where we defined
\beq
\Sigma^2_\lin \equiv \frac{1}{6\pi^2} \int_0^\Lambda\!\d q\, \Pnw(q) \left[1 - j_0\left(q\,\omegalin\right) + 2 j_2\left(q\,\omegalin\right)\right] , \label{eq:linearDampingScale}
\eeq
with the spherical Bessel function of the first kind $j_n(x)$. We note that~$\Sigma_\lin$ is independent of the wavenumber~$k$ unless an implicit dependence is introduced by taking $\Lambda=\Lambda(k)$ as mentioned above. However, the damping scale~$\Sigma_\lin$ of course depends on redshift via the broadband power spectrum, $\Pnw(q) = \Pnw(q,z)$.

\paragraph{Logarithmic features.}
In contrast to linear features, logarithmically-spaced oscillations in Fourier space do not have a simple intuitive interpretation in real space. However, we can proceed with the calculation of the damping without additional caveats because the derivation of~\eqref{eq:1loop} is valid for any oscillations in the matter power spectrum. In Appendix~\ref{app:nonlinearDampingCalculation}, we show how the operator $\cosh\hskip-1pt\left(\q\cdot\nabla_{\k}\right)$ acts on the wiggle component of the linear matter power spectrum in the presence of logarithmic features, which we introduced in~\eqref{eq:logarithmicFeatures} and~\eqref{eq:pwx}. In consequence, the one-loop wiggle spectrum can be computed to be
\beq
\Pw_{1\text{-loop}}(k) = -k^2 \left[ \Sigma^2_\Log (k)\Pw_\Log(k) + \hat{\Sigma}^2_\Log(k)\, \frac{\Pnw(k)}{\omegalog} \frac{\d \delta P_\zeta^\Log(k)}{\d \log k} \right] ,	\label{eq:p1loopf}
\eeq
where we introduced
\begin{align}
\Sigma^2_\Log(k)		&\equiv \frac{1}{4\pi^2} \int_0^\Lambda\! \d q\, \Pnw(q) \int_{-1}^{1}\! \d \mu \, \mu^2 \left\{1 - \cos\!\left[\omegalog \log\!\left(1-\frac{q\,\mu}{k}\right)\right]\right\} ,	\label{eq:cossigma}	\\
\hat{\Sigma}^2_\Log(k)	&\equiv - \frac{1}{4\pi^2} \int_0^\Lambda\! \d q\, \Pnw(q) \int_{-1}^{1}\! \d\mu \, \mu^2\, \sin\!\left[\omegalog \log\!\left(1-\frac{q\,\mu}{k}\right) \right] ,	\label{eq:sinsigma}
\end{align}
with $\mu = \hat{q}\cdot\hat{k}$. In contrast to the damping scales of the BAO~signal,~$\Sigma_\BAO$, and of linear features,~$\Sigma_\lin$, which are constant in~$k$, the damping factors~$\Sigma_\Log$ and $\hat{\Sigma}_\Log$ are scale dependent (but similarly redshift dependent). Moreover, the one-loop wiggle power spectrum is also no longer directly proportional to the oscillatory power spectrum at linear order. We however note that~\eqref{eq:cossigma} and~\eqref{eq:sinsigma} are only valid if $q \ll k$. In this limit and for large enough values of~$\omegalog$, these expressions can be simplified into a form similar to~\eqref{eq:linearDampingScale} for linear features. Since we can always decompose a logarithmic feature in a basis of linear oscillations, this also conforms with our expectation to recover this result in the appropriate limit. For now, we however choose to keep the calculation general and will discuss these limits in detail below.

\vskip4pt
The crucial aspect of both one-loop results~\eqref{eq:1looplin} and~\eqref{eq:p1loopf} is that they correct the linear power spectrum by $\mathcal{O}(1)$-terms for a wide range of parameter space and wavenumbers, exactly as in the case of the standard BAO~spectrum. In other words, we have $\Pw_{1\text{-loop}}(k) \approx \mathcal{O}(1)\, \Pw_\text{tree-level}(k)$ for $k \in \SIrange[range-phrase={,}, open-bracket=[, close-bracket=]]{0.1}{0.3}{\hPerMpc}$ because $k^2\, \Sigma^2 \approx \mathcal{O}(1)$ in this range. This indicates that the perturbative treatment is insufficient. Fortunately, it is possible to compute the leading-order correction of long modes to the wiggle power spectrum at all orders in perturbation theory.

\subsubsection{Infrared Resummation}
The infrared~(IR) resummation of the large-scale bulk flows that damp the BAO~signal has been studied in various ways~\cite{Crocce:2007dt, Creminelli:2013poa, Baldauf:2015xfa, Blas:2016sfa, Senatore:2017pbn, Ivanov:2018gjr, Lewandowski:2018ywf}. Here, we follow the approach of~\cite{Blas:2016sfa}, in which the class of loop diagrams that are most IR-enhanced are first identified and then resummed into the nonperturbative effect, the well-known exponential BAO~damping. Their $L$-loop diagram is given by
\beq
\Pw_{L\text{-loop},\,\mathrm{LO}}(k) = \frac{1}{L!} \prod_{i=1}^L \left[ \frac{1}{2} \int^\Lambda\! [\d q_i]\, \Pnw(q_i) \mathcal{D}_{\q_i} \mathcal{D}_{-\q_i} \right] \Pw(k)\, ,	\label{eq:lthloop}
\eeq
where the subscript `LO' indicates that the leading-order IR-enhanced loops are taken into account. Furthermore, we introduced the notation $[\d q] = \d^3 q$ and defined
\beq
\mathcal{D}_{\q_i} \Pw(k) = \frac{\q_i\cdot \k}{q_i^2} \left(\Pw(|\k+\q\hskip1pt|)-\Pw(k)\right) = \frac{\q_i\cdot \k}{q_i^2} \left.\left(\ee^{\q_i\cdot\nabla_{\k'}}-1\right)\Pw(k')\right|_{k'=k} .
\eeq
It is easy to verify that we exactly recover~\eqref{eq:1loop} for $L=1$. Since the rest of the calculation depends on the form of~$\Pw(k)$, we discuss the linearly- and logarithmically-spaced oscillations again in turn.

\paragraph{Linear features.}
By employing~\eqref{eq:1looplin}, it is straightforward to compute the wiggle power spectrum of linear features at $L^\mathrm{th}$ order,
\beq
\Pw_{L\text{-loop},\,\mathrm{LO}}(k) = \frac{\left(-k^2 \Sigma_\lin^2\right)^L}{L!}\,\Pw_\lin(k)\, .
\eeq
Resumming these terms to all orders, we obtain
\beq
\Pw_{\mathrm{IR},\,\mathrm{LO}}(k) = \sum_{L=0}^\infty \Pw_{L\text{-loop},\,\mathrm{LO}}(k) = \ee^{-k^2 \Sigma_\lin^2} \Pw_\lin(k)\, .	\label{eq:finallim}
\eeq
This result is a generalization of the BAO~expression with $r_s \rightarrow \omegalin$ (cf.\ e.g.~\cite{Blas:2016sfa}). Since the value of~$\Sigma_\lin$ however saturates for $\omegalin \gtrsim \SI{75}{\Mpc}$, we can simply use the BAO~damping scale, $\Sigma_\lin \approx \Sigma_\BAO$. We can therefore factor out the damping and write the full matter power spectrum as
\beq
P_m(k) \approx \Pnw_m(k) + \ee^{-k^2 \Sigma_\BAO^2} \left[\Pw_\BAO(k)+\Pw_\lin(k)\right] ,	\label{eq:linresum}
\eeq
which constitutes a simple generalization of the known result for the standard BAO~signal.

\paragraph{Logarithmic features.}
It is slightly less trivial to compute the expression of the $L^\mathrm{th}$-order loop for logarithmic features with arbitrary frequency~$\omegalog$. We proceed via induction by computing the first few orders and then deriving the general formula. In this way, the IR-resummed wiggle power spectrum is found to be
\beq
\Pw_{\mathrm{IR},\,\mathrm{LO}}(k) = \ee^{-k^2 \Sigma^2_\Log(k)} \cos\hskip-1pt\left(k^2 \hat{\Sigma}^2_\Log(k)\right) \Pw_\Log(k) - \ee^{-k^2 \Sigma^2_\Log(k)} \sin\hskip-1pt\left(k^2 \hat{\Sigma}^2_\Log(k)\right) \frac{\Pnw(k)}{\omegalog} \frac{\d \delta P_\zeta^\Log(k)}{\d \log k}\, .	\label{eq:final}
\eeq
We refer to Appendix~\ref{app:nonlinearDampingCalculation} for further details on this calculation.

While this expression provides the leading nonperturbative damping for logarithmic features, we find by explicit calculation that generically $\hat{\Sigma}_\Log \ll \Sigma_\Log$. This can be understood by noticing that the integrands in~\eqref{eq:cossigma} and~\eqref{eq:sinsigma} can be expanded in $\omegalog (q \mu/k)$ and $\omegalog (q \mu/k)^2$, respectively. The fact that the integrals get their largest contributions from $q\ll k$ explains the hierarchy between~$\Sigma_\Log$ and~$\hat{\Sigma}_\Log$. As a result, when $k^2 \hat{\Sigma}_\Log$ is large enough to be important, the signal is already exponentially suppressed. It is therefore a good approximation to set $\hat \Sigma_\Log \approx 0$ for our analysis choice of $\omegalog \geq 10$ (see below), as confirmed by Fig.~\ref{fig:dampingValidation} in Appendix~\ref{app:forecasts}. We therefore get
\beq
\Pw_{\mathrm{IR},\,\mathrm{LO}}(k) \approx \ee^{-k^2 \Sigma^{2}_\Log(k)} \Pw_\Log(k)\, .	\label{eq:finalapprox}
\eeq
Moreover, it is straightforward to show that in the limit $q/k \ll 1$, $\Sigma_\Log(k)$~approaches the functional form of~$\Sigma_\lin$ with the substitution $\omegalin\rightarrow \omegalog/k$,\footnote{This result has also been independently derived in~\cite{Vasudevan:2019ewf} using $\omegalog \gg 1$ as an expansion parameter.}
\beq
\Sigma_\Log(k) \approx \Sigma_\lin|_{\omegalin=\omegalog/k} = \frac{1}{6\pi^2} \int_0^\Lambda\!\d q\, \Pnw(q) \left[1 - j_0\left(\frac{\omegalog}{k}q\right) + 2 j_2\left(\frac{\omegalog}{k}q\right)\right] ,
\eeq
with~$\Sigma_\lin$ given by~\eqref{eq:linearDampingScale}. Therefore, whenever $\omegalog/k \gtrsim \SI{75}{\Mpc}$, we can use the same approximation as in the linear case, $\Sigma_\Log(k) \approx \Sigma_\BAO$. As we also show in Appendix~\ref{app:forecasts}, this approximation is good enough for the scales and frequencies of interest in an analysis of BOSS~data.\footnote{These findings are also confirmed by analyses of $N$-body simulations performed in~\cite{Vlah:2015zda}.} We can therefore further approximate~\eqref{eq:finalapprox} and write the full nonlinear power spectrum as
\beq
P_m(k) \approx \Pnw_m(k) + \ee^{-k^2 \Sigma_\BAO^2}\left[\Pw_\BAO(k)+\Pw_\Log(k)\right] ,	\label{eq:logresum}
\eeq
i.e.\ in the same way as for linear features in~\eqref{eq:linresum}.

\vskip4pt
Before concluding the discussion of the theoretical damping calculation, a couple of remarks are in order regarding subleading corrections to~\eqref{eq:linresum} and~\eqref{eq:logresum}:
\begin{itemize}
	\item We computed the leading-order IR-resummed power spectrum. It has however been shown that there are subleading contributions which improve the fit to $N$-body simulations in the case of the featureless BAO~signal~\cite{Vlah:2015zda, Blas:2016sfa, Senatore:2017pbn, Ivanov:2018gjr, Lewandowski:2018ywf}. Since we do not employ the theoretically computed results, but fit the damping scale in our data analysis below, we can neglect these corrections for both the BAO~signal (as in the standard BAO~analyses), and the primordial linear and logarithmic features.
	
	\item We have not computed the damping of the mixed BAO-feature term of~\eqref{eq:pw}, $\Pw_\BAO(k) \, \delta P_\zeta^X$. The size of this contribution is of order $A_\BAO \times A_X \approx 0.05 \times 0.01$ and, therefore, contributes less than per mil to the matter power spectrum. For linearly-spaced oscillations, we checked that it is again a good approximation for $\omegalin \gtrsim 2\hskip1ptr_s \approx \SI{300}{\Mpc}$ to also factor out the exponential BAO~damping for the mixed term. We expect this to also be the case for logarithmic features. Consequently, we implement the mixed term in the nonlinear matter power spectrum as follows:
	\beq
	P_m(k) = \Pnw_m(k) + \ee^{-k^2 \Sigma_\BAO^2} \left[\Pw_\BAO(k) + \Pw_X(k) + \Pw_\BAO(k)\, \delta P_\zeta^X(k) \right] . \label{eq:mixedresum}
	\eeq
\end{itemize}
Finally, from now on, we switch to the notation for the BAO~damping scale that has been adopted in data analyses, $\Sigma^2_\BAO \to \Sigma_\mathrm{nl}^2/2$. With this, we are ready to implement~\eqref{eq:mixedresum} and search for primordial feature models in observational LSS~data.

%%%%%%%%%%%%%%%%
\section{Feature Search in Large-Scale Structure}
\label{sec:featureSearch}
%%%%%%%%%%%%%%%%

In this section, we introduce and establish our search for features in the BOSS~dataset. We propose an analysis in which the amplitude and frequency of the linear and logarithmic features can be constrained. Moreover, we check the validity of our approach on mock data and compare the results with the expected constraining power obtained in forecasts.

\subsection{BOSS DR12 Dataset and Analysis Pipeline}
\label{sec:pipeline}

The approach of our analysis, which we introduce in the following, is very general and we expect it to apply to a wide range of surveys. Having said this, some of its aspects are particular to~BOSS, such as the validity of some of the employed approximations, and should therefore be revisited in future analyses.

\vskip4pt
Our analysis is based on the BAO~pipeline of~\cite{Beutler:2016ixs} which utilizes \texttt{nbodykit}~\cite{Hand:2017pqn}. We use the commonly employed density field reconstruction procedure~\cite{Eisenstein:2006nk} to reduce the damping scale caused by gravitational evolution and move information from higher-order statistics back to the power spectrum~\cite{Schmittfull:2015mja}.\footnote{We showed in~\textsection\ref{sec:largeScaleNonlinearities} that features are damped by large-scale modes in the same way as the BAO~signal. In consequence, reconstruction will remove the damping of features in the same way, as encoded by the post-reconstruction value of the damping scale~$\Sigma_\mathrm{nl}$.} We then measure the galaxy power spectrum following the steps described in~\cite{Beutler:2016ixs, Beutler:2016arn}. The corresponding covariance matrix is obtained by measuring the power spectrum monopole in 999 mock catalogs (see~\textsection\ref{sec:mocks} for more details on these mock catalogs). To extract the BAO~(and potential feature) signal, we marginalize over the smooth galaxy broadband power spectrum,
\beq
\Pnw_g(k) = B^2 \Pnw(k) F(k,\Sigma_s) + A(k)\, ,
\eeq
with five polynomial terms
\beq
A(k) = \frac{a_1}{k^3} + \frac{a_2}{k^2} + \frac{a_3}{k} + a_4 + a_5k^2\, .	\label{eq:poly}
\eeq
Here, the bias parameter~$B$ is used to marginalize over the power spectrum amplitude, $\Pnw(k)$ is the linear no-wiggle power spectrum model without any BAO~signal and
\beq
F(k,\Sigma_s) = \frac{1}{(1 + k^2\Sigma_s^2/2)^2}
\eeq
is the velocity damping term arising from the nonlinear velocity field. Finally, the standard BAO~signal and the oscillatory features left after the marginalization described above are modeled as
\beq
P_g(k) = \Pnw_g(k) \left\{1 + \left[O(k/\alpha) + \delta P_\zeta^X(k) + O(k/\alpha)\, \delta P_\zeta^X(k) \right] \ee^{-k^2\Sigma^2_\mathrm{nl}/2}\right\} ,
\eeq
where $O(k) \equiv \Pw_\BAO(k)/\Pnw(k)$ is the standard linear BAO~spectrum, $\alpha$~is the associated isotropic scaling parameter and $\Sigma_\mathrm{nl}$~is the nonlinear damping scale, which we keep as a free parameter. While we left the redshift dependence to be implicit, this model for the galaxy power spectrum is appropriately evaluated in each redshift bin. The fiducial $\Lambda$CDM~cosmology is taken to be the same as in~\cite{Beutler:2016ixs}, with matter density $\Omega_m = 0.31$, physical baryon density $\omega_b = 0.022$, amplitude of linear matter fluctuations on \SI{8}{\hPerMpc}~scales $\sigma_8 = 0.824$, scalar spectral index $\ns = 0.96$ and Hubble constant $H_0 = \SI{67.6}{\kilo\meter\per\second\per\Mpc}$. The linear and logarithmic features are contained in the relative primordial spectrum~$\delta P_\zeta^X(k)$, with $X=\lin,\Log$, that was introduced in~\eqref{eq:linearFeatures} and~\eqref{eq:logarithmicFeatures}.

\begin{figure}
	\centering
	\includegraphics{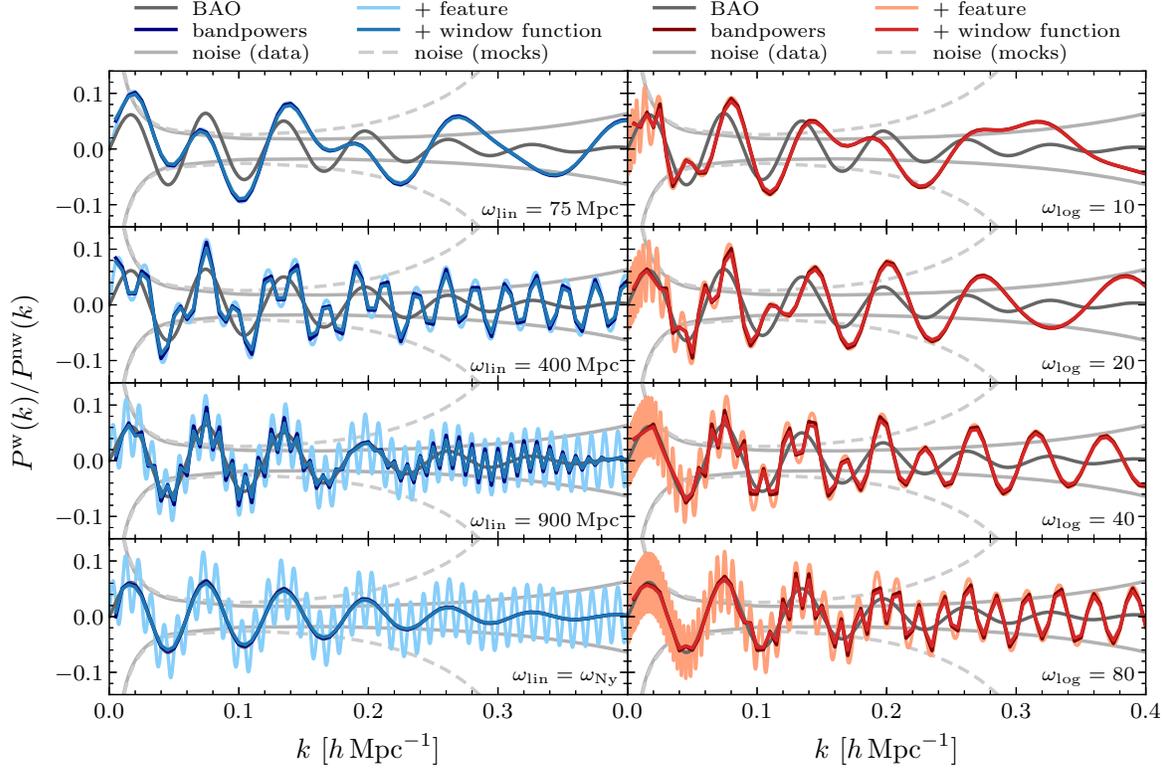}\vspace{-8pt}
	\caption{Illustration of the impact of the finite volume observed in~BOSS on the imprint of primordial features in the linear wiggle spectrum for linearly-~(\textit{left}) and logarithmically-spaced~(\textit{right}) oscillations with $\Asin_X=0.05$ and $\Acos_X=0$. The frequency of the standard featureless BAO~spectrum (dark gray) is small enough to be essentially unaffected by the effects of the bandpower estimation and the window function. The larger the frequency of the features (light colors), the larger the effect of bandpass filtering with $\Delta k=\SI{0.005}{\hPerMpc}$ (medium colors). Including an approximation of the window function (dark colors; see Appendix~\ref{app:forecasts} for details) further smooths the primordial wiggles. In particular at the Nyquist frequency, which is given by $\omega_\mathrm{Ny}=\SI{929}{\Mpc}$ for the BOSS~DR12 dataset used in this work, the feature oscillations are completely aliased with only the featureless BAO~spectrum remaining. The light gray lines indicate the estimated noise curves for the high-redshift bin of~BOSS including the nonlinear exponential damping which differs between the BOSS~data (solid) and mock catalogs (dashed). Finally, we note that the comparison of the left- and right-hand panels together with these noise curves allows for another way to estimate the reliable frequency range for the logarithmic features that is complementary to Fig.~\ref{fig:logsep}.\vspace{-6pt}}
	\label{fig:omegaLinLogComparison}
\end{figure}

\vskip4pt
While the above model is described in terms of a continuous wavenumber~$k$, cosmological experiments can only access a finite number of modes due to their finite survey volume. The associated fundamental mode~$k_f$ is given by the largest scale included in the dataset and is naturally setting a resolution limit on the oscillation frequency that we can measure. In practice, this is realized through the survey window function which introduces couplings between modes separated by the fundamental mode or less~\cite{Beutler:2013yhm, Wilson:2015lup, Beutler:2016arn, Beutler:2018vpe}. In addition, we only measure the power spectrum in discrete bandpowers~$P_i$, which average wavenumbers~$\vec{k}$ with $k \in [k_i-\Delta k/2, k_i+\Delta k/2)$, where the bandwidth~$\Delta k$ is a choice of the analysis. As a consequence, a signal with an (effective) linear frequency above the Nyquist frequency, $\omega_\mathrm{Ny}=\pi/\Delta k$,\footnote{The fundamental Nyquist frequency of a survey is determined by~$k_f$, but for $\Delta k \gtrsim k_f$, it is the bandwidth that sets the limiting scale in an analysis.} will be aliased and is therefore out of reach. Figure~\ref{fig:omegaLinLogComparison}%
highlights the effects that the finite bandwidth and the window function have on the power spectrum, and illustrates why they limit the range of frequencies that are accessible in an LSS~analysis. 

\vskip4pt
In the next sections, we apply this analysis pipeline to the Baryon Oscillation Spectroscopic Survey~(BOSS), which was part of SDSS-III~\cite{Dawson:2012va, Eisenstein:2011sa} and used the SDSS~multi-fibre spectrographs~\cite{Smee:2012wd, Bolton:2012hz} at the \SI{2.5}{m}~Sloan Telescope~\cite{Gunn:2006tw} of the Apache Point Observatory in New Mexico. We employ the final version of this dataset, known as data release~12~(DR12)~\cite{Reid:2015gra}, which contains spectroscopic redshifts of \num{1198006}~galaxies. The survey covered \SI{10252}{deg^2} of the sky, divided in two patches called the North Galactic Cap~(NGC) and the South Galactic Cap~(SGC), and a redshift range of $0.2 - 0.75$. Following the main BOSS~analysis~\cite{Alam:2016hwk}, we split this redshift range into two (independent) redshift bins given by $0.2 < z < 0.5$ (`low-$z$') and $0.5 < z < 0.75$ (`high-$z$'). While the standard BOSS~analysis uses $\Delta k = \SI{0.01}{\hPerMpc}$~\cite{Beutler:2016ixs}, we employ a bandwidth of $\Delta k = \SI{0.005}{\hPerMpc}$, which is close to the fundamental mode of~BOSS, to maximize the feature frequency range accessible in this dataset. This limits our analysis to $\omegalin \leq \omega_\mathrm{Ny} \approx \SI{929}{\Mpc}$, but we conservatively take $\omegalin \leq \SI{900}{\Mpc}$.

\vskip4pt
We analyze this dataset by producing a Markov Chain Monte Carlo~(MCMC) with a modified version of \texttt{emcee}~\cite{ForemanMackey:2012ig} which includes the Gelman\,\&\,Rubin convergence criterion~\cite{Gelman:1992zz} with scale parameter $\epsilon < 0.04$. Since the inflationary signal under consideration is isotropic, we focus on the power spectrum monopole and perform our analysis with one isotropic BAO~parameter~$\alpha$ per redshift bin. Moreover, we treat the low- and high-redshift bins of BOSS~DR12 independently. Since we use separate broadband marginalization parameters for the~NGC and the~SGC, we fit for a total of 18~free parameters per redshift bin:
\beq
\alpha, \omega_X, \Acos_X, \Asin_X; B^\mathrm{NGC}, B^\mathrm{SGC}, \Sigma_s, \Sigma_\mathrm{nl}, a^\mathrm{NGC}_n, a^\mathrm{SGC}_n\, .
\eeq
We impose flat priors on all parameters, including the feature frequencies which are sampled within $\SIrange[range-phrase={,}, open-bracket=[, close-bracket=]]{100}{900}{\Mpc}$ and $[10,80]$ for linearly- and logarithmically-spaced oscillations, respectively. These ranges are motivated by our discussion of small- and large-scale nonlinearities in Section~\ref{sec:theory} (see also Fig.~\ref{fig:logsep}). Given that the primordial feature parameters~$\omega_X$, $\Asin_X$ and~$\Acos_X$ are independent of the redshift bin, we combine them when inferring bounds on these models while marginalizing over the other parameters (see Appendix~\ref{app:analysisDetails}).

\subsection{Forecasting Methodology for BOSS}
\label{sec:forecasting}

To estimate the expected level of sensitivity, validate and cross-check the described analysis of BOSS~data, we perform two types of forecasts: based on the Fisher information matrix and based on the likelihood itself. The Fisher forecasts have to be used with care, but provide useful guidelines over a large range of possible parameters and experimental configurations since they are relatively fast to compute. On the other hand, the likelihood-based forecasts present a more direct picture of the sensitivity and, in particular, allow to study the effects of noisy data and injected signals. In addition, we checked a number of theoretical approximations using Fisher forecasts (see Appendix~\ref{app:forecasts}).

\vskip4pt
Our general forecasting methodology and modeling is based on~\cite{Baumann:2017gkg}, with some modifications that are detailed in Appendix~\ref{app:forecasts}. Since we are only interested in oscillatory features (and not broadband effects), our forecasts directly employ the relative wiggle spectrum $\mathcal{O}_{\!g}(k) \equiv \Pw_g(k)/\Pnw_g(k)$ and not the galaxy power spectrum~$P_g(k)$ as the observable. After marginalizing over the bias~$B$ and the polynomial coefficients~$a_n$ of~\eqref{eq:poly}, we expect these two to be identical. Working with~$\mathcal{O}_{\!g}(k)$ removes much of the degeneracy with these broadband parameters and makes the forecasts more reliable (in particular for the Fisher matrix). The Fisher information matrix~$F_{ij}$ is typically defined as the average curvature of the log-likelihood~$\log\!\L(\vec{\theta})$ around a fiducial point in parameter space spanned by~$\vec{\theta}$. In our BAO~forecasts, we will generally use $\vec{\theta} \equiv \{\alpha, \Asin_X, \Acos_X \}$, with $X = \lin,\Log$, where~$\alpha$ is the standard BAO~parameter, and~$\Asin_X$ and~$\Acos_X$ are the respective feature amplitudes. As in~\cite{Baumann:2017gkg}, we employ a conservative broadband marginalization scheme. Since the inverse Fisher matrix is the covariance matrix for a Gaussian likelihood, the Cram\'er-Rao bound, $\sigma(\theta_i) \geq \sqrt{(F^{-1})_{ii}}$, provides a lower limit on the marginalized constraints, with equality commonly assumed for Fisher forecasts.

The likelihood-based forecasts are based on the same modeling and have previously been utilized successfully in~\cite{Baumann:2017gkg, Baumann:2018qnt}. In this type of forecast, we compute the likelihood function~$\L(\vec{\theta})$ on a grid in parameter space, given a specific fiducial (`data') spectrum computed for a fixed set of parameters~$\vec{\theta}_\mathrm{fid}$. When specified, the fiducial model includes a random realization of the noise to mimic the scatter of experimental data points due to shot noise and cosmic variance. In this case, we talk about ``noisy forecasts'' which will be useful in our estimates of the probability of experimental noise mimicking the presence of oscillatory features. This is in contrast to the ``noiseless forecasts'', for which the experimental effects are only captured by the covariance matrix, as commonly employed. (We emphasize that the latter forecasts are not cosmic variance limited.) Except where noted otherwise, all of the following BOSS~forecasts are based on this likelihood-based approach.

Finally, we note that we produce two sets of forecasts as in~\cite{Baumann:2018qnt}: one for the comparison to the mock catalogs and another to compare to the results from the actual BOSS~data. This is due to the fact that the mock catalogs have a known problem of overdamping the BAO~spectrum which results in an approximately 30\% weaker signal for the traditional BAO~analysis~\cite{Beutler:2016ixs}. When comparing to mocks, we use a (post-reconstruction) nonlinear damping scale of $\Sigma_\mathrm{nl} \approx \SI{7}{\hPerMpc}$, while we employ the standard (redshift-dependent) values otherwise (see~\cite{Baumann:2017gkg}). The gray noise curves in Fig.~\ref{fig:omegaLinLogComparison}, which include the nonlinear damping terms, indicate this difference and we can anticipate that the bounds on the feature amplitudes will be stronger on the data than~in~the~mocks.

\subsection{Validation on Mock Catalogs and in Forecasts}
\label{sec:mocks}

To validate our analysis pipeline, we first perform an analysis on the MultiDark Patchy mock catalogs~\cite{Kitaura:2015uqa}, which mimic the galaxy clustering behavior observed in~BOSS. These mock data have been produced using approximate gravity solvers and analytical-statistical biasing models. The catalogs have been calibrated to an $N$-body-based reference sample extracted from one of the BigMultiDark simulations~\cite{Klypin:2014kpa}, which was performed using \texttt{GADGET-2}~\cite{Springel:2005mi} with $\num{3840}^3$ particles in a volume of $(\SI{2.5}{\per\h\Gpc})^3$ assuming a $\Lambda$CDM~cosmology with $\Omega_m = 0.307115$, $\Omega_b = 0.048206$, $\sigma_8 = 0.8288$, $\ns = 0.9611$ and $H_0 = \SI{67.77}{\kilo\meter\per\second\per\Mpc}$. The mock catalogs use halo abundance matching to reproduce the observed BOSS two- and three-point clustering measurements~\cite{Rodriguez-Torres:2015vqa}. This technique is applied as a function of redshift to reproduce the BOSS~DR12 redshift evolution. Note that these are the same mock catalogs that we use to derive the covariance matrix of our analysis as mentioned above. In addition, we remark that the mock catalogs do not contain any inflationary features since they assume a featureless primordial power spectrum (see Appendix~\ref{app:forecasts} for a check with injected signals).

\vskip4pt
We apply the described MCMC~analysis pipeline to 100~NGC and SGC~Patchy mock catalogs for the low- and high-redshift bins. Since nearby feature frequencies are correlated due to the finite range of wavenumbers used in the analysis, $\SIrange{0.01}{0.3}{\hPerMpc}$, we bin the samples of the resulting Markov chains with widths of $\Delta\omegalin=\SI{10}{\Mpc}$ and $\Delta\omegalog=1.0$. These values have been obtained from the Markov chains by estimating a scale-independent correlation length of the feature frequencies~$\omega_X$. For a number of different purposes, we condense the Markov chains into the following three statistical quantities shown in Fig.~\ref{fig:mocksForecastComparison}: %
\begin{figure}
	\centering
	\includegraphics{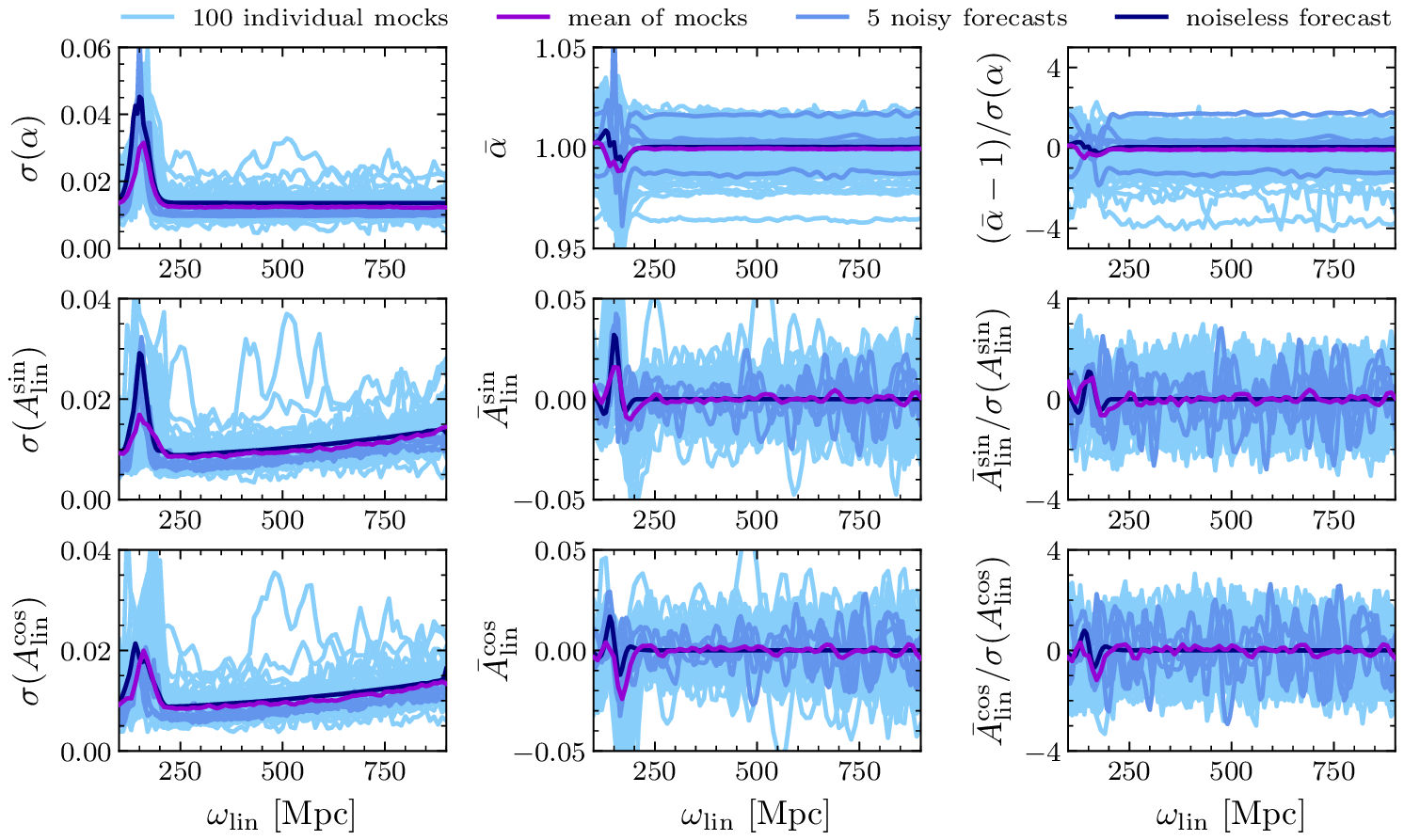}
	\includegraphics{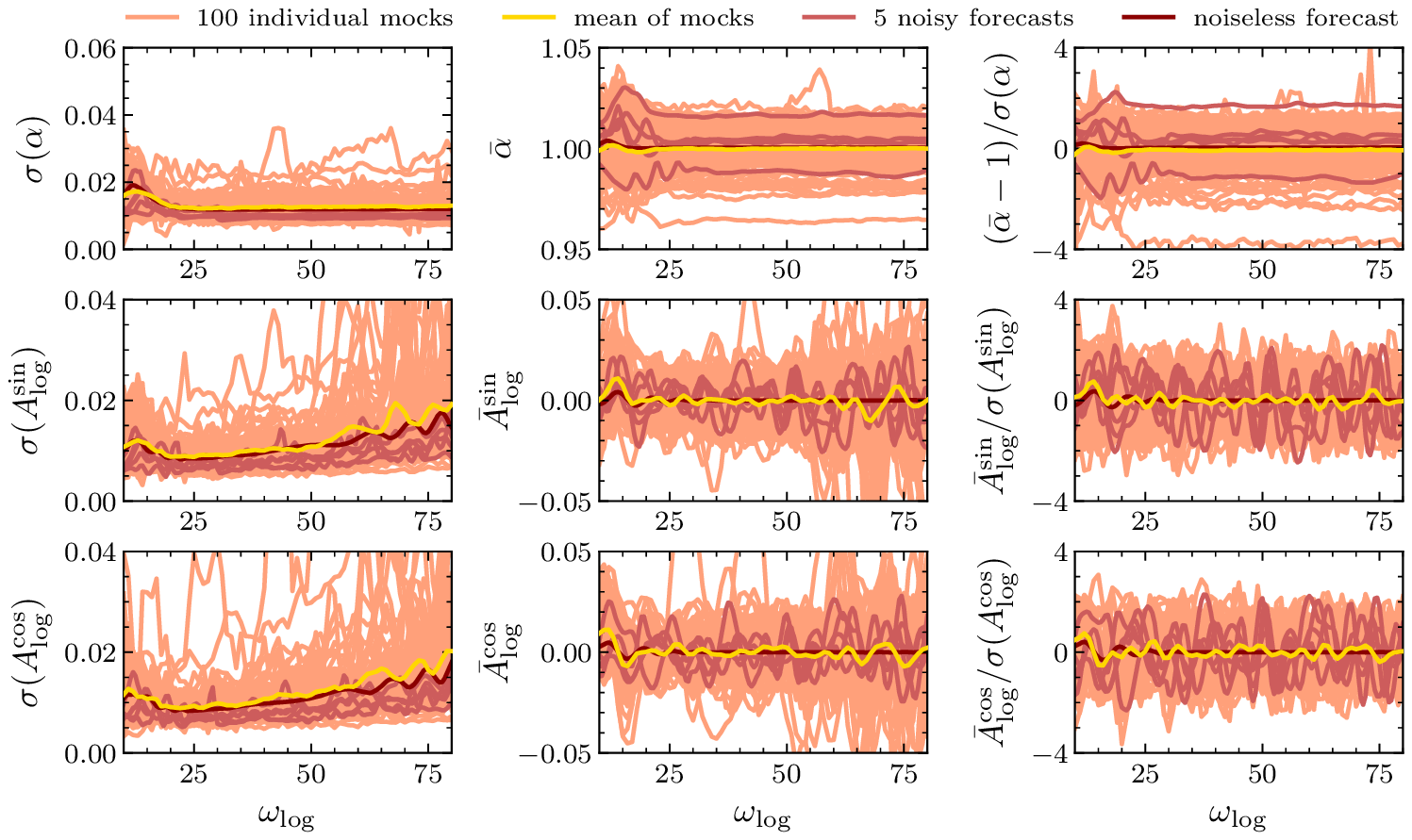}
	\caption{Comparison of the mock analysis with likelihood-based forecasts for linear~(\textit{top}) and logarithmic features~(\textit{bottom}) in the high-redshift bin. We present the standard deviation~$\sigma(\theta_i)$, mean~$\bar\theta_i$ and significance~$\bar\theta_i/\sigma(\theta_i)$ for the BAO~parameter~$\alpha$ and the amplitudes~$\Asin_X$ and~$\Acos_X$ as a function of the frequency~$\omega_X$. We display these quantities for the individual mock catalogs together with their mean, showing very good agreement with the noisy and noiseless forecasts.}
	\label{fig:mocksForecastComparison}
\end{figure}
the mean value~$\bar\theta_i$, the variance~$\sigma^2(\theta_i)$ and their ratio, the significance~$\bar\theta_i/\sigma(\theta_i)$, for the parameters $\vec{\theta} = \{\alpha, \Asin_X, \Acos_X\}$ as a function of the frequency~$\omega_X$. Apart from providing a validation of important parts of our analysis pipeline, the comparison of these quantities to those obtained in forecasts serves as a check of both the mock analysis leading to results within the expected sensitivity and the forecasts being suitable to compare to the data as well as to perform additional checks. Moreover, the significance provides a metric that helps to quantify any possible detections of features in the data analysis, for instance.

\vskip4pt
The results from the analysis of the low-redshift mocks are presented in Fig.~\ref{fig:mocksForecastComparison}, with the results from the high-redshift bin being similar. The middle panels of the mean values clearly show that our pipeline results in an unbiased estimation of both the BAO~parameter~$\alpha$ and the feature amplitudes given that the mocks are generated from a featureless primordial power spectrum. For linear features, the larger variance and non-zero mean values around $\omegalin=\SI{150}{\Mpc}$ indicate the expected degeneracy between the primordial features and the standard BAO~spectrum with a sound horizon of $r_s \approx \SI{150}{\Mpc}$. (Note that the degeneracy is not perfect because the BAO~spectrum is not a perfect sine oscillation, but contains a $k$-dependent amplitude and phase shift.) As expected, we also observe that the BAO~parameter~$\alpha$ is independent of the primordial parameters away from the scale of the sound horizon reproducing the constraints in the standard BOSS~analysis~\cite{Beutler:2016ixs}. The fact that the constraints on the feature amplitudes become (slightly) weaker with growing frequency can be attributed to the finite-survey effects of bandpowers and window function discussed above (cf.\ Appendix~\ref{app:forecasts}). Finally, the right column shows that one typically finds a $2\sigma$~fluctuation in some frequency bins for any given mock catalog, or for a given frequency bin for some of the 100~mocks. This should not be surprising given the roughly 80~sampled (but partly correlated) frequencies. Furthermore, the fact that we do not find many greater-than-$3\sigma$ fluctuations is consistent with the statistical expectations.

While the fluctuations seen in individual mocks are consistent with the variance inferred from the posterior, we would also like to know if this variance is consistent with the expectations for this type of survey. For this purpose, we turn to the likelihood-based forecasts whose results are shown in direct comparison in Fig.~\ref{fig:mocksForecastComparison}. We see that the mean values of $\sigma(\theta_i)$ and $\bar\theta_i$ are in excellent agreement with the noiseless forecasts across the entire mock catalogs. Furthermore, when a specific realization of noise is added to these forecasts, one finds the fluctuations in both the mean and the variance are consistent with the fluctuations observed in the mocks. (As expected, the mean of the noisy forecasts approaches the noiseless forecasts in the limit of many realizations.) This therefore further establishes that the observed fluctuations in the significances are entirely generated by and consistent with the experimental noise of cosmic variance and shot noise of~BOSS. In other words, these fluctuations occur because we fit random noise. Together with the extensive checks using forecasts presented in Appendix~\ref{app:forecasts}, this establishes our feature search in the clustering of galaxies and we can now turn to the BOSS~DR12~data.

%%%%%%%%%%%%%%%%
\section{First Large-Scale Structure Constraints}
\label{sec:dataAnalysis}
%%%%%%%%%%%%%%%%

In this section, we discuss the constraints on primordial feature models that we infer from the BOSS~DR12 galaxy power spectrum. Furthermore, we compare and combine these novel bounds with those obtained from current Planck CMB~data. We conclude this section with estimates of the future sensitivity of cosmological observations of the~CMB and~LSS.

\subsection{Limits on Features from BOSS DR12}

\begin{figure}
	\centering
	\includegraphics{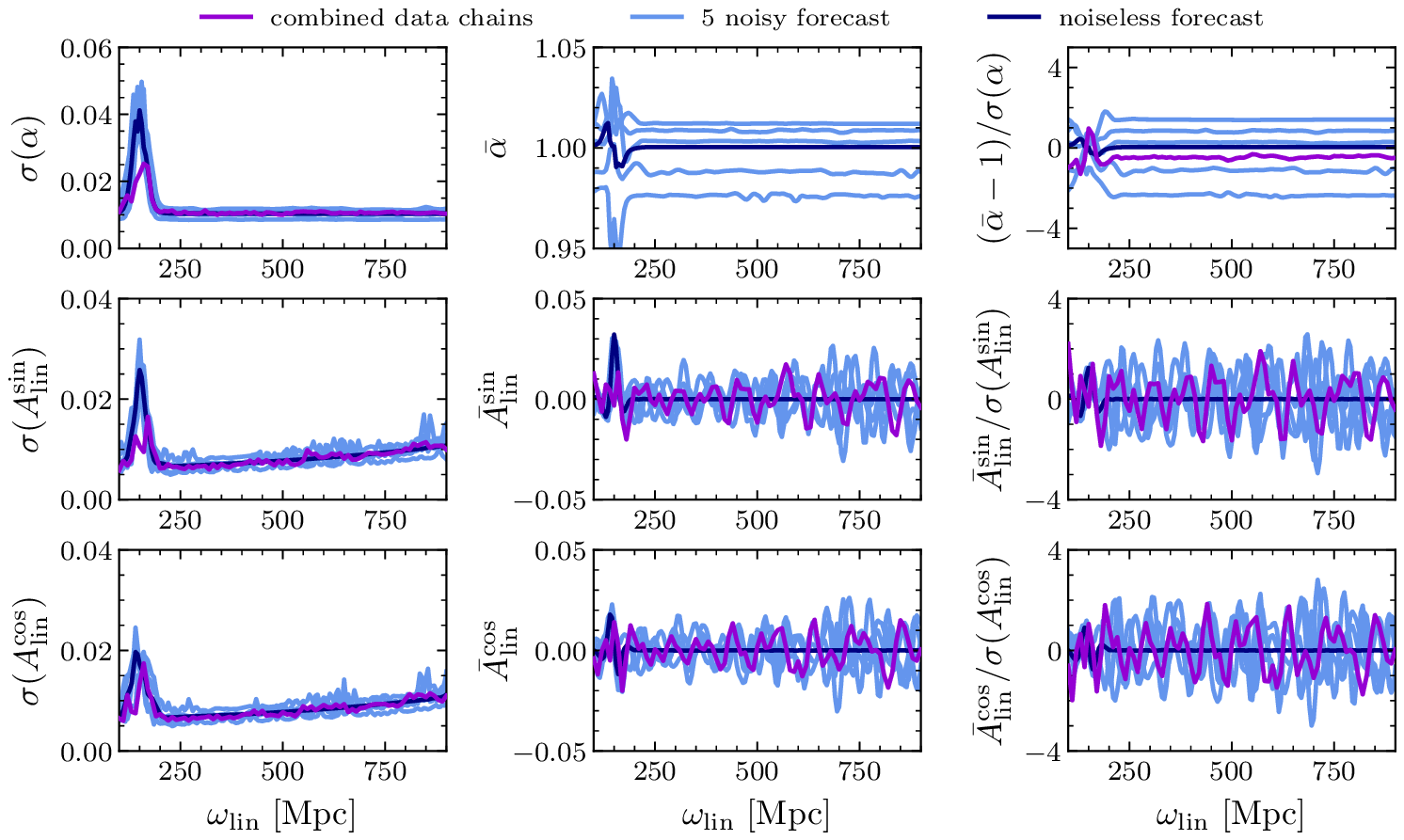}
	\includegraphics{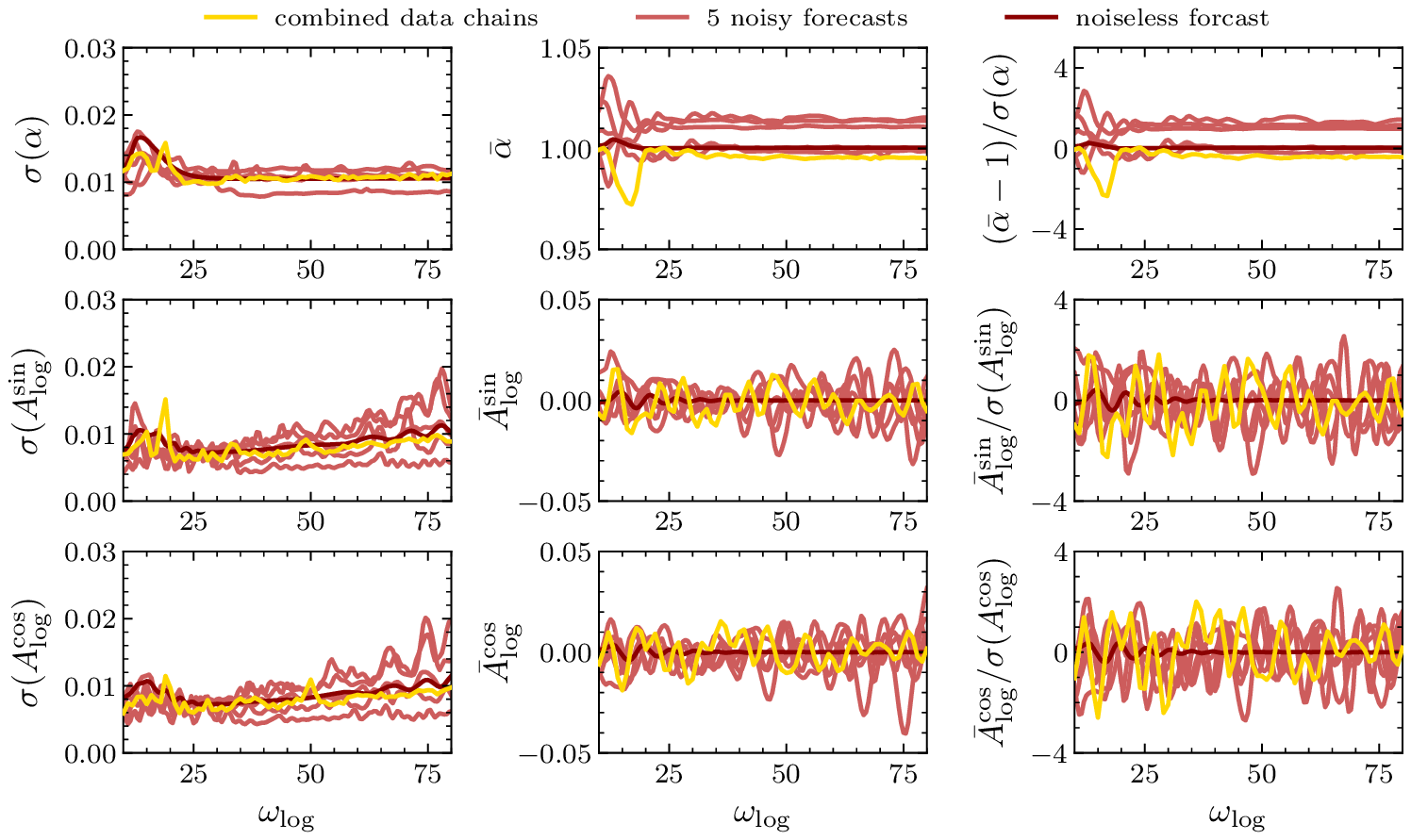}
	\caption{Comparison of the analysis of the BOSS~DR12~dataset with likelihood-based forecasts for linear~(\textit{top}) and logarithmic~(\textit{bottom}) features in the low-redshift bin with the same quantities as in Fig.~\ref{fig:mocksForecastComparison}. We again observe very good agreement.}
	\label{fig:dataForecastComparison}
\end{figure}

We now apply our analysis and forecasting pipeline to the BOSS~DR12 dataset. Figure~\ref{fig:dataForecastComparison}%
shows the posterior distributions derived from the Monte Carlo Markov chains of the low-redshift bin in terms of the same characterizing statistical quantities as in Fig.~\ref{fig:mocksForecastComparison} for the mock catalogs. When comparing these results inferred from the data chains with likelihood-based forecasts, we again find very good agreement for the low-redshift bin and similar results for the high-redshift bin. We reiterate that these forecasts differ from those in Fig.~\ref{fig:mocksForecastComparison} especially in the value of the nonlinear damping scale. Since the smaller damping scale in the data leads to a larger signal-to-noise ratio and considerably extends the range of wavenumbers contributing to the feature search, we observe a smaller variance, less scatter in the mean values and a smaller (but statistically consistent) number of greater-than-$2\sigma$~fluctuations than in the mocks. The fact that the inferred significances in the third column of Fig.~\ref{fig:dataForecastComparison} agree well with those found in the noisy forecasts indicates that we do not have any significant detection of a feature, but rather that the data analysis is consistent with fitting experimental noise. We note that the oscillations in $\bar{A}_\Log^Y$, $Y=\sin,\cos$, that are visible towards smaller~$\omegalog$ in the noiseless forecasts, arise due to interference of the logarithmic feature spectrum with the BAO~spectrum in the range $k \sim \SIrange{0.1}{0.2}{\hPerMpc}$. The noisy forecasts show however that this does not impact our BOSS~analysis.

\vskip4pt
Having established the reliability and robustness of our data analysis in the amplitude parameterization of~$(\Asin_X,\Acos_X)$, we want to infer the constraints on the overall feature amplitude~$A_X$ while marginalizing over the phase~$\phase_X$. Since we do not find any significant detections (see also Appendix~\ref{app:analysisDetails}), we are mainly interested in deriving limits on the presence of primordial features which is why we take $\phase_X \in [0,2\pi)$ and the amplitude to be positive semi-definite:
\beq
A_X = \sqrt{\left(\Asin_X\right)^2 + \left(\Acos_X\right)^2}\, .
\eeq
In this way, we can directly infer the upper limits on~$A_X$ at 95\%~c.l.\ from the Markov chains of the low- and high-redshift bin, respectively. To derive constraints from the entire BOSS~DR12~data, we combine the two sets of Markov chains by multiplying the binned posterior distributions. In this process, we neglect a possible correlation between the BAO~parameter~$\alpha$ and the feature amplitudes~$\Asin_X$ and~$\Acos_X$. As previously noted, this correlation is however small away from feature frequencies around the BAO~scale and taking this correlation into account would only strengthen the inferred bounds. We refer to Appendix~\ref{app:analysisDetails} for an extensive discussion and further details.

\begin{figure}
	\centering
	\includegraphics{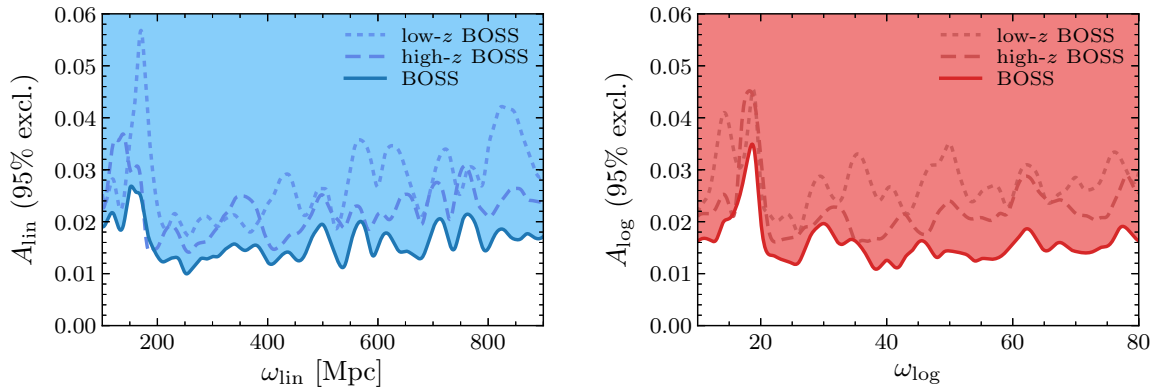}
	\caption{Upper limits on the feature amplitude~$A_X$ at 95\%~c.l.\ as a function of the feature frequency~$\omega_X$ for linear~(\textit{left}) and logarithmic oscillations~(\textit{right}), $X=\lin,\Log$, from BOSS~DR12~data alone. The dotted and dashed lines show the bounds that are separately inferred from the low- and high-redshift bins, while the solid line indicates the current limits from~LSS by combining the two BOSS~redshift bins.}
	\label{fig:constraints_baoOnly}
\end{figure}

We present the resulting first constraints on linear and logarithmic features from large-scale structure data alone, i.e.\ without the inclusion of any other external datasets or information, in Fig.~\ref{fig:constraints_baoOnly}. %
Our analysis limits the amplitude of these primordial feature models,~$\Alin$ and~$\Alog$, to be less than one to two percent of the primordial scalar amplitude~$\As$ in the range of feature frequencies accessible with~BOSS. Moreover, we do not find any significant detections of features as expected from Fig.~\ref{fig:dataForecastComparison} (see also Appendix~\ref{app:analysisDetails}).

\subsection{Comparison with Planck CMB~Bounds}

\begin{figure}
	\centering
	\includegraphics{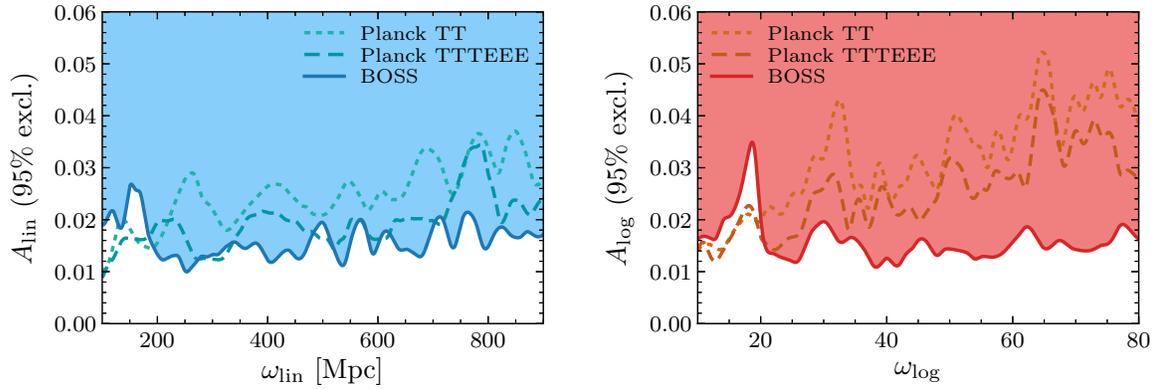}
	\caption{Comparison of the 95\%~upper limits on the feature amplitudes~$A_X$, $X=\lin,\Log$, from~LSS and the~CMB for linear~(\textit{left}) and logarithmic features~(\textit{right}). The solid lines indicate our new BOSS-only results and are identical to the solid lines of Fig.~\ref{fig:constraints_baoOnly}. The bounds from Planck~2015 temperature~(dotted) and temperature+polarization data~(dashed) are for the first time displayed as a function of feature frequency as well. Beyond those frequencies which show a degeneracy with the standard BAO~spectrum, the BOSS~data are able to improve over the~CMB.}
	\label{fig:constraints_bao+cmb}
\end{figure}

While we present the first limits on feature models from~LSS alone, constraints have been inferred from CMB~observations for more than a decade (cf.\ e.g.~\cite{Martin:2003sg, Pahud:2008ae, Adshead:2011jq, Meerburg:2011gd, Dvorkin:2011ui, Peiris:2013opa, Planck:2013jfk, Meerburg:2013cla, Meerburg:2013dla, Easther:2013kla, Miranda:2013wxa, Fergusson:2014tza, Ade:2015lrj, Akrami:2018odb}). It is therefore interesting to compare the deduced constraints. While the frequency coverage is wider in the~CMB, our LSS-only bounds interestingly improve the limits derived from current CMB~data by up to a factor of~2.3 and~3.1 for $\omegalin\gtrsim\SI{200}{\Mpc}$ and $\omegalog\gtrsim20$, respectively. This is illustrated in Fig.~\ref{fig:constraints_bao+cmb}, %
which directly compares the constraints on the feature amplitudes from our BOSS~analysis with those deduced from current CMB temperature~(TT), and temperature and polarization~(TTTEEE) data released by the Planck collaboration in~2015~\cite{Adam:2015rua, Aghanim:2015xee} (see Appendix~\ref{app:cmbAnalysis} for details on these limits).\footnote{We show the constraints from both~TT and~TTTEEE since the Planck collaboration had labeled the results employing high-multipole polarization data as preliminary in~2015. Having said that, the available information on feature models released by the collaboration seems to have remained fairly stable between their 2015~and 2018~releases~\cite{Ade:2015lrj, Akrami:2018odb}. We will therefore use the polarization data when deriving the joint constraints below.} Since the common focus of previous analyses was on the best-fit points or the likelihood improvement, we note that the limits on the feature amplitudes from the~CMB have not been shown as a function of frequency before.

The improvements of our LSS~bounds over those from Planck are primarily the consequence of two effects. First, the number of signal-dominated modes over the employed range of wavenumbers in~BOSS and~Planck are roughly comparable (approximately $\kmax^3 V$ and $\lmax^2$, respectively). Second, the imprint of high-frequency oscillations in the CMB power spectra is suppressed relative to that in the matter power spectrum, as shown in Fig.~\ref{fig:cmbLSScomparison} of Appendix~\ref{app:cmbAnalysis}. In combination, the signal-to-noise ratio of a high-frequency feature is somewhat larger in~BOSS than in~Planck which leads to a more stringent constraint.

\vskip4pt
Finally, we can infer the best current limits on primordial linearly- and logarithmically-oscillating feature models by combining the~BOSS and Planck~data.\footnote{Combined analyses of CMB~and LSS~data have previously been explored in~\cite{Hu:2014hra, Hunt:2015iua, Benetti:2016tvm, Zeng:2018ufm} by employing measurements of the linear matter power spectrum over a limited range of wavenumbers without nonlinear modeling.} These joint constraints are derived in Appendix~\ref{app:cmbAnalysis} and shown in Fig.~\ref{fig:constraints_joint}. %
\begin{figure}
	\centering
	\includegraphics{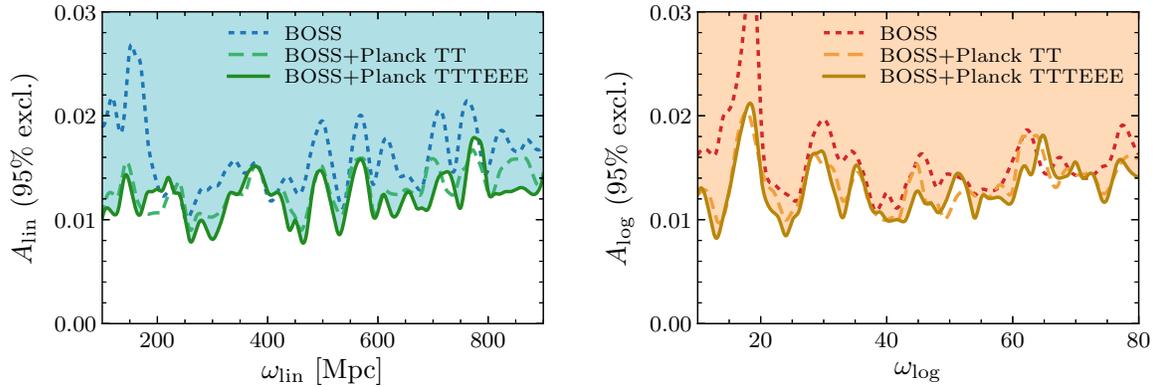}
	\caption{Joint BOSS and Planck upper limits at 95\%~c.l.\ on the linear~(\textit{left}) and logarithmic~(\textit{right}) feature amplitudes~$A_X$, $X=\lin,\Log$. The best current constraints come from a combination of BOSS~DR12 and Planck~2015~TTTEEE data (solid). We also show the BOSS+Planck~TT~(dashed) results and include the BOSS-only bounds~(dotted) for comparison.}
	\label{fig:constraints_joint}
\end{figure}
As expected, we observe that these bounds are dominated by and, therefore, closely follow our limits from galaxy clustering data of~BOSS except at smaller frequencies. Generally speaking, the bounds on features in the discussed range of frequencies~$\omega_X$ are now established at the one-percent level relative to the primordial power spectrum.

\subsection{Future LSS and CMB Constraints}
\label{sec:futureConstraints}

With the discussed improvements in the constraints on primordial features inferred from~BOSS over those derived from Planck~CMB~data, it is timely to ask how these bounds will evolve with future~CMB and LSS~surveys. To this end, we performed Fisher matrix forecasts for upcoming, planned and futuristic experiments. We extend previous LSS~forecasts (cf.\ e.g.~\cite{Chen:2016zuu, Chen:2016vvw, Ballardini:2016hpi, Xu:2016kwz, Fard:2017oex, Palma:2017wxu, LHuillier:2017lgm, Ballardini:2017qwq}) in a number of ways, in particular by taking the effects of nonlinearities, bandpowers and window functions into account, and (conservatively) marginalizing over further uncertainties in the broadband power spectrum. Furthermore, we compare the reach of LSS~surveys to that of future CMB~missions. In this section, we focus on linear features since most other features can be easily decomposed into a basis of linear oscillations. Before discussing the results of these forecasts, we briefly summarize our approach and refer to Appendices~\ref{app:forecasts} and~\ref{app:cmbAnalysis} for further details.

\vskip4pt
For our future LSS~forecasts, we use the relative wiggle spectrum $\mathcal{O}_{\!g}(k) \equiv \Pw_g(k)/\Pnw_g(k)$ as the observable up to $\kmax=\SI{0.5}{\hPerMpc}$ based on~\cite{Baumann:2017gkg} as outlined in Appendix~\ref{app:forecasts}, including the effects of nonlinearities, bandpowers and window function.\footnote{In this way, we also find that our choice of $\kmax=\SI{0.3}{\hPerMpc}$ in the described analysis captures essentially all the information on features available in the BOSS~DR12 dataset.} To estimate the sensitivity of the~CMB, we directly follow the methodology of~\cite{Baumann:2017gkg}, employing perfectly delensed temperature and polarization power spectra. The fiducial point in both cases is a featureless $\Lambda$CDM~cosmology consistent with the Planck measurements~\cite{Ade:2015xua, Aghanim:2018eyx}. After computing the Fisher matrices in the amplitude parameterization, we obtain the forecasted 95\%~upper limits on~$\Alin$ by randomly sampling from the associated Gaussian distributions and applying the same procedure as in our BOSS~analysis (see Appendices~\ref{app:forecasts} and~\ref{app:analysisDetails}).

\begin{figure}
	\centering
	\includegraphics{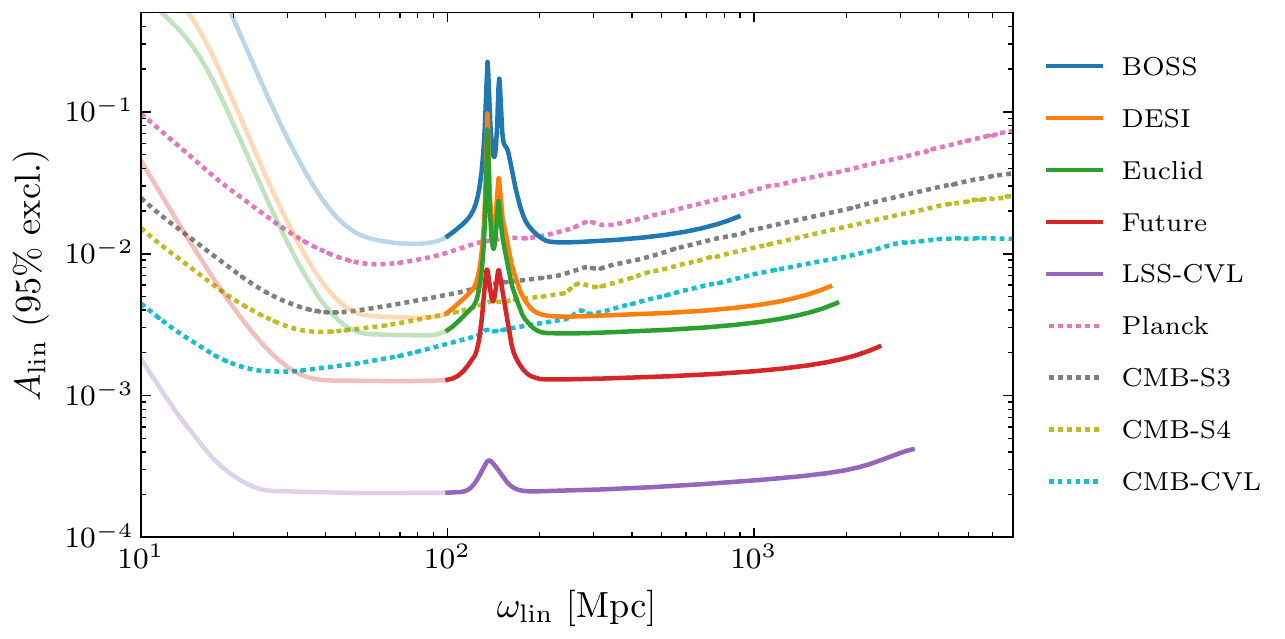}
	\caption{Forecasted sensitivity for the ``feature spectrometer'' of linear features. The potential reach of various LSS~(solid) and CMB~(dashed) experiments to constrain the feature amplitude~$\Alin$ at a confidence level of~95\% is presented as a function of their frequency~$\omegalin$. We refer to the main text (and Appendix~\ref{app:forecasts}) for the details regarding the experiments and note that the positive-semi-definite nature of~$\Alin$ is taken into account in the displayed estimates. Large-scale structure surveys have the potential to improve over the~CMB by more than one order of magnitude, while the~CMB will always dominate the reach in feature frequency. As we discussed in Section~\ref{sec:theory}, the LSS~forecasts for $\omegalin\lesssim\SI{100}{\Mpc}$ should be treated with care since these frequencies might be affected by the effects of nonlinear gravitational evolution on small scales and be generally more sensitive to the details of signal modeling.}
	\label{fig:futureForecast}
\end{figure}

\vskip4pt
The resulting forecasted sensitivity of several~LSS and CMB~experiments is illustrated in Fig.~\ref{fig:futureForecast}. %
Apart from~BOSS and~Planck, we included the planned surveys DESI~\cite{Aghamousa:2016zmz}, Euclid~\cite{Laureijs:2011gra} and~CMB-S3 as an umbrella for the multiple upcoming CMB~missions~\cite{Benson:2014qhw, Henderson:2015nzj, Ade:2018sbj}. In addition, we show the potential reach of CMB-S4~\cite{Abazajian:2016yjj} (or, similarly, PICO~\cite{Hanany:2019lle}) and a `Future' LSS~experiment which is assumed to map about \num{e8}~objects up to redshift $z=3$ over half of the sky. To get a sense for the theoretically possible limits, we also forecast a half-sky, cosmic-variance-limited LSS~survey with $z\leq6$ and $\kmax=\SI{0.75}{\hPerMpc}$ (`LSS-CVL'),\footnote{A similar performance could in principle be achieved by a \SI{21}{cm}~intensity mapping survey~\cite{Obuljen:2017jiy, Ansari:2018ury, Bandura:2019uvb}.} and a CMB~experiment that measures the temperature and polarization spectra to the cosmic variance limit up to multipoles of $\lmax^T=3000$ and $\lmax^P=5000$ on 75\%~of the sky (`CMB-CVL'). We note that the peak in the forecasted limits around $\omegalin = r_s \approx \SI{150}{\Mpc}$ is due to the (approximate) degeneracy of the feature with the BAO~spectrum over the signal-dominated range of wavenumbers as previously discussed in~\textsection\ref{sec:mocks}.

The possibly most notable aspect of these forecasts is that the coming generation of surveys, in particular~DESI and Euclid, are projected to be more sensitive than a cosmic-variance-limited CMB~experiment over a substantial range of frequencies. In this precise sense, large-scale structure will permanently surpass the~CMB in sensitivity. Equally significant is the potential for future LSS~observations to increase their constraining power on feature models. The new BOSS~limit presented in this work and the forecasts for LSS-CVL leave approximately two orders of magnitude that could be achievable with a suitably designed survey. As mentioned previously, it will however be necessary to revisit some of the aspects of the analysis that we employed on BOSS~data to credibly achieve such sensitivities.

The improvements seen in future surveys come primarily from two factors: smaller shot noise and higher redshifts. The constraining power of a survey is dominated by the number of signal-dominated $k$-modes. Most of these modes are at large wavenumbers, but are limited by the shot noise of the survey. The significant increase in the number density of objects available in upcoming surveys substantially increases the number of modes and drives the improvements in sensitivity. In addition, the larger redshift range of these observations means that the nonlinear damping is reduced, increasing the size of the signal at higher wavenumbers.\footnote{Recent advances in density field reconstruction (see e.g.~\cite{Wang:2017jeq, Schmittfull:2017uhh, Hada:2018fde, Schmidt:2018bkr, Birkin:2018nag, Sarpa:2018ucb, Modi:2019hnu, Zhu:2019gzu}) also have the potential to effectively further decrease the nonlinear damping scale. An increase of the reconstruction efficiency from about 50\% to 100\% would lead to an improvement in the forecasted sensitivity to~$\Alin$ of factors in excess of 3, 2 and 1.5 for BOSS, DESI and LSS-CVL, for instance, and the possibility of gaining signal-to-noise for larger values of~$\kmax$. This is a considerable improvement compared to the results shown in Fig.~\ref{fig:futureForecast} and substantially larger than what we expect for the BAO~frequency, i.e.\ the search for primordial features might give new motivation to develop more efficient reconstruction techniques.} Furthermore, future surveys will also benefit from larger survey volumes which can be seen clearly in the larger range of feature frequencies~$\omegalin$ that are accessible. This is because increasing the volume allows for finer $k$-bins, which results in a larger Nyquist frequency~$\omega_\mathrm{Ny}$.

\vskip4pt
In summary, LSS~bounds on features are currently competitive with and will surpass those from the~CMB (present and future) over an increasing range of frequencies. Large-scale structure observations have a significantly larger sensitivity over their available frequency range due to the large number of modes. Furthermore, the transfer of primordial power to the matter power spectrum is more efficient than for the~CMB, which leads to a larger intrinsic signal (see Appendix~\ref{app:cmbAnalysis} for a more detailed discussion). On the other hand, the~CMB can cover a wider range of frequencies than will be accessible even with futuristic LSS~surveys.

%%%%%%%%%%%%%%%%
\section{Conclusions}
\label{sec:conclusions}
%%%%%%%%%%%%%%%%

In this paper, we explored the impact of large-scale structure data on the search for primordial features in the power spectrum. We showed theoretically that such analyses are promising since they are not limited by the small-scale nonlinearities of structure formation and the exponential damping caused by large-scale bulk flows can be reliably computed (as we explicitly did at leading order for both linear and logarithmic features). We then applied these results to BOSS~DR12 data and found constraints comparable to (but somewhat stronger than) the best limits from Planck. The joint bounds on these models are therefore dominated by the galaxy clustering data. Moreover, we forecast that near-term surveys improve on this result by up to an order of magnitude and could out-perform a cosmic-variance-limited CMB~experiment over a substantial range of feature frequencies.

\vskip4pt
Large-scale structure surveys offer great promise for dramatically improving our understanding of the very early universe. However, to date, these hopes have been largely limited by the modeling uncertainties around the nonlinear scale. In this work, we have however shown that, for the right observables, the statistical power of current surveys is already sufficient to significantly impact our understanding of inflation and beyond.

\vskip4pt
While our emphasis was on primordial features, in particular from an inflationary origin, both the method and the results have significantly broader implications. Any sufficiently sharp feature in the matter power spectrum could be analyzed in this way and could even be decomposed in a basis of linear oscillations. We expect that constraints from~LSS will be competitive with those derived from the~CMB, provided that the signal appears directly in the (dark) matter and is not suppressed by the baryon fraction.

\vskip4pt
Finally, the statistical power of this approach is not limited to the power spectrum and ultimately could be extended to higher-point statistics. Primordial features are known to have associated non-Gaussian signatures (see e.g.~\cite{Chen:2006xjb, Flauger:2010ja, Achucarro:2013cva, Gong:2014spa, Palma:2014hra, Flauger:2016idt}) which should similarly be robust to the complications presented by nonlinear evolution. This presents the unique opportunity not only to perform joint CMB~power and bispectrum analyses~\cite{Fergusson:2014hya, Fergusson:2014tza, Meerburg:2015owa, Akrami:2018odb}, but to also include the respective LSS~observables. Furthermore, the three-dimensionality of galaxy surveys may allow for entirely new types of analyses that exploit the full angular dependence of higher-point correlation functions. The universe has given us the unprecedented power of large-scale structure to answer the most basic questions of our cosmic origins. We contributed a small step towards this ultimate goal.

\vskip20pt
\paragraph{Acknowledgements}
We thank Daniel Baumann, Dick Bond, Raphael Flauger, Mikhail Ivanov, Marcel Schmittfull, Zvonimir Vlah and Matias Zaldarriaga for helpful conversations, and Anagha Vasudevan, Mikhail Ivanov, Sergey Sibiryakov and Julien Lesgourgues for sharing a draft of their work~\cite{Vasudevan:2019ewf} with us. F.\,B.~is a Royal Society University Research Fellow. M.\,B.~acknowledges support from the Netherlands Organisation for Scientific Research~(NWO), which is funded by the Dutch Ministry of Education, Culture and Science~(OCW), under VENI~grant~016.Veni.192.210. D.\,G.~is supported by the US~Department of Energy under grant no.~DE-SC0019035. B.\,W.~gratefully acknowledges support from the Marvin L.\ Goldberger Membership at the Institute for Advanced Study, NSF~Grant PHY-1820775, a Research Studentship Award of the Cambridge Philosophical Society, the Simons Foundation and a Visiting PhD Fellowship of the Delta-ITP consortium, a program of~NWO.

This work is based on observations obtained by the Sloan Digital Sky Survey~III (\mbox{SDSS-III}, \href{http://www.sdss3.org/}{http:/\!/www.sdss3.org/}). Funding for SDSS-III has been provided by the Alfred P.~Sloan Foundation, the Participating Institutions, the National Science Foundation and the US~Department of Energy Office of Science. This research also partly uses observations obtained with the Planck satellite (\href{http://www.esa.int/Planck}{http:/\!/www.esa.int/Planck}), an ESA~science mission with instruments and contributions directly funded by ESA~Member States, NASA and Canada.
Parts of this work were performed using resources provided by the Cambridge Service for Data Driven Discovery~(CSD3) operated by the University of Cambridge Research Computing Service (\href{http://www.csd3.cam.ac.uk/}{http:/\!/www.csd3.cam.ac.uk/}), provided by Dell~EMC and Intel using Tier-2 funding from the Engineering and Physical Sciences Research Council (capital grant EP/P020259/1), and DiRAC~funding from the Science and Technology Facilities Council (\href{http://www.dirac.ac.uk/}{http:/\!/www.dirac.ac.uk/}). The DiRAC~component of~CSD3 was funded by BEIS~capital funding via STFC~capital grants~ST/P002307/1 and~ST/R002452/1, and STFC~operations grant~ST/R00689X/1. DiRAC is part of the National e-Infrastructure.

We acknowledge the use of \texttt{CAMB}~\cite{Lewis:1999bs}, \texttt{CLASS}~\cite{Blas:2011rf}, \texttt{CosmoMC}/\texttt{GetDist}~\cite{Lewis:2002ah}, \texttt{IPython}~\cite{Perez:2007ipy}, \texttt{MultiNest}~\cite{Feroz:2007kg, Feroz:2008xx} and the Python packages \texttt{emcee}~\cite{ForemanMackey:2012ig}, \texttt{Matplotlib}~\cite{Hunter:2007mat}, \texttt{nbodykit}~\cite{Hand:2017pqn} and \texttt{NumPy}/\texttt{SciPy}~\cite{Walt:2011num}.

\clearpage
\appendix
%%%%%%%%%%%%%%%%
\section{Nonlinear Damping of Logarithmic Features} 
\label{app:nonlinearDampingCalculation}
%%%%%%%%%%%%%%%%

The effect of gravitational nonlinearities on the BAO~signal has been considered in various ways and can easily be extended to linear features as we discuss in Section~\ref{sec:theory}. The effects of large-scale gravitational nonlinearities on logarithmically-spaced oscillations have however not been considered previously.\footnote{The authors of~\cite{Vasudevan:2019ewf} independently performed this calculation.} In this appendix, we provide additional details on the computation of the resulting damping of these features, complementing the discussion in~\textsection\ref{sec:largeScaleNonlinearities}. As in the main text, we first detail the perturbative treatment and then resum the infrared contributions to all orders in perturbation theory.

\subsection{Perturbative Treatment}

We have found in~\eqref{eq:1loop} that the effect of long modes on a generic wiggle power spectrum~$\Pw(k)$ at one-loop order implies the action of the derivative operator~$\cosh\hskip-1pt\left({\q\cdot\nabla_\k}\right)$ on~$\Pw(k)$. For logarithmic features, we need to consider $\Pw_\Log(k) = \Pnw(k)\, \Alog \sin\!\left[\omegalog \log(\kappa)+\phaselog\right]$, with $\kappa = k/\kp$. As in the case of linearly-spaced oscillations~\eqref{eq:coshlin}, we neglect small corrections that arise from applying the derivative operators to the smooth envelope~$\Pnw(k)$. Moreover, in order to avoid clutter, we set $\phaselog=0$ in the following, but note that it is straightforward to include the phase in the calculation. The $(2n)^\text{th}$~derivative of the oscillatory part, $\sin\!\left[\omegalog \log(\kappa)\right]$, is given by
\begin{align}
\nabla_{k_{i_1}}\dots\nabla_{k_{i_{2n}}} \sin\!\left[\omegalog \log(\kappa)\right] 
	&= \frac{ \hat{k}_{i_1} \cdots \hat{k}_{i_{2n}}}{2\ii\,k^{2n}} \left[f_n(\omegalog)\,\kappa^{\ii \omegalog}-f_n(-\omegalog) \kappa^{-\ii \omegalog}\right]	\nonumber	\\
	&=\frac{\hat{k}_{i_1} \cdots \hat{k}_{i_{2n}}}{k^{2n}}\left\{\left[f_n(\omegalog)+f_n(-\omegalog)\right] \sin\!\left[\omegalog \log(\kappa)\right]\right.	\\
	&\hphantom{=\frac{\hat{k}_{i_1} \cdots \hat{k}_{i_{2n}}}{k^{2n}}\{\,} \left.-\ii\left[f_n(\omegalog)-f_n(-\omegalog)\right]\cos\!\left[\omegalog \log(\kappa)\right]\right\} ,	\nonumber
\end{align}
where we employed
\beq
\sin (x \log y) = \frac{1}{2 \ii} \left( \ee^{\ii x \log y} -\ee^{-\ii x \log y} \right) = \frac{1}{2\ii} \left( x^{\ii y}- x^{-\ii y}\right) , \quad f_n(\omegalog) \equiv \frac{(\ii \omegalog)!}{(\ii \omegalog-2n)!}\, .
\eeq
We can then perform the sum over~$n$ and get
\begin{align}
\Bigr[\cosh\!\left({\q\cdot\nabla_\k}\right)-1\Bigr] \Pw_\Log(k) 
	&= \left\{\cos \left[\omegalog \log\!\left(1-\frac{q\,\mu}{k}\right)\right]-1\right\} \Pw_\Log(k)	\nonumber	\\
	&\hphantom{=\ } + \left\{\sin \left[\omegalog \log\!\left(1-\frac{q\,\mu}{k}\right)\right]\right\} \left\{\Alog \cos\!\left[\omegalog \log\!\left(\frac{k}{\kp}\right)\right]\right\}\Pnw(k)\nonumber	\\
	&= \left\{\cos \left[\omegalog \log\!\left(1-\frac{q\,\mu}{k}\right)\right]-1\right\} \Pw_\Log(k)	\nonumber	\\
	&\hphantom{=\ }+ \left\{\sin \left[\omegalog \log\!\left(1-\frac{q\,\mu}{k}\right)\right]\right\} \frac{\Pnw(k)}{\omegalog} \frac{\d \delta P_\zeta^\Log(k)}{\d \log k}\, ,	\label{eq:ddop}
\end{align}
where we defined $\mu \equiv \hat{k}\cdot\hat{q}$. It is also useful to consider
\begin{align}
\Bigr[\cosh\!\left({\q\cdot\nabla_\k}\right)-1\Bigr] \frac{\Pnw(k)}{\omegalog} \frac{\d \delta P_\zeta^\Log(k)}{\d \log k} 
	&= \left\{\cos \left[\omegalog \log\!\left(1-\frac{q\,\mu}{k}\right)\right]-1\right\} \frac{\Pnw(k)}{\omegalog} \frac{\d \delta P_\zeta^\Log(k)}{\d \log k}	\nonumber	\label{eq:ddop2}	\\
	&\hphantom{=\ }- \left\{\sin \left[\omegalog \log\!\left(1-\frac{q\,\mu}{k}\right)\right]\right\} \Pw_\Log(k)\, ,
\end{align}
since we need this expression in the following calculation of the IR-resummed damping.

\subsection{Infrared Resummation}

In~\textsection\ref{sec:damping_perturbative}, we showed that $\Pw_{1\text{-loop}} \approx \mathcal{O}(1) \Pw_{\text{tree-level}}$, which suggests that all higher-order terms might be equally important corrections to the linear wiggle power spectrum of logarithmic features. In order to resum these infrared contributions, we need to evaluate all the higher-loop diagrams of~\eqref{eq:lthloop}. In contrast to linear features, there is no straightforward way to write down the $L$-loop contribution based on the $1$-loop result, which is why we proceed by induction. Let us start with the two-loop contribution, $L=2$, to get some intuition:
\begin{align}
\Pw_{2\text{-loop},\mathrm{LO}}(k) 
	&=\frac{1}{8}\int^\Lambda\! \frac{\d^3q_1}{(2\pi)^3} \frac{\d^3 q_2}{(2\pi)^3}\, \Pnw(q_1)\Pnw(q_2) \mathcal{D}_{\q_1} \mathcal{D}_{-\q_1}\, \mathcal{D}_{\q_2} \mathcal{D}_{-\q_2} \Pw(k)	\nonumber	\\
	&= -\frac{k^2}{4} \int^\Lambda\! \frac{\d^3q_1}{(2\pi)^3}\, \Pnw(q_1) \mathcal{D}_{\q_1} \mathcal{D}_{-\q_1} \left[ \Sigma^2_\Log(k) \Pw_\Log(k) + \hat{\Sigma}^2_\Log(k)\, \frac{\Pnw(k)}{\omegalog} \frac{\d \delta P_\zeta^\Log(k)}{\d \log k} \right]	\nonumber	\\
	&= \frac{k^4}{2} \left[\Sigma^2_\Log(k) \left(\Sigma^2_\Log(k) \Pw_\Log(k) + \hat{\Sigma}^2_\Log(k)\, \frac{\Pnw(k)}{\omegalog} \frac{\d \delta P_\zeta^\Log(k)}{\d \log k}\right) \right.	\nonumber	\\
	&\hphantom{=\frac{k^4}{2}\ \ \,} +\left.\hat{\Sigma}^2_\Log(k) \left( \Sigma^2_\Log(k) \frac{\Pnw(k)}{\omegalog} \frac{\d \delta P_\zeta^\Log(k)}{\d \log k} - \hat{\Sigma}^2_\Log(k) \Pw_\Log(k) \right) \right]	\nonumber	\\
	&= \frac{k^4}{2} \left[ \left( \Sigma^4_\Log(k) - \hat{\Sigma}^4_\Log(k)\right) \Pw_\Log(k) + 2\Sigma^2_\Log(k)\, \hat{\Sigma}^2_\Log(k)\, \frac{\Pnw(k)}{\omegalog} \frac{\d \delta P_\zeta^\Log(k)}{\d \log k}\right] ,
\end{align}
where we used~\eqref{eq:ddop} and~\eqref{eq:ddop2}. Here, we should note that the operator~$\mathcal{D}_{\q}\, \mathcal{D}_{-\q}$ does not act on~$\Sigma^2_\Log(k)$ or~$\hat{\Sigma}^2_\Log(k)$ since it only acts on the wiggle power spectra~\cite{Blas:2016sfa}. The three-loop term can be derived along the same lines to be
\begin{align}
\Pw_{3\text{-loop},\mathrm{LO}}(k) 
	= -\frac{k^6}{6} \Biggr[& \left( \Sigma^6_\Log(k) - 3\Sigma^2_\Log(k)\, \hat{\Sigma}^4_\Log(k) \right) \Pw_\Log(k)	\nonumber	\\
							&+\left( 3\Sigma^4_\Log(k) \hat{\Sigma}^2_\Log(k) - \hat{\Sigma}^6_\Log(k)\right) \frac{\Pnw(k)}{\omegalog} \frac{\d \delta P_\zeta^\Log(k)}{\d \log k} \Biggr]\, .
\end{align}
Considering the one-, two- and three-loop contributions, it becomes apparent that the structure at $L^\text{th}$~order is given by
\begin{align}
\Pw_{L\text{-loop},\mathrm{LO}}(k) 
	=\frac{(\ii k)^{2L}}{L!} \Biggr\{& \frac{1}{2} \hskip-1pt\left[ \left(\Sigma^2_\Log(k) + \ii \hat{\Sigma}^2_\Log(k)\right)^{\!L}\!\! + \left(\Sigma^2_\Log(k) - \ii \hat{\Sigma}^2_\Log(k)\right)^{\!L} \right]\! \Pw_\Log(k)	\nonumber	\\
									 &\!\!+\frac{1}{2\ii} \hskip-1pt\left[ \hskip-1pt \left(\Sigma^2_\Log(k) + \ii \hat{\Sigma}^2_\Log(k)\right)^{\!L}\!\! - \left(\Sigma^2_\Log(k) - \ii \hat{\Sigma}^2_\Log(k)\right)^{\!L} \right] \!\frac{\Pnw(k)}{\omegalog} \frac{\d \delta P_\zeta^\Log(k)}{\d \log k}\Biggr\} .
\end{align}
The IR-resummed wiggle power spectrum of~\eqref{eq:final} is then obtained by resumming all the loops.

\clearpage
%%%%%%%%%%%%%%%%
\section{Large-Scale Structure Forecasts} 
\label{app:forecasts}
%%%%%%%%%%%%%%%%

We employ a suite of likelihood- and Fisher-based forecasts in particular to validate and cross-check our analysis pipeline, and investigate the potential reach of future surveys. In this appendix, we collect further details regarding these LSS~forecasts~(\textsection\ref{app:baoForecasts}) and collect the utilized experimental specifications~(\textsection\ref{app:lssSpecs}). Furthermore, we provide additional checks of our feature search~(\textsection\ref{app:checks}) as well as supplementary information for the forecasts of future experiments~(\textsection\ref{app:futureLSS}).

\subsection{Forecasting with the Wiggle Spectrum}
\label{app:baoForecasts}

As previously stated, the wiggle~spectrum is the main observable in our forecasting pipeline, which was developed in~\cite{Baumann:2017gkg} for the standard BAO~spectrum. In the following, we summarize its main aspects and introduce further advances which especially include the use of bandpowers and the convolution with a window function. These components are not required in a wide range of applications, such as light relics, but are important to reliably predict the sensitivity to (highly-)oscillating features. 

\vskip4pt
We use two types of forecasts in this work, which are based either on the Fisher information matrix~$F_{ij}$ or on the likelihood function~$\L$ itself. The former are computationally efficient and are therefore very useful in particular to cover a large space of parameters and experimental specifications. However, they only allow to access the standard deviation around a fixed fiducial point assuming smooth noise and have to also be taken with care given the involved approximations. We therefore only employ these forecasts to estimate the sensitivity of future surveys and for a limited number of tests. The likelihood-based forecasts come with a larger computational cost, but are much more versatile. For instance, we can not only obtain the standard deviations, but can also extract the mean values which allows us to estimate significances and provides more direct comparisons with MCMC~analyses. In addition, it is possible to inject random noise realizations and/or artificial feature signals. For these reasons, the majority of forecasts in this work are of the latter type. In the following, we first discuss the Fisher methodology, since it is commonly employed, and especially highlight modifications to the standard approach. We then build on this pipeline and introduce the likelihood-based forecasts.

\subsubsection{Fisher Matrix Forecasts}
Focusing on the oscillatory part of the power spectrum, the Fisher matrix of a galaxy survey with multiple (independent) redshift bins~$z$ can generally be approximated by~\cite{Baumann:2017gkg}\hskip1pt\footnote{This is based on the standard Fisher matrix for galaxy surveys of~\cite{Tegmark:1997rp} which employs the galaxy power spectrum~$P_g(k,\mu)$ as the observable.}
\beq
F_{lm} = \sum_{z,\, k_i} \frac{\Delta k\, k_i^2}{(2\pi)^2} \int_{-1}^1\! \frac{\d\mu}{2}\, \frac{D_z(k_i,\mu)^2 }{(1+D_z(k_i,\mu) \, O_z(k_i,\mu))^2} \frac{\partial \mathcal{O}_z(k_i,\mu)}{\partial \theta_l} \frac{\partial \mathcal{O}_z(k_i,\mu)}{\partial \theta_m} \,V_\mathrm{eff}(k_i,\mu;z)\, ,	\label{eq:baoFisherMatrix}
\eeq
where~$\mu$ is the cosine between the wavevector~$\k$ and the line-of-sight, $\mathcal{O}_z(k,\mu) = \Pw_z(k,\mu)/\Pnw_z(k,\mu)$ is the (linear) relative anisotropic wiggle spectrum, $D_z(k,\mu)$~is the nonlinear damping function and $V_\mathrm{eff}(k,\mu)$~is the effective volume. In a featureless universe, the wiggle spectrum is simply the BAO~spectrum, while it may also contain a feature signal in our case. Since we assume isotropic clustering,\footnote{In other words, we take the limit of a spherically-averaged clustering measurement. This is motivated by the fact that the primordial information that we are interested in is strictly isotropic and most of the information in~BOSS is contained in the monopole power spectrum.} this quantity is given by
\beq
\mathcal{O}_z(k,\mu) = \mathcal{O}_z(k) \equiv B_z(k) \left\{ O(k/q;z) + \big[1 + O(k/q;z)\big] \delta P_\zeta(k) \right\} + A_z(k)\, ,
\eeq
with the linear BAO~spectrum $O(k;z) = \Pw_\BAO(k;z)/\Pnw(k;z)$ being evaluated at the rescaled wavenumber $k/q = D_V^\mathrm{fid}(z)/D_V(z)\,k$. This rescaling with the radial BAO~dilation $D_V \propto (D_A^2/H)^{1/3}$ is necessary because the wavenumbers~$k$ are derived from the measured angles and redshifts in a survey using the angular diameter distance~$D_A^\mathrm{fid}(z)$ and Hubble rate~$H^\mathrm{fid}(z)$ in a fiducial cosmology.\footnote{In contrast to the data analysis, we do not additionally rescale by the fiducial ratio of the sound horizon in our forecasts since we recompute the BAO~spectrum~$O(k)$ for different cosmologies using \texttt{CLASS}~\cite{Blas:2011rf}.} Moreover, we introduced the free functions~$B_z(k)$ and~$A_z(k)$ which are taken to be smooth polynomials in~$k$ and distinct in each redshift bin, $\sum_m b_{m,z} k^{2m}$ and $\sum_m a_{n,z} k^n$, with $m=0,\ldots,3$ and $n=0,\ldots,4$. By marginalizing over these functions with fiducial values $a_{n,z}=0$, $b_{0,z}=1$ and $b_{m\neq0,z}=0$ in our forecasts, we effectively discard any information in the observable that might be affected by nonlinearities, biasing or observational systematics so that we only use a robust signal of the primordial features and the standard BAO~imprint. Finally, the nonlinear damping and effective volume are implemented as
\beq
D_z(k,\mu) \approx \ee^{-k^2 \Sigma_\mathrm{nl}(z)^2/2}\, ,	\qquad
V_\mathrm{eff}(k,\mu;z) \approx \left[\frac{\bar{n}_g(z) \, P_g(k,\mu;z)}{\bar{n}_g(z) \, P_g(k,\mu;z)+1} \right]^2 V_z \, ,
\eeq
where we assumed a constant nonlinear damping scale, $\Sigma_\mathrm{nl}(k,\mu;z) \approx \Sigma_\mathrm{nl}(z)$, and position independence of the comoving number density of galaxies, $n_g(\r\hskip1pt) \approx \bar{n}_g = \mathrm{const}$, in each redshift bin. Furthermore, the survey volume in a given redshift bin with spherical geometry is denoted by~$V_z$ and the fiducial galaxy power spectrum by~$P_g(k)$ which in particular includes the linear galaxy bias. We note that we implicitly assumed in~\eqref{eq:baoFisherMatrix} that the feature spectrum is nonlinearly damped in the same way as the BAO~spectrum (cf.\ \textsection\ref{sec:largeScaleNonlinearities}). (We reiterate that this is a brief summary and all details can be found in~\cite{Baumann:2017gkg}, including the modeling of the galaxy power spectrum, the nonlinear damping scale and the effects of reconstruction.)

\vskip4pt
We have already written the Fisher matrix~\eqref{eq:baoFisherMatrix} as a sum over discrete wavenumbers since the finite size of a galaxy survey introduces both a minimum accessible wavenumber\hskip1pt\footnote{The minimum wavenumber, or fundamental mode, which is available in a survey with a spherical geometry is, in principle, given by the survey volume~$V$ according to $\kmin = 2\pi [3V/(4\pi)]^{-1/3}$.}~$\kmin$ and a minimum binning width in Fourier space given by the fundamental mode,~$\Delta k\geq\kmin$. For many current applications, the width~$\Delta k$ has become small enough so that the power spectrum~$P(k)$ is smooth in a given band $[k_i - \Delta k/2, k_i + \Delta k/2]$ and we can approximate it by bandpowers $P_i \approx P(k_i)$. However, highly-oscillating primordial features introduce a significant variation within any such band so that we have to compute the finite-size bandpowers according to
\beq
P_i = \frac{1}{\Delta k} \int_{k_i - \Delta k/2}^{k_i + \Delta k/2}\!\d k\, P(k)\, .	\label{eq:bandpower}
\eeq
The bandpower-averaged wiggle spectrum, which contains both the BAO~and the feature spectra, is then given by $\mathcal{O}_i = \left(P_i^{}-\Pnw_i\right)/\Pnw_i$, for instance. To illustrate the effect of this averaging procedure (see also Fig.~\ref{fig:omegaLinLogComparison}), the bandpass-filtered primordial power spectrum~\eqref{eq:primordialSpectrum} with linear features is given by
\beq
P_{\zeta,i}^\lin \approx P_{\zeta,0}(k_i) \left[ 1 + \mathrm{sinc}(\omegalin\,\Delta k/2)\, \delta P_\zeta^\lin(k_i)\right] ,	\label{eq:primordialBandpowers}
\eeq
where $\mathrm{sinc}(x) = \sin(x)/x$ and we assumed $P_{\zeta,0}(k) \approx P_{\zeta,0}(k_i)$ for $k\in [k_i - \Delta k/2, k_i + \Delta k/2]$. This implies that the oscillatory features are suppressed unless $\omegalin \Delta k \ll 2$, or $\omegalin \ll 2/\Delta k \approx \SI{600}{\Mpc}$ for $\Delta k = \SI{0.005}{\hPerMpc}$. For logarithmic features, we could decompose the oscillations into linear features in a given band and arrive at an analogous conclusion.

The second effect of a finite survey volume that we have to take into account is the convolution of the power spectrum with the window function. This is of course directly related to the bandpass filtering in reality although we separate them here for convenience. Whereas the former averages the power spectrum over the wavevectors~$\k$ in a given band, the window function introduces a coupling between otherwise independent wavenumbers. For an all-sky survey with redshift range $[z_-, z_+]$ and effective redshift~$\bar{z}$, the spherical top-hat window function is
\beq
W_{\bar{z}}(\x) = W_{\bar{z}}(r,\theta,\phi) = \frac{1}{V_{\bar{z}}} \left[ \Theta(d_+ - r) - \Theta(d_- -r) \right] ,
\eeq
where $V_{\bar{z}} = 4\pi (d_+^3 - d_-^3)/3$ is the bin volume, $d_\pm \equiv d_c(z_\pm) = \int_0^{z_\pm} \d z\, c/H(z)$ are the comoving distances to the edges of the survey (or, equivalently, redshift bin) and~$\Theta(x)$ is the Heaviside step function. In practice, we however do not have access to the full sky, but only to a fraction~$\fsky<1$. For the purpose of our forecasts, we therefore include an incomplete sky by restricting the integration over the azimuthal angle~$\phi$: 
\beq
W_{\bar{z}}(\x) = \frac{1}{V_{\bar{z}}} \left[ \Theta(d_+ - r) - \Theta(d_- -r) \right] \Theta(2\pi\fsky - \phi) \, .
\eeq
In this case, the Fourier transform of the window function is radial,\footnote{Although this is an idealized form of the window function, we explicitly checked that forecasts employing the actual NGC~and SGC~window functions of~BOSS lead to consistent results.}
\beq
W_{\bar{z}}(\k) = W_{\bar{z}}(k) = \frac{3}{d_+^3 - d_-^3} \left[ \frac{d_+^3}{k d_+} j_1(k d_+) - \frac{d_-^3}{k d_-} j_1(k d_-) \right] ,	\label{eq:windowFunction}
\eeq
with the spherical Bessel function of the first kind~$j_n(x)$. By restricting the power spectrum to finite-size bandpowers~$P_i$ and using the fact that the window function~\eqref{eq:windowFunction} is radial, we can rewrite the convolved power spectrum, which is generally given by
\beq
P^c(k,\bar{z})	 = \int\!\frac{\d^3 k'}{(2\pi)^3}\, P(k',\bar{z})\, W_{\bar{z}}^2(\k-\k')\, ,
\eeq
in terms of a matrix equation:
\beq
P_i^c = w_{ij} P_j\, ,	\qquad w_{ij}(\bar{z}) = \frac{k_j^2\, \Delta k}{(2\pi)^2} \int_{-1}^1\! \d y\, W_{\bar{z}}^2\!\left(\sqrt{k_i^2 + k_j^2 - 2 k_i k_j y}\right) ,	\label{eq:windowConvolution}
\eeq
which we can evaluate numerically for all $k_i,k_j$ of an LSS~survey. As in the case of bandpowers, we again decompose the convolved spectrum~$P_i^c$ in its smooth and oscillatory components according to~\eqref{eq:powerSpectrumDecomposition}. To summarize, the main extensions to the Fisher forecasting methodology of~\cite{Baumann:2017gkg} based on the wiggle spectrum are given in~\eqref{eq:bandpower} and~\eqref{eq:windowConvolution}.

\subsubsection{Likelihood-Based Forecasts}
We also implemented forecasts based on the likelihood function~$\L(\vec{\theta})$ itself, as previously reported in~\cite{Baumann:2017gkg, Baumann:2018qnt}. While the modeling of the observables and covariances is the same as in the Fisher analyses, we directly evaluate the likelihood function~$\L(\vec{\theta})$ on a grid in the parameter space\hskip1pt\footnote{We note that it is advantageous to employ the amplitude parameterization in terms of~$(\Asin_X, \Acos_X)$ over the phase parameterization in both the Fisher- and likelihood-based forecasts for a few reasons. For instance, we can use a fiducial featureless power spectrum, the likelihood in~$(\Asin_X, \Acos_X)$ is close to Gaussian and we do not have to deal with the rather flat posterior in the phase~$\phase_X$.} of~$\vec{\theta} = (\alpha_z, \Asin_X, \Acos_X)$ as follows:
\beq
-2\log\L(\vec{\theta}) = \chi^2(\vec{\theta}) = \sum_{z,\,k_i}\Delta k\, k_i^2 \left[\mathcal{O}_z(k_i; \vec{\theta})-\tilde{\mathcal{O}}_z(k_i)\right]^T C^{-1}(k_i,z) \left[\mathcal{O}_z(k_i; \vec{\theta})-\tilde{\mathcal{O}}_z(k_i)\right] .	\label{eq:logLike}
\eeq
Here, we used the theoretical (`model') wiggle~spectrum~$\mathcal{O}_z(k_i;\vec{\theta})$, the fiducial (`data') spectrum~$\tilde{\mathcal{O}}_z(k_i)$ and the inverse covariance~$C^{-1}_z(k_i)$ of the respective experiment. The latter is computed as in the Fisher matrix~\eqref{eq:baoFisherMatrix} and includes the (white) instrumental noise contribution, cosmic variance and the exponential nonlinear damping. We note that all spectra are generally bandpass-filtered and convolved with the window function as discussed above, $\mathcal{O}_i^c = (P_i^c-P_i^{\mathrm{nw}\hskip-1pt,c})/P_i^{\mathrm{nw}\hskip-1pt,c}$, which we have however omitted in~\eqref{eq:logLike} for ease of notation.

The model spectrum~$\mathcal{O}_z$ varies over the considered parameter space and is defined as
\beq
\mathcal{O}_z(k; \vec{\theta}; a_i, b_i) = B_z(k) \left\{ O_\mathrm{fid}(k/\alpha_z,z) + \left[ 1 + O_\mathrm{fid}(k/\alpha_z,z) \right] \delta P_\zeta(k) \right\} + A_z(k)\, ,	\label{eq:theoreticalBAOspectrum}
\eeq
where $O_\mathrm{fid}$ is the linear BAO~spectrum of the fiducial cosmology, $\alpha_z = \alpha(z)$ is the isotropic BAO~parameter, and $A_z(k) = \sum_{i=0}^n a_{i,z} k^i$ and $B_z(k) = \sum_{j=0}^m b_{j,z} k^{2j}$ are the same polynomial `broadband' polynomials as above, where six terms with $m=n=2$ turn out to be sufficient. We marginalize over these terms by minimizing~$\chi^2$ of~\eqref{eq:logLike} for these parameters, i.e.\ $\chi^2(\vec{\theta}) = \min_{a_n,b_m}\!\big\{\chi^2(\vec{\theta};a_n,b_m)\big\}$.

The data spectrum~$\tilde{\mathcal{O}}_z$ is computed by evaluating~\eqref{eq:theoreticalBAOspectrum} for a fiducial set of parameters~$\vec{\theta}_\mathrm{fid}$ (which can include non-zero feature amplitudes), with $B_z(k) = 1$ and $A_z(k) = 0$. In addition to the smooth data with the experimental uncertainties being simply captured by the covariance matrix, we also perform forecasts with `noisy data'. In this case, we obtain the data spectrum by randomly picking the value of~$\tilde{\mathcal{O}}_i$ from a one-dimensional Gaussian distribution function with mean~$\tilde{\mathcal{O}}_i$ and variance~$C(k_i)$. This therefore simulates the scatter of the actual measurement due to the expected noise of an experiment (including sample variance) as captured by the covariance matrix. We can include this in our forecasts in order to estimate how likely it might be that features are found in the noise instead of the data or, in other words, that the noise mimics the presence of oscillatory features. In the main text, this constitutes an important check of the mock and data analyses, and provides an estimate of the actual significance of possible feature signals.

Having computed the likelihood function~$\L(\vec{\theta})$ over all of parameter space in which it is non-negligible, we then infer the predicted posterior distribution~$p(\theta_l)$ of a parameter~$\theta_l$ by marginalizing over all other parameters~$\theta_{m \neq l}$. Since the one-dimensional posteriors for $\alpha$, $\Asin_X$ and~$\Acos_X$ are very close to Gaussian, we finally obtain the mean~$\bar{\theta}_l$ and standard deviation~$\sigma(\theta_l)$ through a Gaussian fit to~$p(\theta_l)$.

\subsection{Experimental Specifications}
\label{app:lssSpecs}

We do not only build on the signal modeling of~\cite{Baumann:2017gkg}, but also its characterization of the LSS~surveys (which was derived from~\cite{Font-Ribera:2013rwa}). In general, we can characterize a cosmological galaxy survey by the following quantities: redshift range, sky coverage, linear galaxy bias~$b$ per redshift bin and number (density) of objects~$N_g$~($\bar{n}_g$) in each redshift bin. Here, we neglect the redshift error in spectroscopic surveys since it is usually small compared to the damping scales, but would need to take it into account for photometric observations. For planned experiments, such as~DESI and Euclid, we use specific values (see Appendix~B of~\cite{Baumann:2017gkg}), with Tables~\ref{tab:specsBOSS}%
\begin{table}
	\centering
	\subfloat[Forecasts for BOSS~DR12~data.]{
		\begin{tabular}{S[table-format=2.4] S[table-format=2.3] S[table-format=2.3] S[table-format=2.3] S[table-format=2.4] S[table-format=2.3] S[table-format=2.2]}
				\toprule
			{$\bar{z}$}	& {$z_\mathrm{min}$}	& {$\zmax$}	& {$b$}	& {$\num{e3}\,\bar{n}_g\ [\si{\h\tothe{3}\per\Mpc\tothe{3}}]$}	& {$V\ [\si{\per\h\tothe{3}\Gpc\tothe{3}}]$}	& {$\Sigma_\mathrm{nl}\ [\si{\MpcPerh}]$}	\\
				\midrule[0.065em]
			0.350		& 0.20					& 0.50	& 1.63	& 0.275	& 2.20	& 4.6	\\
			0.625		& 0.50					& 0.75	& 1.88	& 0.142	& 4.19	& 4.4	\\
				\bottomrule
		\end{tabular}
	}\\
	\subfloat[Forecasts for BOSS~mock catalogs.]{
		\begin{tabular}{S[table-format=2.4] S[table-format=2.3] S[table-format=2.3] S[table-format=2.3] S[table-format=2.4] S[table-format=2.3] S[table-format=2.2]}
				\toprule
			{$\bar{z}$}	& {$z_\mathrm{min}$}	& {$\zmax$}	& {$b$}	& {$\num{e3}\,\bar{n}_g\ [\si{\h\tothe{3}\per\Mpc\tothe{3}}]$}	& {$V\ [\si{\per\h\tothe{3}\Gpc\tothe{3}}]$}	& {$\Sigma_\mathrm{nl}\ [\si{\MpcPerh}]$}	\\
				\midrule[0.065em]
			0.350		& 0.20					& 0.50	& 2.04	& 0.275	& 2.20	& 7.0	\\
			0.625		& 0.50					& 0.75	& 2.34	& 0.142	& 4.19	& 7.0	\\
				\bottomrule
		\end{tabular}
	}
	\caption{Basic specifications for~BOSS (inspired by~\cite{Beutler:2016ixs} as detailed in~\cite{Baumann:2017gkg}) with a sky area of $\Omega=\SI{10252}{deg^2}$ resulting in roughly \num{1.2e6}~objects in a volume of about~\SI{6.4}{\per\h\tothe{3}\Gpc\tothe{3}}. We separately list the characteristic quantities employed when comparing to (a)~the BOSS~DR12 data and (b)~the corresponding mock catalogs since they differ in the linear bias~$b$ and the (post-reconstruction-equivalent) nonlinear damping scale~$\Sigma_\mathrm{nl}$ as discussed in the main text.}
	\label{tab:specsBOSS}
\end{table}
and~\ref{tab:specsDESI}%
\begin{table}
	\centering
	\begin{tabular}{l S[table-format=2.4] S[table-format=2.3] S[table-format=1.3] S[table-format=1.3] S[table-format=1.4] S[table-format=1.3] S[table-format=1.2] S[table-format=1.3] S[table-format=1.3]}
		\toprule
		$\bar{z}$													& 0.05		& 0.15	& 0.25	& 0.35	& 0.45		& 0.65	& 0.75	& 0.85	& 0.95		\\
		\midrule[0.065em]
		$b$ 														& 1.40		& 1.48	& 1.55	& 1.61	& 1.67		& 2.05	& 1.71	& 1.71	& 1.53		\\
		$\num{e3}\,\bar{n}_g\ [\si{\h\tothe{3}\per\Mpc\tothe{3}}]$	& 38.8		& 15.7	& 3.96	& 0.883	& 0.0992	& 0.591	& 1.31	& 0.920	& 0.779		\\
		$V\ [\si{\per\h\tothe{3}\Gpc\tothe{3}}]$					& 0.0357	& 0.229	& 0.563	& 0.985	& 1.45		& 2.41	& 2.86	& 3.28	& 3.66		\\
		\bottomrule
	\end{tabular}\\[4pt]
	\begin{tabular}{l S[table-format=1.3] S[table-format=1.3] S[table-format=1.3] S[table-format=1.3] S[table-format=1.3] S[table-format=1.3] S[table-format=1.4] S[table-format=1.4] S[table-format=1.4]}
		\toprule
		$\bar{z}$													& 1.05	& 1.15	& 1.25	& 1.35	& 1.45	& 1.55	& 1.65		& 1.75		& 1.85		\\
		\midrule[0.065em]
		$b$ 														& 1.45	& 1.48	& 1.47	& 1.47	& 1.69	& 1.68	& 2.27		& 2.45		& 2.47		\\
		$\num{e3}\,\bar{n}_g\ [\si{\h\tothe{3}\per\Mpc\tothe{3}}]$	& 0.466	& 0.398	& 0.387	& 0.180	& 0.133	& 0.110	& 0.0387	& 0.0197	& 0.0208	\\
		$V\ [\si{\per\h\tothe{3}\Gpc\tothe{3}}]$					& 4.00	& 4.30	& 4.56	& 4.79	& 4.98	& 5.14	& 5.28		& 5.39		& 5.48		\\
		\bottomrule
	\end{tabular}
	\caption{Basic specifications for DESI (derived from~\cite{Aghamousa:2016zmz} as explained in~\cite{Baumann:2017gkg}), covering a sky area $\Omega=\SI{14000}{deg^2}$ and resulting in roughly \num{2.7e7}~objects in a volume of about~\SI{59}{\per\h\tothe{3}\Gpc\tothe{3}}.}
	\label{tab:specsDESI} 
\end{table}
updating the employed parameterizations of~BOSS and~DESI. For more distant surveys, we assume a constant number density~$\bar{n}_g$ for a given total number of objects~$N_g$ and a linear bias of $b(z=0)=1$. Our `Future' LSS~survey contains $N_g=\num{e8}$ objects distributed over half the sky up to redshift~$\zmax=3$. The experiment referred to as `LSS-CVL' is cosmic variance limited on all employed scales and is designed to survey half of the sky for $z\leq6$. In our forecasts for~BOSS, we generally take the maximum wavenumber to be $\kmax=\SI{0.3}{\hPerMpc}$ to coincide with the choice in the data analysis. All other (Fisher) forecasts use $\kmax=\SI{0.5}{\hPerMpc}$, except for `LSS-CVL' for which we choose $\kmax=\SI{0.75}{\hPerMpc}$ since further extending the range of wavenumbers would likely yield only minor improvements in sensitivity due to the exponential damping.

\subsection{Additional Tests of the Pipeline}
\label{app:checks}

Given the described forecasting pipeline, we can provide additional insights into our primordial feature search and discuss some of the tests that we performed. In the following, we study the impact of the approximations in the theoretical damping calculation on the BOSS~constraints, revisit the impact of the finite-volume effects and, in particular, test whether injected feature signals can be detected in the analysis.

\subsubsection{Check of Damping Assumptions}
\label{app:dampingCheck}
When computing the nonlinear damping of the linear and logarithmic oscillations from large-scale bulk flows in~\textsection\ref{sec:largeScaleNonlinearities}, we made a number of simplifying approximations which allowed us to use a single damping scale, the standard BAO~damping scale~$\Sigma_\BAO$, in our data analysis. We can explicitly check the validity of these approximations in Fisher forecasts that generalize~\eqref{eq:baoFisherMatrix} to include the full resummed expressions for the linear and logarithmic spectra of~\eqref{eq:finallim} and~\eqref{eq:final}, and compare with the approximate formulas of~\eqref{eq:linresum} and~\eqref{eq:logresum}, respectively. 

\vskip4pt
In order to perform this test, we need to numerically evaluate the three damping scales of~\eqref{eq:linearDampingScale}, \eqref{eq:cossigma} and~\eqref{eq:sinsigma}, while choosing an appropriate value of the cutoff scale~$\Lambda$ which separates long modes~$q$ from other wavenumbers. The crucial point of the approximations is the fact that all the computations are strictly valid in the regime of $q/k\ll 1$, i.e.\ a separation of long and short modes. The cutoff~$\Lambda$ therefore needs to be smaller than the wavenumbers~$k$ of interest. At the same time, however, all long modes within the support of the feature also experience a damping effect. This is the reason why it is sensible to take $\Lambda = \epsilon k$ for some $\epsilon \ll 1$ (we employ $\epsilon=0.5$).\footnote{We note that the logarithmic damping factors~$\Sigma_\Log$ and~$\hat{\Sigma}_\Log$ are not well defined in the limit $\Lambda \rightarrow k$ because the argument of the logarithms in~\eqref{eq:cossigma} and~\eqref{eq:sinsigma} approaches zero. This is precisely the limit in which the computation becomes invalid since it is based on the separation of long and short modes. Interestingly, this is not the case for the BAO~damping factor~$\Sigma_\BAO$, whose value asymptotes for $\Lambda \gtrsim \SI{0.5}{\hPerMpc}$ and can be integrated to $\Lambda \to +\infty$ without significantly affecting the value of~$\Sigma_\BAO$, even though the validity of the nonlinear damping calculation breaks down at $\Lambda \sim k$~\cite{Baldauf:2015xfa}.} This choice leads to all damping scales, including~$\Sigma_\BAO$, to be effectively $k$-dependent, $\Sigma_X \to \Sigma_X(k)$. Having said that, it is important to remark once again that any dependence of these quantities on the specific choice of the cutoff indicates that next-to-leading-order effects should be taken into account (see e.g.~\cite{Blas:2016sfa} for the case of the standard BAO~signal). Since we fit $\Sigma_\BAO = \mathrm{const}$ in the data analysis (as is standard), we also compute this damping scale for a $k$-independent cutoff. Motivated by the maximum wavenumber of $\kmax=\SI{0.3}{\hPerMpc}$, we take $\Lambda=\SI{0.15}{\hPerMpc}$ in this case.

\begin{figure}
	\centering
	\includegraphics{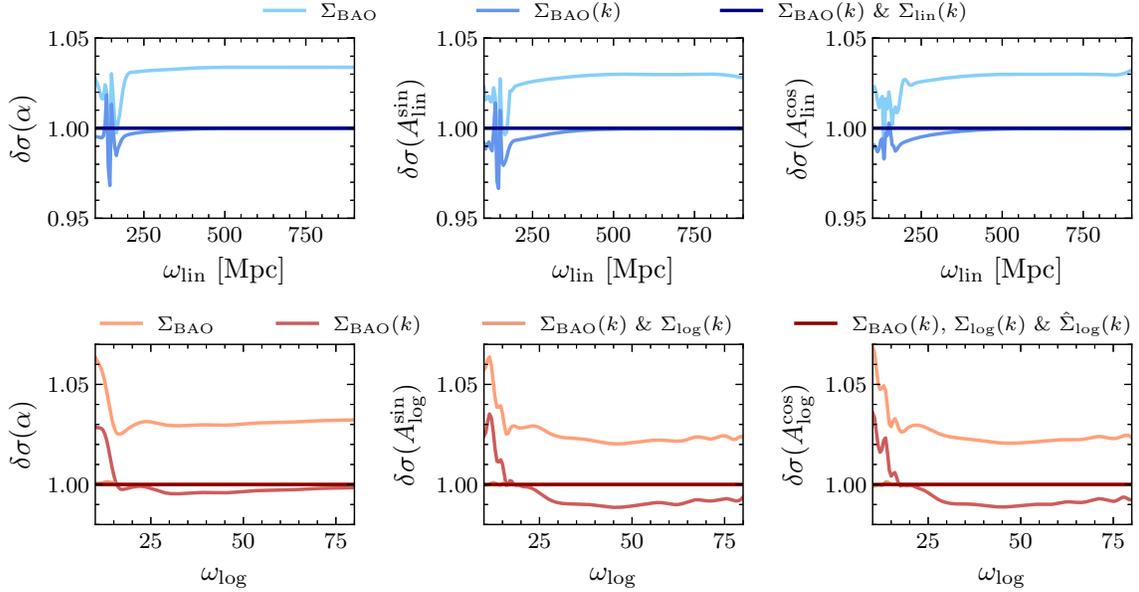}
	\caption{Impact of the various approximations to the theoretical damping scales on the BOSS~constraints for linear~(\textit{top}) and logarithmic features~(\textit{bottom}). We display the relative difference of the Fisher-forecasted standard deviation, $\delta\sigma = \sigma/\sigma_\mathrm{full}$, for the BAO~parameter~$\alpha$ and the feature amplitudes~$\Asin_X$ and~$\Acos_X$, where~$\sigma_\mathrm{full}$ is obtained using the full theoretical result. In the considered parameter space, the constraints are essentially unaffected by~$\hat{\Sigma}_\Log(k)$. Here, we used the effective post-reconstruction damping scales inferred at $z=0$.}
	\label{fig:dampingValidation}
\end{figure}

\vskip4pt
Figure~\ref{fig:dampingValidation}%
shows the effect of the various approximations on the estimated constraints of~BOSS. We note that we evaluate the damping scales at redshift $z=0$ for simplicity, given that the redshift dependence is the same for all damping terms. This however also means that we effectively exaggerate the employed damping scales and the actual impact on the constraints is even smaller than shown. Even with this conservative choice, we can deduce that all of our assumptions are valid in the context of the BOSS~DR12 dataset. To be more specific, assuming $\hat{\Sigma}_\Log(k) \approx 0$ has basically no visible impact on the constraints in the displayed parameter space of interest in this work, as expected. In addition, approximating $\Sigma_\lin(k),\,\Sigma_\Log(k) \approx \Sigma_\BAO(k)$ only results in sub-percent variations to the constraints for~$\omegalin$ away from the BAO~scale and $\omegalog \gtrsim 20$, and differences at the few-percent level for $\omegalog \in [10,20]$. Finally, taking~$\Sigma_\BAO$ to be constant instead of computing it with a $k$-dependent cutoff penalizes the constraints by roughly~3\% for all linear and logarithmic frequencies. This implies that all of the employed approximations are justified in the context of the BOSS~DR12 dataset and the upper limits that we infer in Section~\ref{sec:dataAnalysis} are in fact conservative. Nevertheless, the constraints inferred in future surveys will likely benefit from using the theoretically-computed forms of the damping scales~$\Sigma_\lin$ and~$\Sigma_\Log$.

\subsubsection{Impact of Finite-Volume Effects}
\begin{figure}
	\centering
	\includegraphics{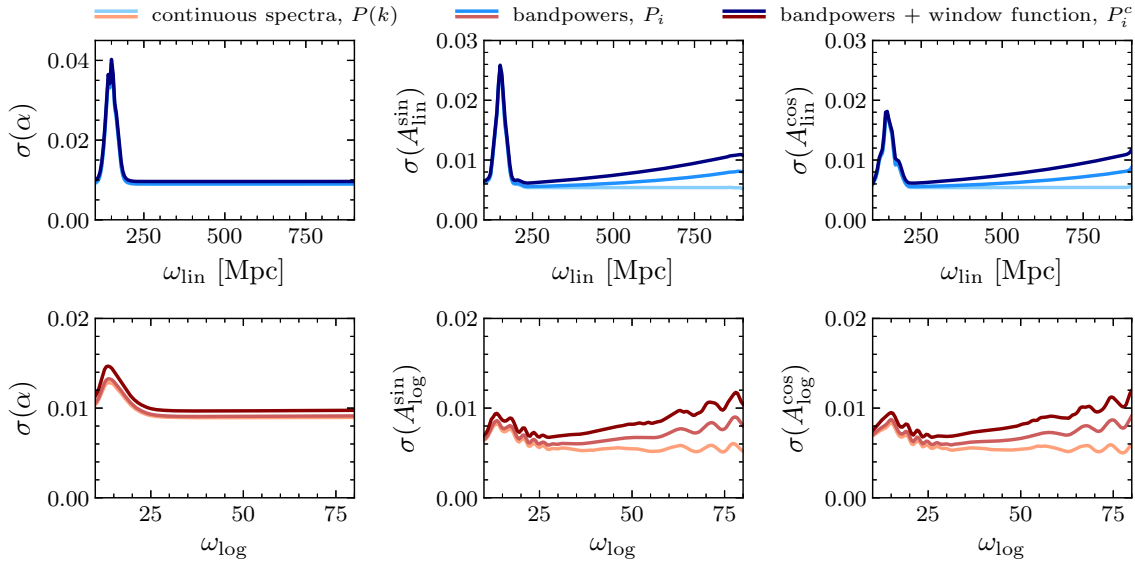}
	\caption{Illustration of the effects of the finite survey volume of~BOSS on the sensitivity to linear~(\textit{top}) and logarithmic features~(\textit{bottom}). We compare the likelihood-based constraints on the BAO~parameter~$\alpha$, and the feature amplitudes~$\Asin_X$ and~$\Acos_X$ when using continuous spectra,~$P(k)$, bandpass-filtered spectra,~$P_i$, and window-function-convolved spectra,~$P_i^c$.}
	\label{fig:bandpowerWindowFunctionConstraints}
\end{figure}

Our ability to search for highly-oscillating features is limited by the fact that we have only access to a finite cosmic volume, as we discussed in the main text. Apart from introducing a cutoff at the Nyquist frequency due to aliasing, the impact of finite-size bandpowers and the window function has to be taken into account. We illustrate the consequences of these effects on the sensitivity of~BOSS in Fig.~\ref{fig:bandpowerWindowFunctionConstraints}. %
While the constraints on the BAO~parameter~$\alpha$ are essentially unchanged, as expected given the BAO~scale of~\SI{150}{\Mpc}, we observe a gradual decrease in sensitivity to the feature amplitudes for larger frequencies~$\omega_X$. In consequence, we would overestimate the constraining power of~BOSS by up to a factor of two if we neglected the finite volume of the survey.

\vskip4pt
These results can be easily understood in the context of Fig.~\ref{fig:omegaLinLogComparison} which shows the impact of the finite-size effects on the spectra themselves. If we could employ continuous spectra~$P(k)$, a given primordial signal would have the same amplitude independent of the feature frequency in the analysis, resulting in the same sensitivity on all parameters (except for the interference with the BAO~signal). Since the amplitude effectively decreases for larger~$\omega_X$ when bandpass filtering the power spectrum [proportional to $\mathrm{sinc}(\omegalin\,\Delta k/2)$ according to~\eqref{eq:primordialBandpowers} for linear features], the constraints gradually weaken and the feature model becomes essentially unconstrained at the Nyquist frequency. Convolving the bandpowers additionally with the window function of the survey couples otherwise independent modes which leads to an additional reduction in the amplitude and, consequently, the sensitivity. Finally, the frequency of the standard BAO~signal (or equivalently the survey volume) is large enough so that the BAO~spectrum and ultimately the constraints on~$\alpha$ are barely affected.

\subsubsection{Detection of Injected Signals}
Our likelihood-based forecasts also allow us to test whether we would be able to detect a feature signal if it was present in the data. This is an important check of our analysis pipeline that we cannot perform on mock catalogs because their underlying primordial spectrum is featureless. Since the results of the forecasting pipeline are consistent with both the mock and data analyses, we can still reliably perform a search for injected signals.

\begin{figure}
	\centering
	\includegraphics{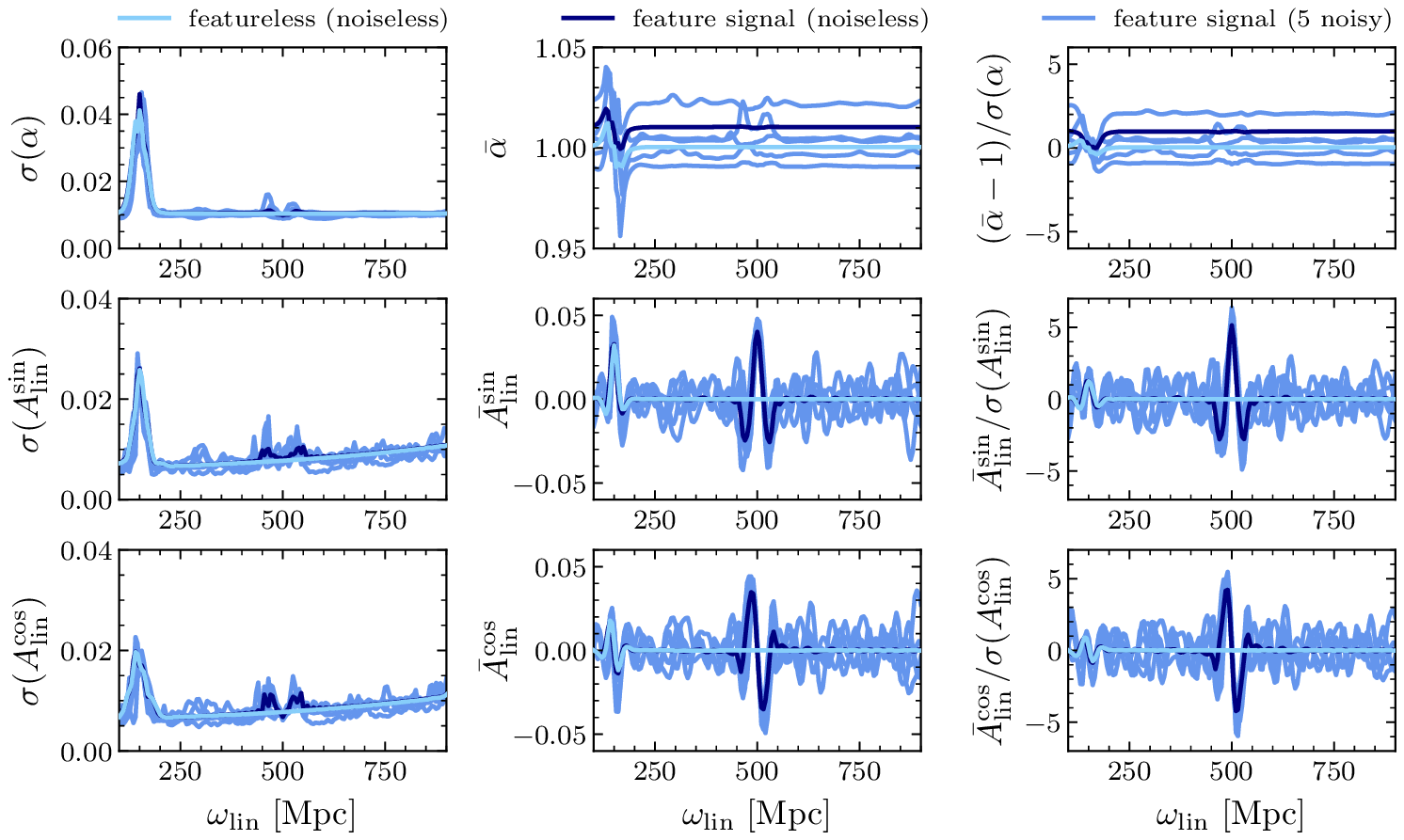}
	\includegraphics{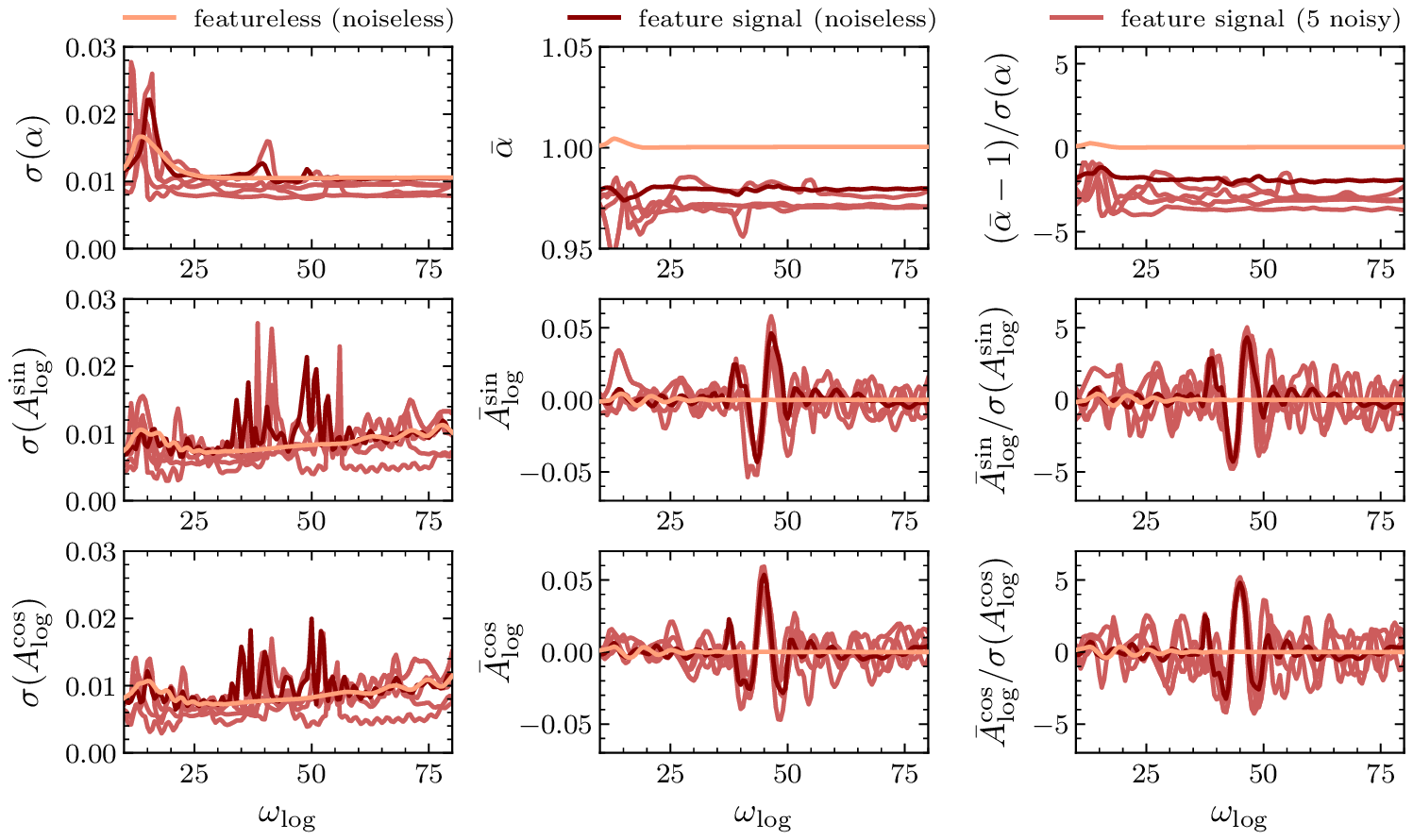}
	\caption{Detection of feature signals in likelihood-based forecasts of the low-redshift bin for linearly-~(\textit{top}) and logarithmically-spaced oscillations~(\textit{bottom}). This test uses artificially injected signals with $(\alpha,\omegalin,\Alinsin,\Alincos) = (1.01, \SI{500}{\Mpc}, 0.04, 0)$ and $(\alpha,\omegalog,\Alogsin,\Alogcos) = (0.98, 45, 0, 0.05)$, which can be reliably inferred with the expected significances. For comparison, we also show the results for a featureless spectrum with $\alpha=1$ as employed in the main text.}
	\label{fig:injectedSignals}
\end{figure}

\vskip4pt
We performed this test for a wide range of parameters. In Figure~\ref{fig:injectedSignals}, %
we show the representative results for linearly- and logarithmically-spaced oscillations characterized by $(\alpha,\omegalin,\Alinsin,\Alincos) = (1.01, \SI{500}{\Mpc}, 0.04, 0)$ and $(\alpha,\omegalog,\Alogsin,\Alogcos) = (0.98, 45, 0, 0.05)$. These parameters were chosen to produce a roughly $5\sigma$~signal for a single redshift bin in the center of our frequency range. Here, we display the standard deviation and mean values inferred from the marginalized likelihood function (and the significance of any signal) of the low-$z$ bin, as in Fig.~\ref{fig:dataForecastComparison} for the featureless cosmology, but note that the results are as consistent and positive in the high-redshift bin.

For the linear features in the top row of Fig.~\ref{fig:injectedSignals}, we first of all see that the posterior of the BAO~parameter~$\alpha$ is barely affected by the injected feature signal. In addition, the underlying value of~$\alpha$ is correctly recovered within the noise-related scatter. While the standard deviations of the feature amplitudes are hardly affected, their mean values clearly show the characteristic signal around $\omegalin=\SI{500}{\Mpc}$: $\bar{A}_\lin^\mathrm{sin}$~and~$\bar{A}_\lin^\mathrm{cos}$ peak/vanish at the injected value and frequency, and approach zero away from it in an oscillatory fashion. This is due to the fact that features with neighboring frequencies interfere with the signal and can also be fit with different amplitudes since we only probe a limited range of wavenumbers. Nevertheless, the shape of the signal in the sine and cosine amplitudes clearly picks out the true value. Furthermore, the noise-induced scatter in the mean values is essentially absent around the injected signal, while it is consistent with the featureless case away from it. Given these observations, it is also evident that the significance of the signal is reproduced at the expected value (with some small variations in a given noise realization).

The injected logarithmic signal can be extracted with a similar level of confidence. We again observe the same characteristic behavior of the mean values around the injected feature frequency $\omegalog=45$. Since we employed a primordial cosine instead of sine feature, the roles of~$\Alogsin$ and~$\Alogcos$ are naturally reversed and correctly captured. In contrast to the linear oscillations, however, the standard deviations show additional variations and the mean values exhibit a slightly more pronounced `ringing' across the $\omegalog$-range. Given the noise levels of~BOSS, this however does not have a significant impact on the detectability of a primordial signal with a large enough amplitude.

For both types of feature models, we find similar results over a wide range of frequencies. As could be expected, it however becomes somewhat harder to extract signals with small values of~$\omega_X$ due to the interference with the standard BAO~signal and associated effects. Nevertheless, we should be able to extract even these oscillations from the data due to their overall signature. We can therefore conclude that we should be able to detect any primordially imprinted oscillatory feature with a large enough amplitude in our analysis pipeline.

\subsection{Forecasts for Future LSS Surveys}
\label{app:futureLSS}

\begin{figure}
	\centering
	\includegraphics{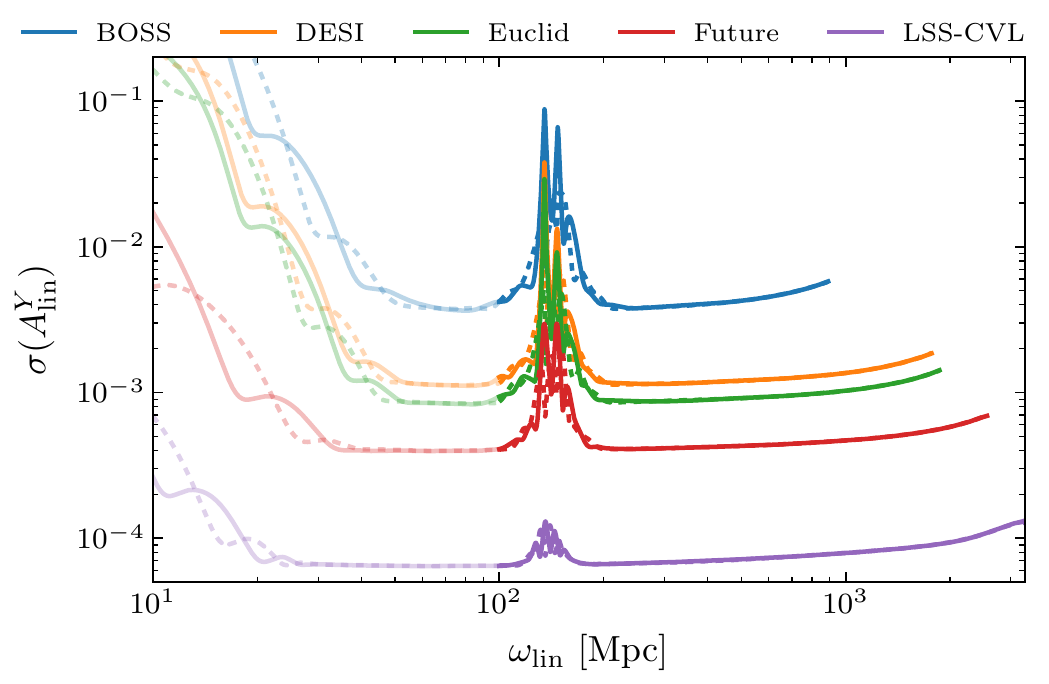}
	\caption{Fisher forecasts of the sensitivity of future LSS~surveys (see~\textsection\ref{app:lssSpecs} for details regarding the employed survey specifications) to the primordial feature amplitudes in the amplitude parametrization, $\Alin^Y$, with $Y=\mathrm{sin},\mathrm{cos}$. The constraints on~$\Alinsin$ are shown in solid lines, while the standard deviation~$\sigma(\Alincos)$ is displayed with dashed lines. Due to the possible impact of small-scale nonlinearities and a reduced damping from large-scale bulk flows, the forecasts for $\omegalin\lesssim\SI{100}{\Mpc}$ should be taken with care.}
	\label{fig:lin_future_Asin+Acos_lss}
\end{figure}

We do not only consider currently available data, but we also employ Fisher forecasts in~\textsection\ref{sec:futureConstraints} to estimate the sensitivity of future LSS~surveys to primordial features. Since large classes of feature models can be expressed in a basis of linear oscillations, we focus on the ``feature spectrometer''. As in the rest of this work, we initially work in the parameter space spanned by the isotropic BAO~parameter~$\alpha$ and the feature amplitudes~$\Alinsin$ and~$\Alincos$, fiducially taken to be~$\alpha=1$ and $\Alinsin=\Alincos=0$. Since we use a total of nine polynomial broadband parameters ($a_{m\leq4,z}$, $b_{m\leq3,z}$) and compute the Fisher matrices for a given frequency~$\omegalin$, these forecasts contain twelve parameters per redshift bin. Summing the broadband-marginalized Fisher matrices, we obtain the forecasted standard deviations~$\Alinsin$ and~$\Alincos$ displayed in Fig.~\ref{fig:lin_future_Asin+Acos_lss}. %
Apart from the well-known degeneracy with the BAO~scale, we observe that the constraints on~$\Alinsin$ and~$\Alincos$ are basically identical for $\omegalin\gtrsim\SI{250}{\Mpc}$, but oscillate around a common mean value for smaller frequencies. This is as expected and exemplifies again that the sine and cosine feature terms are essentially independent modes for large enough frequencies~$\omegalin$.

\vskip4pt
To turn these constraints into limits on the overall feature amplitude~$\Alin$ while retaining the correlations between the parameters, we draw random samples from a Gaussian distribution whose covariance matrix is given by the inverse Fisher matrix. Since the amplitude~$\Alin$ is positive semi-definite, which implies that the mean of~$\Alin$ can only fluctuate upwards from zero, we also repeatedly take the mean values from Gaussian distributions with zero mean and covariance given by the same inverse Fisher matrix. Finally, we can compute the 95\%~confidence limits on~$\Alin$ by similar means as in our BOSS~analysis above (see Appendix~\ref{app:analysisDetails}). In this way, we obtain the forecasted bounds of Fig.~\ref{fig:futureForecast}. To conclude, we remark that these constraints are likely conservative since we employed the same constant damping scale for both the BAO~and the feature spectra~(cf.~\textsection\ref{app:dampingCheck}).

\clearpage
%%%%%%%%%%%%%%%%
\section{Further Details on the BOSS Analysis}
\label{app:analysisDetails}
%%%%%%%%%%%%%%%%

We employ the amplitude parametrization of the feature models in our analysis and forecasting pipelines since the posterior distributions of~$\Asin_X$ and~$\Acos_X$, $X=\lin,\Log$ are close to Gaussian (unlike the phase~$\phase_X$). Since the phase of the primordial features is not expected to carry much information about the inflationary epoch (at least in the pre-discovery era), we are ultimately interested in the constraints on the overall feature amplitude~$A_X$. In this appendix, we describe our method to combine the two BOSS~redshift bins and infer the reported upper limits from the Monte Carlo Markov chains, including some checks~(\textsection\ref{app:upperLimits}). Moreover, we outline how we determine whether the data exhibits any statistically significant detections of features~(\textsection\ref{app:detections}).

\subsection{From Posteriors to Upper Limits}
\label{app:upperLimits}

The analysis pipeline of~\textsection\ref{sec:pipeline} results in Markov chains that provide samples from the marginalized posterior distribution as a function of~$\Asin_X$ and~$\Acos_X$ in each feature frequency bin. It is useful to consider constraints on the two-dimensional parameter space of these feature amplitudes as constraints in the complex plane. From this perspective, we are interested in computing the upper limits on the absolute value of the complex amplitude $A_X = \sqrt{(\Asin_X)^2 + (\Acos_X)^2}$ for which there is however no unique procedure. Since the feature phase is not an independent parameter, the upper limit is actually not a single number, but depends on the phase~$\phase_X$. This is important because the maximum posterior point will in general not be at $\Asin_X=\Acos_X=0$ in the presence of noise. Given that we marginalize over the feature phase, it is important to keep in mind that a uniform prior on~$A_X$ and~$\phase_X$ corresponds to a non-uniform prior in the \mbox{$\Asin_X$-$\Acos_X$}~plane and vice versa.

\vskip4pt
Our method of compressing the available information considers circles in the \mbox{$\Asin_X$-$\Acos_X$}~plane centered at the origin that enclose a given probability or, equivalently, a fraction of all Monte Carlo samples. We therefore define the upper limit on~$A_X$ at a given confidence level as the radius of the respective circle. For the separate~MCMCs of the low- and high-redshift bins, this means that we compute the
amplitude~$A_X$ for each sample and rank-order the resulting values. The upper limit is then given by the value of~$A_X$ at the desired confidence limit percentile.

\begin{figure}
	\centering
	\includegraphics{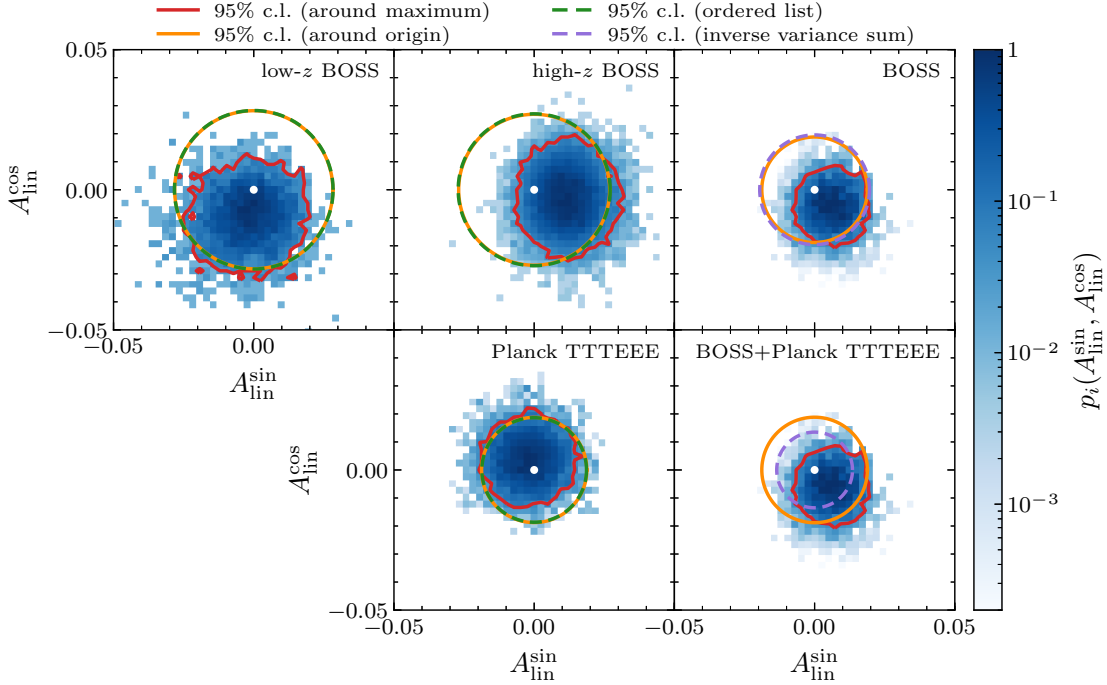}
	\caption{Illustration of our method to infer 95\%~confidence limits on the feature amplitude~$A_X$ from Markov chain samples of $(\Asin_X,\Acos_X)$. The top panels show the pixelated posteriors for the \mbox{low-$z$}~(\textit{left}) and the high-$z$ bins~(\textit{middle}), and the joint pixelated BOSS~posterior~(\textit{right}) for the frequency bin centered at $\omegalin=\SI{700}{\Mpc}$. In the bottom panels, the pixelated posterior for the respective Planck~TTTEEE samples~(\textit{middle}) and the joint posterior for~BOSS and Planck~(\textit{right}) are displayed. The red contours enclose the pixelated 95\%~confidence region around the maximum posterior point. The solid~(orange) and dashed~(green)~circles enclose 95\%~of the total probability around the origin $A_X=0$ (marked by the white dot) as obtained from the pixelated posterior and an ordered list, respectively. The agreement between these circles demonstrates that the error introduced by pixelization is negligible. For~BOSS and the joint~BOSS+Planck constraint, the dashed~(purple)~circle shows the constraint when combining the separate low-$z$, high-$z$ and CMB confidence limits by adding inverse variances, which demonstrates that the non-Gaussianity of the likelihood has a non-negligible effect on the inferred upper limit.}
	\label{fig:featureAmplitudeInference}
\end{figure}

\vskip4pt
We are however not only interested in the constraints from a single redshift bin, but want to compute joint limits from both BOSS~redshift bins (or of~LSS and the~CMB). Although running a joint~MCMC would be the formally correct statistical approach, it would result in the simultaneous variation of 33~parameters (and even more for a joint analysis with the~CMB), which would be computationally more complex and expensive. This is why we proceed as follows:
\begin{enumerate}
\item We bin the samples of a single~MCMC in the \mbox{$\Asin_X$-$\Acos_X$}~plane. This results in a pixelated posterior distribution~$p_i(\Asin_X,\Acos_X)$.

\item We obtain the joint posterior by multiplying the pixelated posteriors, $\prod_i p_i(\Asin_X,\Acos_X)$.

\item We measure the probability~$P(A_X)$ enclosed in a circle centered at the origin of the \mbox{$\Asin_X$-$\Acos_X$}~plane as a function of its radius and obtain the inverse (interpolated) function~$\tilde{A}_X(P)$.
\end{enumerate}
In this way, the 95\%~confidence limit is then given by $\tilde{A}_X(P=0.95)$, for instance. We illustrate this approach in Fig.~\ref{fig:featureAmplitudeInference}%
for one feature frequency bin. This figure shows that the phase-independent limits are necessarily less constraining than those centered at the maximum posterior value since they also enclose low-likelihood regions away from the maximum. Nevertheless, the described method allows us to correctly infer the quantity that we are interested in, the maximum value of the feature amplitude~$A_X$ that is allowed by the data for any phase~$\phase_X$. The comparison of the two circles for the joint posteriors also demonstrates that compressing the confidence limits into a single upper limit for a given single dataset and subsequently combining them by summing the inverse variances would result in a significant error on the inferred upper limits from joint probes.

\vskip4pt
Having outlined our procedure, a few comments are in order. As a consequence of working in the two-dimensional plane spanned by~$\Asin_X$ and~$\Acos_X$, we assumed that the feature amplitudes are completely uncorrelated with any of the other parameters, in particular the BAO~parameter~$\alpha$. We explicitly confirmed this assumption by computing the three-dimensional (Gaussian) covariance matrix in each frequency bin to estimate the correlation coefficient~$\rho$ in forecasts and on data. For linear (logarithmic) features in~BOSS, we find that~$|\rho|$ is consistent with zero, but approaches significant values (up to about~0.5) for $\omegalin\lesssim200$ ($\omegalog\lesssim30$), as expected due to the interference with the standard BAO~signal. Since this effect is minimal and including these correlations would only strengthen the bounds, the deduced bounds are conservative, albeit slightly suboptimal because we are effectively assuming a different set of non-amplitude parameters in each redshift bin.

We also check that the pixelation does not introduce numerical artifacts due to the choice of too small or too large pixel sizes. The former could lead to a biased estimate because the posterior distribution becomes noisy, whereas the latter might artificially smooth the posterior. To mitigate these possibilities, while including all samples in the analysis, we separately set the pixel size in each frequency bin. For this purpose, we sampled~$\Asin_X$ and~$\Asin_X$ in about 100~pixels over the range given by $\pm1.2\,\mathrm{max}\{|\Asin_X|,|\Acos_X|\}$. For a single MCMC, we find that this choice results in virtually the same confidence limits as when inferring them from a rank-ordered list of samples (while combining the latter in a Gaussian way leads to suboptimal joint constraints).

\begin{figure}
	\centering
	\includegraphics{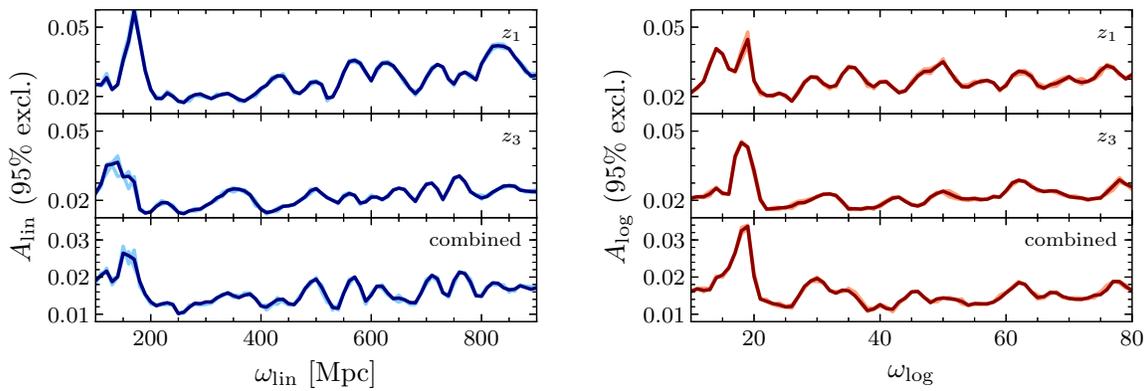}
	\caption{Convergence test of the BOSS~analyses for linear~(\textit{left}) and logarithmic features~(\textit{right}). The constraints inferred from splitting the Markov chains into two independent halves (light colors) are compared to those derived from all Markov chains (dark color). Note that the former bounds are barely visible under the latter due to the high level of convergence.}
	\label{fig:constraintsCheck}
\end{figure}

Finally, it is important to check that the sampling noise due to the inherently finite length of the Markov chains does not affect the constraints. We therefore test the convergence of our analysis by splitting the chains into several independent parts. Figure~\ref{fig:constraintsCheck}%
shows that the Markov chains are converged and do not show evidence for numerical noise. As a consequence, we can also report that the shape of the constraints as a function of frequency is robust and inherent to the data.

\subsection{Upper Limits or Detections?}
\label{app:detections}

So far, we have only discussed the inference of upper limits from the data. Of course, any analysis should also allow for the possibility of detecting a signal. Our method of determining detections at a given confidence level is again based on the pixelated posterior distributions. 

We start by drawing the two-dimensional contours that enclose the desired confidence limit. We then declare a detection if the origin is excluded from this contour, i.e.\ if the white dot in Fig.~\ref{fig:featureAmplitudeInference} is outside the red contour. This is determined as follows. First, we rank-order the pixelated likelihood values and sort them from the most to the least likely pixel. For each value in this list, we then compute the cumulative probability and map the cumulative probability to the pixel likelihood by an interpolating spline. The value of the pixel likelihood at which the cumulative probability reaches $P$ finally determines the contour level at which the total probability~$P$ will be enclosed (assuming a unimodal distribution that falls off monotonically away from the peak). 

\vskip4pt
We calculated the number of~95\% and 99.7\%~confidence limit (corresponding to $2\sigma$ and~$3\sigma$) detections on mocks and on data. We can confirm that detections at the 95\%~c.l.\ occur in roughly 5\%~of the mocks for each feature bin, except around the BAO~scale, where we find a modest excess in the number of detections. At the $3\sigma$-level, we find no detections in our data. We note that a small number of detections would have been consistent with the look-elsewhere effect since we sample many independent frequencies. Since we do not find any such detection, there is no need to quantify this.

\clearpage
%%%%%%%%%%%%%%%%
\section{CMB Analysis and Forecasts} 
\label{app:cmbAnalysis}
%%%%%%%%%%%%%%%%

The focus of this work is the first analysis of primordial features in LSS~data alone. Given the long history of searches in the CMB~anisotropies, it is however natural to compare (and combine) our newly-inferred bounds from the BOSS~DR12 dataset to those derived from current Planck data. In this appendix, we outline the performed CMB~data analysis~(\textsection\ref{app:planckAnalysis}) and discuss the effects of the different transfer of primordial power onto the large-scattering surface and the~LSS~(\textsection\ref{app:lssCMBcomparison}). Moreover, we provide details on our joint LSS and CMB~bounds~(\textsection\ref{app:jointAnalysis}), and comment on the CMB~Fisher forecasts~(\textsection\ref{app:futureCMB}).

\subsection{Analysis of Planck Data}
\label{app:planckAnalysis}

The phenomenological feature models of~\eqref{eq:linearFeatures} and~\eqref{eq:logarithmicFeatures} have been searched for in CMB~data for quite some time, including the Planck collaboration~\cite{Planck:2013jfk, Ade:2015lrj, Akrami:2018odb}. These analyses have however focused on reporting the best-fit points, and/or the likelihood improvements and significances of possible signals as a function of feature frequency~$\omega_X$. Since any possible signals have not been significant to date (in particular after taking the look-elsewhere effect into account~\cite{Fergusson:2014hya, Fergusson:2014tza, Meerburg:2015owa, Akrami:2018odb}), we are interested in studying the entire parameter space of features. We therefore want to report the frequency-dependent constraints on the feature amplitudes~$A_X$, as we did in the BOSS~analysis.

\vskip4pt
Following the analyses by the Planck collaboration~\cite{Ade:2015lrj, Akrami:2018odb}, we first run \texttt{MultiNest}~\cite{Feroz:2007kg, Feroz:2008xx} with a modified version of \texttt{CAMB}~\cite{Lewis:1999bs}.\footnote{Due to the highly-oscillatory nature of the primordial feature spectrum, in particular for logarithmic features at large scales, we have to run~\texttt{CAMB} with increased accuracy settings which were checked to resolve all oscillations.} Since we also fix the foreground and nuisance parameters to their best-fit values~\cite{Ade:2015xua}, we vary a total of nine parameters: the six standard $\Lambda$CDM~parameters (physical baryon and cold dark matter fractions~$\omega_b$ and~$\omega_c$, angular size of the sound horizon~$\theta_s$, logarithm of the primordial scalar amplitude~$\ln(10^{10}\As)$, scalar spectral index~$\ns$ and optical depth~$\tau$) and three feature parameters ($\omega_X$, $\Asin_X$ and~$\Acos_X$). We employ wide flat priors on all parameters, including the feature frequencies, $\omegalin \in [0.5,1005]$ and $\omegalog \in [0.1,101]$. We note that the~CMB is also sensitive to models with larger frequencies~$\omega_X$, but we restricted ourselves to a range around the region available to~BOSS.

From these \texttt{MultiNest}~runs, we compute the mean values and covariance matrices of the nine parameters in bins of $\Delta\omegalin=100$ and $\Delta\omegalog=10$. To effectively increase the number of samples, we then run standard~MCMCs with four chains using \texttt{CosmoMC}~\cite{Lewis:2002ah} in these frequency bins starting from the computed covariance, with the priors chosen to enclose the one-dimensional $5\sigma$~ranges. Having acquired enough samples and a Gelman\,\&\,Rubin convergence criterion~\cite{Gelman:1992zz} with scale parameter generally given by $\epsilon \lesssim 0.01$, we implicitly marginalize over the $\Lambda$CDM~parameters and compute the 95\%~upper limits on~$A_X$ as described in Appendix~\ref{app:analysisDetails} for the BOSS~analysis in one redshift bin. For convenience, we also use the same binning in the feature frequency, $\Delta\omegalin=\SI{10}{\Mpc}$ and $\Delta\omegalog=1$, although the correlation length differs (e.g.\ $\Delta\omegalin\approx\SI{26}{\Mpc}$ was estimated in the Planck~TT analysis of~\cite{Fergusson:2014tza}).

\begin{figure}
	\centering
	\includegraphics{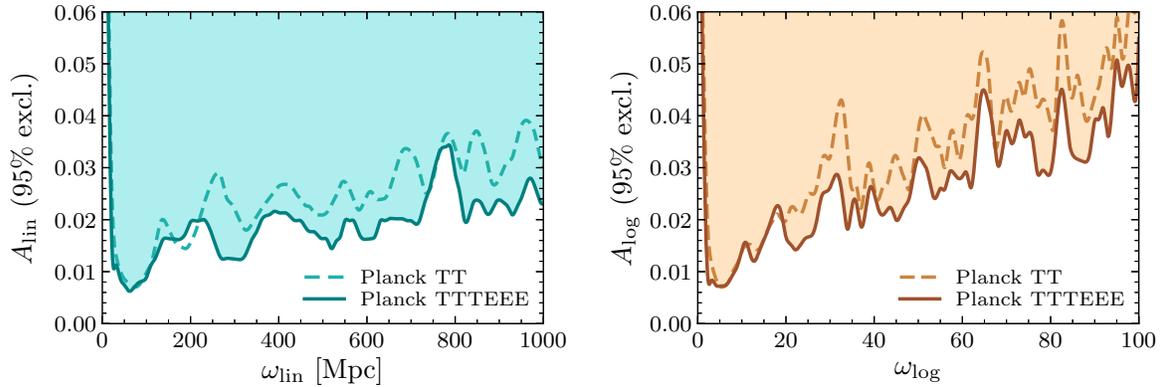}
	\caption{Upper limits on the feature amplitude~$A_X$, $X=\lin,\Log$, at 95\%~c.l.\ as a function of the frequency~$\omega_X$ for linear~(\textit{left}) and logarithmic features~(\textit{right}) from Planck~2015 CMB~data. The main analysis employs temperature and polarization data~(TTTEEE, solid), while the analysis without high-multipole polarization data~(TT, dashed) leads to slightly weaker bounds.}
	\label{fig:cmbConstraints}
\end{figure}

\begin{figure}
	\centering
	\includegraphics{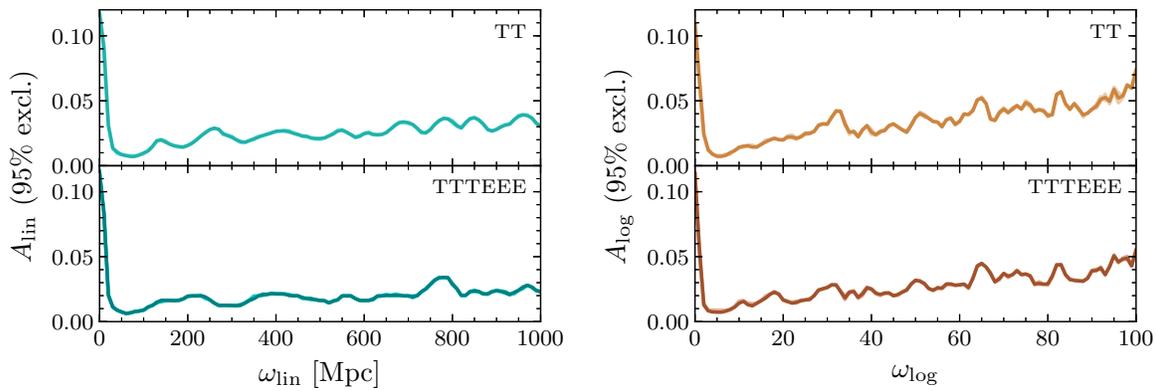}
	\caption{Convergence test of the Planck~2015~TT and~TTTEEE analyses for linear~(\textit{left}) and logarithmic features~(\textit{right}). The bounds derived from all Markov chains are shown in dark colors, whereas those inferred from splitting them into two independent halves are shown in light colors, but are barely visible as a result of the excellent convergence.}
	\label{fig:cmbConstraintsCheck}
	\end{figure}

\vskip4pt
Given the preliminary status of the Planck~2015 polarization data,\footnote{The comparison of the published~2015 and 2018~results on primordial features suggests little changes. We therefore expect our results employing Planck~2015 polarization to be consistent with those derived by the Planck collaboration in~\cite{Akrami:2018odb}.} we run this pipeline on two sets of Planck~2015 likelihood combinations~\cite{Aghanim:2015xee}:
\begin{itemize}
	\item `Planck~TT': low-$\ell$ ($2\leq\ell\leq29$) \texttt{commander} temperature and polarization data, and unbinned high-$\ell$ $\texttt{Plik}$ half-mission temperature cross-spectra data with $\lmax^T=2508$,
	\item `Planck~TTTEEE': low-$\ell$ \texttt{commander} and unbinned high-$\ell$ $\texttt{Plik}$ half-mission temperature and polarization cross-spectra data with $\lmax^T=2508$ and $\lmax^E=1996$.
\end{itemize}
We emphasize that we use the unbinned likelihoods to have access to all measured multipoles~$\ell$ without averaging over $\ell$-bins. This way, we obtain the bounds on the feature amplitudes~$A_X$ displayed in Fig.~\ref{fig:cmbConstraints}. %
We see that the constraints only degrade significantly for very small frequencies and are basically unaffected by the polarization data at small~$\omega_X$. Over the rest of parameter space, the full dataset yields slightly stronger bounds. Finally, Figure~\ref{fig:cmbConstraintsCheck}%
illustrates the excellent convergence of the CMB~Markov chains for all frequencies and both sets of data.

\subsection{Transfer of Feature Power}
\label{app:lssCMBcomparison}

We have already discussed the experimental reasons for the better sensitivity of~BOSS to features than Planck (or, more generally, future LSS~surveys compared to CMB~observations) in the main text. In the following, we shed additional light on this by studying the signal of primordial oscillations that is imprinted in the observables of the~CMB and~LSS.

\begin{figure}
	\centering
	\includegraphics{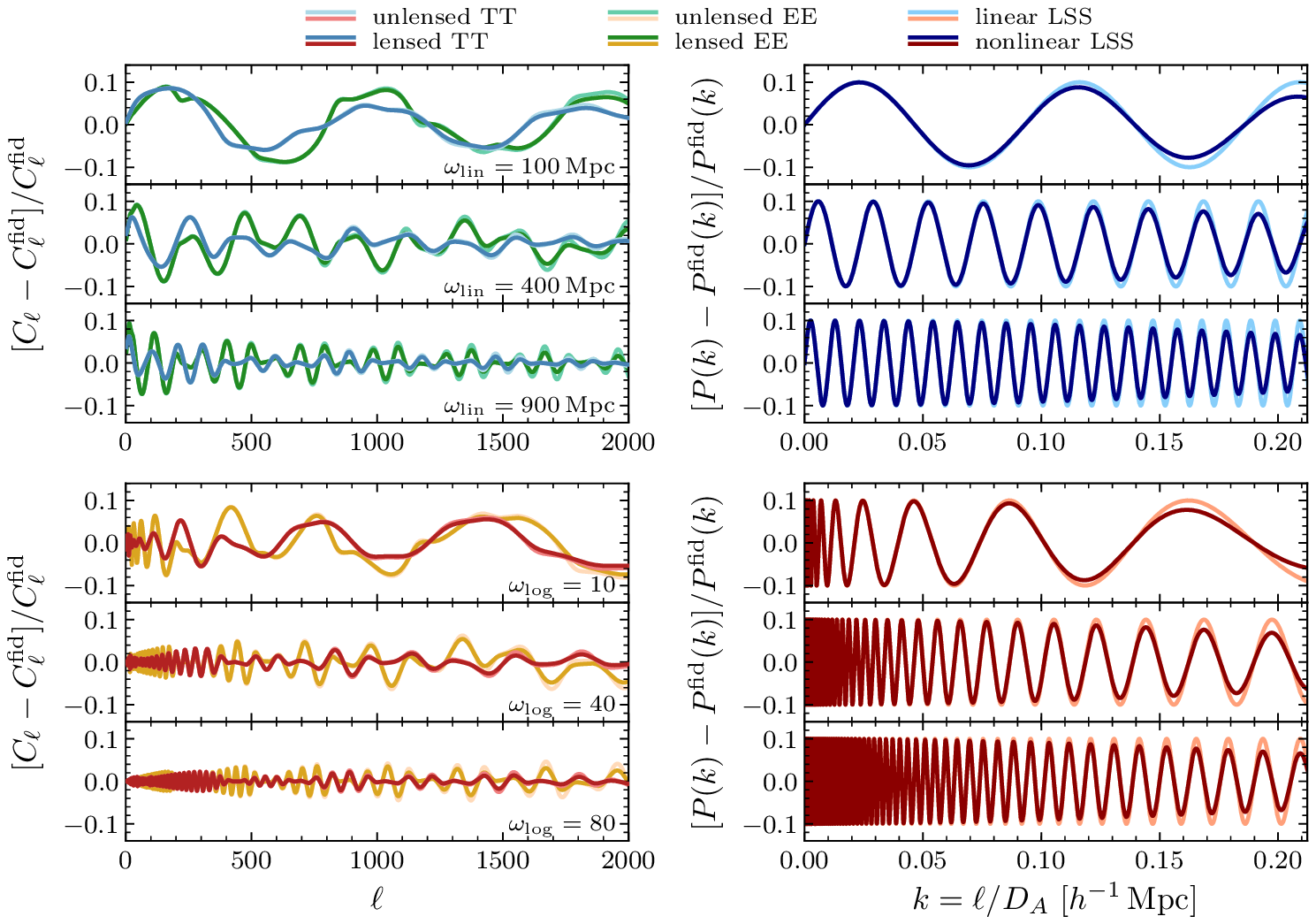}
	\caption{Imprint of primordial features in the CMB~and LSS~power spectra for a set of linear~(\textit{top}) and logarithmic~(\textit{bottom}) frequencies~$\omega_X$, $X=\lin,\Log$. We compare the relative contribution of features to the unlensed and lensed temperature~(TT) and E-mode polarization~(EE) power spectra~$C_\ell$, with the contribution to the linear and nonlinear matter power spectrum~$P(k)$. The fiducial spectra, which are denoted by the superscript~`fid', are computed in a standard featureless $\Lambda$CDM~cosmology, which is then augmented by a feature with amplitudes~$\Asin_X=0.1$ and~$\Acos_X=0$ for illustrative comparison. We display the same range of scales for the observables, linking multipoles~$\ell$ and wavenumbers~$k$ via the flat-sky approximation, $\ell = D_A k$, where~$D_A$ is the angular diameter distance to the last-scattering surface. Finally, we note that we neglected survey-related effects for both the~CMB and~LSS.}
	\label{fig:cmbLSScomparison}
\end{figure}

\vskip4pt
A comparison of the size and shape of the features in these cosmological measurements is displayed in Fig.~\ref{fig:cmbLSScomparison}. %
We show both the lensed and unlensed auto-spectra of temperature and E-mode polarization for the~CMB, and the matter power spectrum in linear and nonlinear theory, i.e.\ without and with the exponential damping caused by large-scale nonlinearities. In all cases, we can clearly see the primordial oscillations with the given frequencies. We however observe a few notable differences between the imprint of features in these quantities. For small frequencies~$\omega_X$, the signature in the~CMB is comparably similar to the signature in~LSS, but with a sinusoidal oscillation that is slightly distorted. Having said this, the amplitude of the feature contribution decreases significantly in the~CMB for larger frequencies.\footnote{As we illustrated in Fig.~\ref{fig:omegaLinLogComparison}, the finite-volume effects present in galaxy surveys also lead to some suppression of the primordial signal in LSS~observations (cf.\ Fig.~\ref{fig:bandpowerWindowFunctionConstraints} for the resulting impact on the constraints). This suppression is however not shown in Fig.~\ref{fig:cmbLSScomparison} because it is a survey-dependent effect (similar to the beam in CMB~measurements, for instance) that will be less and less important for future LSS~measurements at these frequencies due to their much larger observational volume.} Since this effect is additionally more pronounced in the temperature than in the polarization spectrum, we deduce that it is predominantly the CMB~transfer functions, especially the projection from the three-dimensional cosmic volume to the two-dimensional CMB~sky, that wash out the primordial oscillations.

In the temperature power spectrum, the primordial feature signal becomes suppressed by more than an order of magnitude towards larger frequencies and wavenumbers. Since the Planck~measurement has to overcome this smaller signal in comparison to our BOSS~observations, the constraints turn out to be somewhat worse for larger frequencies despite the more accurate measurement ($\ell\lesssim1600$ is cosmic variance limited~\cite{Aghanim:2015xee}). We note that the slight difference in the employed range of scales in our BOSS~measurement, $\kmax=\SI{0.3}{\hPerMpc}$, compared to $\lmax^T=2508 \approx \SI{0.27}{\hPerMpc}$ can likely be neglected, but will become important for future surveys with a larger reach in wavenumbers.

Finally, it is also evident from Fig.~\ref{fig:cmbLSScomparison} that future CMB~missions will in particular benefit from improved polarization measurements. Apart from the larger signal that survives in the spectrum due to the sharper transfer function compared to temperature, this remaining signature is also partly complementary as can be in particular seen for the highly-oscillating logarithmic features.

\subsection{Joint CMB and LSS Analysis}
\label{app:jointAnalysis}

In the main text, we inferred the first LSS-only constraints on primordial features and compared them to the current bounds from the~CMB as derived above. Having obtained Monte Carlo Markov chains for these observables, we can also consistently combine them to obtain the best current limits. In the following, we elaborate on our computation of these joint constraints.

\vskip4pt
We start by converting the CMB~Markov chains into the same parameter space as the BOSS~analysis. This means that we keep the three feature parameters~$\omega_X$, $\Asin_X$ and~$\Acos_X$, but reduce the six $\Lambda$CDM~parameters to the two isotropic BAO~parameters~$\alpha_z$ evaluated at the effective redshifts of the two BOSS~bins, $z = 0.38\text{ and }0.51$, where 
\beq
\alpha_z = \left(\frac{H^\mathrm{fid}(z)}{H(z)}\right)^{\!1/3}\left(\frac{D_A(z)}{D_A^\mathrm{fid}(z)}\right)^{\!2/3} \frac{r_s^\mathrm{fid}}{r_s}\, ,
\eeq
with the fiducial BOSS~cosmology (see~\textsection\ref{sec:pipeline}). Ideally, we would combine the frequency-binned samples in the four-dimensional space of $\{\alpha_{0.38}, \alpha_{0.51}, \Asin_X, \Acos_X\}$. This is in principle possible by generalizing the approach discussed in Appendix~\ref{app:analysisDetails} for the BOSS~analysis, but a very large number of chain samples would be required to reliably cover this parameter space. Since we are not interested in constraints on the BAO~parameters, we therefore proceed by independently marginalizing the low-$z$ BOSS~chains, the high-$z$ BOSS~chains and the Planck CMB~chains over~$\alpha_z$. Having reduced the parameter space to the two feature amplitudes, we can directly follow our procedure of combining the two BOSS redshift bins as outlined in Appendix~\ref{app:analysisDetails}, but including the CMB data as a third pixelated likelihood. By repeating this for the~TT and the TTTEEE~Markov chains, we obtain the 95\%~confidence limits shown in Fig.~\ref{fig:constraints_joint}.

\vskip4pt
As a consequence of marginalizing over the BAO~parameters, we neglect any possible correlations between~$\alpha_z$ and the feature parameters. We already discussed in Appendix~\ref{app:analysisDetails} that this assumption renders our limits overly conservative, but also checked its impact for the CMB~data. By inferring the four-dimensional (Gaussian) covariance matrix in each frequency bin, we find that the TT-only analysis shows correlations of $|\rho| < 0.5$, while the addition of polarization data further reduces this correlation coefficient. We therefore expect our approximate joint analysis to result in the same bounds as the full analysis except around the frequencies that interfere with the BAO~scale. This is also confirmed using Fisher forecasts that lead to essentially the same forecasted limits except around the scale of the sound horizon where our analysis is suboptimal at the ten-percent level.

Instead of neglecting the correlations with~$\alpha_z$, we could have also assumed the (three-dimensional) almost Gaussian posterior distributions inferred in the BOSS~analysis to be exactly Gaussian. With this approximation, it would be possible to impose the low-$z$ and high-$z$ BOSS~constraints as Gaussian priors on the CMB~analysis by importance sampling its Markov chains.\footnote{Importance sampling the two sets of BOSS~chains with Gaussian CMB~priors and then combining these in the approach of Appendix~\ref{app:analysisDetails} would double-count part of the CMB~information.} We tested this possibility, but found that vanishing $\alpha_z$ correlations are a better assumption than the Gaussian approximation.

\subsection{Forecasts for Future CMB Surveys}
\label{app:futureCMB}

\begin{figure}
	\centering
	\includegraphics{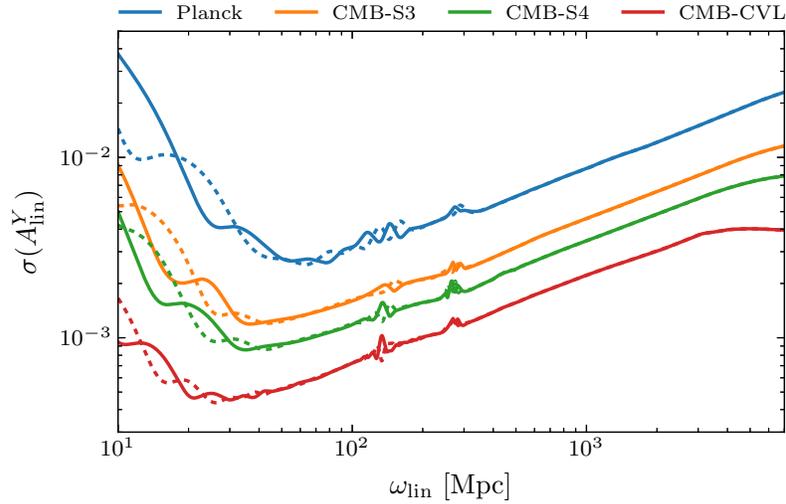}
	\caption{Fisher forecasts of the sensitivity of future CMB~experiments (as defined in~\cite{Baumann:2017gkg}) to the primordial feature amplitudes in the amplitude parametrization, $\Alin^Y$, $Y=\mathrm{sin},\mathrm{cos}$. The constraints on~$\Alinsin$ are shown in solid lines, while the standard deviation~$\sigma(\Alincos)$ is displayed with dashed lines.}
	\label{fig:lin_future_Asin+Acos_cmb}
\end{figure}

In addition to the analysis of current CMB~data from Planck, we also estimate the sensitivity of future CMB~experiments to (linear) feature models in~\textsection\ref{sec:futureConstraints}. (As explained, most other types of features can be decomposed in a basis of linear oscillations so that constraints can be deduced from our results.) These forecasts directly follow the Fisher methodology and the experimental specifications of~\cite{Baumann:2017gkg}. The fiducial point is a featureless $\Lambda$CDM~cosmology consistent with the Planck measurements~\cite{Ade:2015xua, Aghanim:2018eyx}, i.e.\ we in particular take $\Alinsin=\Alincos=0$. Since we compute the constraints as a function of feature frequency~$\omegalin$ within a $\Lambda$CDM~universe, the Fisher information matrices are eight-dimensional. By employing perfectly delensed temperature and polarization power spectra, we infer the most optimistic bounds on~$\Alinsin$ and~$\Alincos$ which we present in Fig.~\ref{fig:lin_future_Asin+Acos_cmb}. %

As can be understood from the additional smoothing of the oscillations in the lensed compared to the unlensed spectra in Fig.~\ref{fig:cmbLSScomparison}, the forecasted sensitivities are worse when using lensed spectra. The degradation of the constraints depends on the experiment and feature frequency, but may be up to about~20\% and~50\% for Planck and the CMB-S3 missions, respectively. However, not delensing the spectra could lead to constraints on the feature amplitudes~$\sigma(\Alin^Y)$, $Y=\sin,\cos$, being worse by a factor of two for CMB-S4 and more for a cosmic-variance-limited experiment. We also observe that the feature parameters are independent of the $\Lambda$CDM~parameters (and of one another) for $\omegalin\gtrsim\SI{300}{\Mpc}$. For smaller frequencies, the primordial oscillations interfere with the baryon acoustic oscillations which in particular leads to a degeneracy with the scale of the sound horizon, as has previously been pointed out in the~CMB (see e.g.~\cite{Meerburg:2013cla, Fergusson:2014tza}) and was discussed in the main text for~LSS. Finally, we note that the Nyquist frequency is much larger in the~CMB since the effective cosmic volume extends all the way back to the last-scattering surface.

\clearpage
\phantomsection
\addcontentsline{toc}{section}{References}
\bibliographystyle{utphys}
\bibliography{references}

\end{document}